\documentclass[11pt]{cernrep}
\usepackage{graphicx}
\usepackage{here}
\usepackage{latexsym}
\usepackage{cite}

\renewcommand{\theequation}{\arabic{section}.\arabic{equation}}
\renewcommand{\thesection}{\arabic{section}}
\renewcommand{\thesubsection}{\arabic{section}.\arabic{subsection}}

\def\Journal#1#2#3#4{#1 {\bf #2} (#3) #4}
\def\refjl#1#2#3#4#5#6{\bibitem{#1} #2, \Journal{#3}{#4}{#5}{#6}.}
\def\refbk#1#2#3#4{\bibitem{#1} #2, #3, #4.}

\def\PRL{Phys. Rev. Lett.}

\def\NP{Nucl. Phys.}

\def\PL{Phys. Lett.}
\def\PR{Phys. Rev.}

\def\ZP{Z. Phys.}

\newcommand{\eqn}[1]{(\ref{#1})}
\newcommand{\be}{\begin{equation}}
\newcommand{\ee}{\end{equation}}
\newcommand{\no}{\nonumber}
\newcommand{\bel}[1]{\be\label{#1}}
\newcommand{\ba}{\begin{array}{c}}
\newcommand{\bat}{\begin{array}{cc}}
\newcommand{\bath}{\begin{array}{ccc}}
\newcommand{\ea}{\end{array}}
\newcommand{\beqn}{\begin{eqnarray}}
\newcommand{\eeqn}{\end{eqnarray}}

\newcommand{\bi}{\begin{itemize}}
\newcommand{\ei}{\end{itemize}}

\newcommand{\rms}{\rm\scriptstyle}

\newcommand{\lsim}{~{}_{\textstyle\sim}^{\textstyle <}~}

\newcommand{\toG}{\stackrel{G}{\,\longrightarrow\,}}

\newcommand{\ssb}{\stackrel{\mbox{\rm\scriptsize SSB}}{\longrightarrow}}
\newcommand{\toU}{\stackrel{\mbox{\rm\scriptsize U(1)}}{\longrightarrow}}

\newcommand{\toTheta}{\stackrel{\theta^i=0}{\longrightarrow}}

\def\gap{\;\lower3pt\hbox{$\buildrel > \over \sim$}\;}
\def\lap{\;\lower3pt\hbox{$\buildrel < \over \sim$}\;}
\newcommand{\cL}{{\cal L}}

\newcommand{\cM}{{\cal M}}
\newcommand{\cO}{{\cal O}}
\newcommand{\cP}{{\cal P}}
\newcommand{\cQ}{{\cal Q}}

\newcommand{\cA}{{\cal A}}
\newcommand{\cI}{{\cal I}}

\newcommand{\cJ}{{\cal J}}
\newcommand{\cC}{{\cal C}}
\newcommand{\cCP}{{\cal CP}}
\newcommand{\cT}{{\cal T}}
\newcommand{\cCPT}{{\cal CPT}}
\def\bV{\mathbf{V}}
\def\bM{\mathbf{M}}
\def\bU{\mathbf{U}}
\def\bS{\mathbf{S}}
\def\bH{\mathbf{H}}
\def\bmd{\mathbf{d}}
\def\bmu{\mathbf{u}}
\def\bml{\mathbf{l}}
\def\bmf{\mathbf{f}}

\newcommand{\e}{\mbox{\rm e}}
\def\CR{\nonumber \\ }
\def\gp{g\hskip .7pt\raisebox{-1.5pt}{$'$}}
\def\dop{d\hskip .6pt'}
\def\up{u\hskip .6pt'}
\def\lp{l\hskip .2pt'}
\def\nup{\nu\hskip .7pt'}
\def\sp{s\hskip .7pt'}
\def\bp{b\hskip .6pt'}
\def\sweff{\sin^2{\theta\hskip .2pt^{\rms lept}_{\rms eff}}}
\begin{document}

%%%%%%%%%% TITLE AND ABSTRACT %%%%%%%%%%

\title{THE \ STANDARD \ MODEL \ OF \ ELECTROWEAK \ INTERACTIONS}
\author{A. Pich}
\institute{Departament de F\'{\i}sica Te\`orica, IFIC,
Universitat de Val\`encia--CSIC,\\
%Edifici d'Instituts d'Investigaci\'o,
Apt. Correus 22085, E-46071 Val\`encia, Spain}
\maketitle
\begin{abstract}
Gauge invariance is a powerful tool to
determine the dynamics of the electroweak and strong forces.
The particle content, structure and symmetries of the
Standard Model Lagrangian are discussed.
Special emphasis is given to the many phenomenological tests
which have established this theoretical framework
as the Standard Theory of electroweak interactions.
\end{abstract}

\renewcommand{\theequation}{\arabic{section}.\arabic{equation}}
%\setcounter{equation}{0}
%\setcounter{section}{0}

%%%%%%%%%% INTRODUCTION %%%%%%%%%%

\section{INTRODUCTION}

The Standard Model (SM) is a gauge theory, based on the symmetry group
$SU(3)_C \otimes SU(2)_L \otimes U(1)_Y$,
which describes strong, weak and electromagnetic interactions,
via the exchange of the corresponding spin--1 gauge fields:
8 massless gluons and 1 massless photon for the strong and
electromagnetic interactions, respectively,
and 3 massive bosons, $W^\pm$ and $Z$, for the weak interaction.
The fermionic matter content is given by the known
leptons and quarks, which are organized in a 3--fold
family structure:
\bel{eq:families}
\left[\bat \nu_e & u \\  e^- & \dop \ea \right] \qquad , \qquad
\left[\bat \nu_\mu & c \\  \mu^- & \sp \ea \right] \qquad , \qquad
\left[\bat \nu_\tau & t \\  \tau^- & \bp \ea \right] \qquad ,
\ee
where (each quark appears in 3 different ``colours'')
\bel{eq:structure}
\left[\bat \nu^{}_l & q^{}_u \\  l^- & q^{}_d \ea \right] \quad\equiv\quad
\left(\ba \nu^{}_l \\ l^- \ea \right)_L\; , \;\,
\left(\ba q^{}_u \\ q^{}_d \ea \right)_L\; , \;\, l^-_R\; ,
\;\, q^{\phantom{j}}_{uR}\; , \;\,
q^{\phantom{j}}_{dR}\; ,
\ee
plus the corresponding antiparticles.
Thus, the left-handed fields are $SU(2)_L$ doublets, while
their right-handed partners transform as $SU(2)_L$ singlets.
The 3 fermionic families in Eq.~\eqn{eq:families} appear
to have identical properties (gauge interactions); they only
differ by their mass and their flavour quantum number.

The gauge symmetry is broken by the vacuum,
which triggers the Spontaneous Symmetry Breaking (SSB)
of the electroweak group to the electromagnetic subgroup:
\bel{eq:ssb}
SU(3)_C \otimes SU(2)_L \otimes U(1)_Y \quad \ssb\quad
SU(3)_C \otimes U(1)_{\mathrm{QED}} \, .
\ee
The SSB mechanism generates the masses of the weak gauge bosons,
and gives rise to the appearance of a physical scalar particle
in the model, the so-called ``Higgs''.
The fermion masses and mixings are also generated through the
SSB.

The SM constitutes one of the most successful achievements
in modern physics. It provides a very elegant theoretical
framework, which is able to describe the
known experimental facts in particle physics with high precision.
These lectures provide an introduction to the
electroweak sector of the SM, i.e. the $SU(2)_L \otimes U(1)_Y$ part
\cite{GL:61,WE:67,SA:69,GIM:70}.
The strong $SU(3)_C$ piece is discussed in more detail in
Refs.~\cite{PI:00,SE:04}.
The power of the gauge principle is shown in Section~2,
where the simpler Lagrangians of Quantum Electrodynamics
and Quantum Chromodynamics are derived.
The electroweak theoretical framework is presented in Sections~3 and 4,
which discuss the gauge structure and the SSB mechanism, respectively.
Section~5 summarizes the present phenomenological status and
shows the main precision tests performed at the $Z$ peak.
The flavour structure is discussed in Section~6, where
the knowledge on the quark mixing angles is briefly reviewed
and the importance of $\cCP$ violation tests is emphasized.
A few comments  on open questions, to be investigated at future facilities,
are finally given in the summary.

Some useful but more technical information has been collected in several
appendixes: a minimal amount of quantum field theory concepts are given
in Appendix~A, Appendix~B summarizes the most important algebraic
properties of $SU(N)$ matrices, and a short discussion on gauge anomalies
is presented in Appendix~C.

%%%%%%%%%% GAUGE %%%%%%%%%%

\setcounter{equation}{0}
\section{GAUGE \ INVARIANCE}
\label{sec:gauge}
\subsection{Quantum Electrodynamics}
\label{sec:qed}

Let us consider the Lagrangian describing a free Dirac fermion:
\bel{eq:l_free}
\cL_0\, =\, i \,\overline{\psi}(x)\gamma^\mu\partial_\mu\psi(x)
\, - \, m\, \overline{\psi}(x)\psi(x) \, .
\ee
$\cL_0$ is invariant under {\em global} \
$U(1)$ transformations
\bel{eq:global}
\psi(x) \quad\toU\quad \psi'(x)\,\equiv\,\exp{\{i Q \theta\}}\,\psi(x) \, ,
\ee
where $Q\theta$ is an arbitrary real constant.
The phase of $\psi(x)$ is then a pure convention--dependent
quantity without physical meaning.
However,
the free Lagrangian is no longer invariant if one allows
the phase transformation to depend on the space--time coordinate,
i.e. under {\em local} phase redefinitions $\theta=\theta(x)$,
because
\bel{eq:local}
\partial_\mu\psi(x) \quad\toU\quad \exp{\{i Q \theta\}}\;
\left(\partial_\mu + i Q \,\partial_\mu\theta\right)\,
\psi(x) \, .
\ee
Thus, once a given phase convention has been adopted
at the reference point $x_0$, the same convention must
be taken at all space--time points. This looks very unnatural.

The ``Gauge Principle'' is the requirement that the $U(1)$
phase invariance should hold {\em locally}.
This is only possible if one adds some additional piece to the
Lagrangian, transforming in such a way  as to cancel the
$\partial_\mu\theta$ term in Eq.~\eqn{eq:local}.
The needed modification is completely fixed by the transformation 
\eqn{eq:local}: one introduces a new spin-1
(since $\partial_\mu\theta$  has a Lorentz index)
field $A_\mu(x)$, transforming as
\bel{eq:a_transf}
A_\mu(x)\quad\toU\quad A_\mu'(x)\,\equiv\, A_\mu(x) + {1\over e}\,
\partial_\mu\theta\, ,
\ee
and defines the covariant derivative
\bel{eq:d_covariant}
D_\mu\psi(x)\,\equiv\,\left[\partial_\mu-ieQA_\mu(x)\right]
\,\psi(x)\, ,
\ee
which has the required property of transforming like the field itself:
\bel{eq:d_transf}
D_\mu\psi(x)\quad\toU\quad\left(D_\mu\psi\right)'(x)\,\equiv\,
\exp{\{i Q \theta\}}\,D_\mu\psi(x)\,.
\ee
The Lagrangian
\bel{eq:l_new}
\cL\,\equiv\,
i \,\overline{\psi}(x)\gamma^\mu D_\mu\psi(x)
\, - \, m\, \overline{\psi}(x)\psi(x)
\, =\, \cL_0\, +\, e Q A_\mu(x)\, \overline{\psi}(x)\gamma^\mu\psi(x)
\ee
is then invariant under local $U(1)$ transformations.

The gauge principle has generated an interaction
between the Dirac spinor and the gauge field $A_\mu$,
which is nothing else than the familiar vertex of
Quantum Electrodynamics (QED).
Note that the corresponding electromagnetic
charge $e Q$ is completely arbitrary.
If one wants $A_\mu$ to be a true propagating field, one needs to add
a gauge--invariant kinetic term
\bel{eq:l_kinetic}
\cL_{\rms Kin}\,\equiv\, -{1\over 4}\, F_{\mu\nu}(x)\, F^{\mu\nu}(x)\,,
\ee
where
$F_{\mu\nu}\,\equiv\, \partial_\mu A_\nu -\partial_\nu A_\mu$
is the usual electromagnetic field strength.
A possible mass term for the gauge field, 
$\cL_m = {1\over 2}m^2A^\mu A_\mu$, is forbidden because it would violate
gauge invariance; therefore,
the photon field is predicted to be massless.
Experimentally, we know that $m_\gamma < 6\cdot 10^{-17}$~eV \cite{PDG:04}.

The total Lagrangian in \eqn{eq:l_new} and \eqn{eq:l_kinetic}
gives rise to the well-known Maxwell equations:
\bel{eq:Maxwell_QED}
\partial_\mu F^{\mu\nu} = J^\nu\, ,
\qquad\qquad
J^\nu = - e Q\, \overline{\psi}\gamma^\nu\psi\, ,
\ee
where $J^\nu$ is the fermion electromagnetic current.
From a simple gauge--symmetry requirement, we have deduced
the right QED Lagrangian, which leads to a very  successful
quantum field theory.

\subsubsection{Lepton anomalous magnetic moments}
\label{sec:g-2}

%%%%%%%%%%%%%%%%%%%%%%%%%%%%%%%%%%%%%%%%%%%%%%
%%
%%   FIGURE Anomalous Magnetic Moment
%%
\begin{figure}[tbh]
\begin{center}
\includegraphics[width=9.5cm]{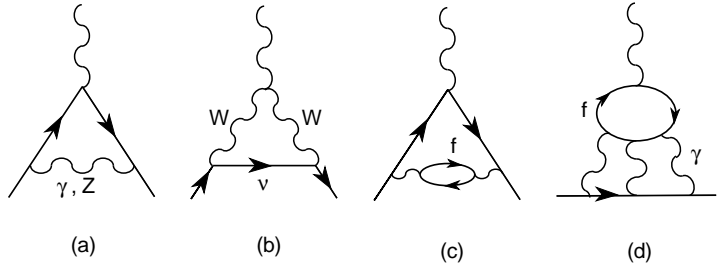}
\caption{Feynman diagrams contributing to the lepton anomalous magnetic moment.}
\label{fig:AnMagMom}
\end{center}
\end{figure}
%%%%%%%%%%%%%%%%%%%%%%%%%%%%%%%%%%%%%%%%%%%%%%%

The most stringent QED test comes from the high-precision
measurements of the $e$ and $\mu$
anomalous magnetic moments \ $a_l\equiv (g^\gamma_l-2)/2\, $,
where \ $\vec\mu_l\equiv g^\gamma_l \,(e/2m_l)\, \vec{S}_l$ \ \cite{PDG:04,BNL:04}:
\bel{eq:a_mu}
a_e = (115 \, 965 \, 218.59\pm 0.38) \,\cdot\, 10^{-11} \, ,
\qquad\qquad
a_\mu = (116 \, 592 \, 080\pm 60) \,\cdot\, 10^{-11} \, .  %%% BNL04
%%% a_\mu = (116 \, 592 \, 030\pm 80) \,\cdot\, 10^{-11} \, .  PDG04
\ee

To a measurable level, $a_e$ arises entirely from virtual electrons and
photons; these contributions are known to $O(\alpha^4)$
\cite{KI:96,KN:02,DM:04}.
The impressive agreement achieved between theory and experiment
has promoted QED to the level of the best theory ever built 
to describe nature.
The theoretical error is dominated by the uncertainty in the
input value of the QED coupling $\alpha\equiv e^2/(4\pi)$.
Turning things around, $a_e$ provides the most accurate
determination of the fine structure constant.
The latest CODATA recommended value has a precision
of $3.3 \times 10^{-9}$ \cite{PDG:04}:
\be
%\alpha^{-1} = 137.035\, 998\, 75 \,\pm\, 0.000\, 000\, 52  %%% Before PDG04
\alpha^{-1} = 137.035\, 999\, 11 \,\pm\, 0.000\, 000\, 46 \, .
\ee

The anomalous magnetic moment of the muon is sensitive to small
corrections from virtual heavier states; compared to $a_e$, they scale
with the mass ratio $m_\mu^2/m_e^2$.
Electroweak effects from virtual $W^\pm$ and $Z$ bosons amount to
a contribution of $(154 \pm 2)\cdot 10^{-11}$
\cite{DM:04},           %%%\cite{CM:01,a_mu_EW}.
which is larger than the present experimental precision.
Thus, $a_\mu$ allows to test the entire SM.
The main theoretical uncertainty comes from strong interactions.
Since quarks have electric charge, virtual quark-antiquark pairs
induce {\it hadronic vacuum polarization} corrections to the
photon propagator (Fig.~\ref{fig:AnMagMom}.c).
Owing to the non-perturbative character of the strong interaction
at low energies,
the light-quark contribution cannot be reliably calculated at
present. This effect can be
extracted from the measurement of the cross-section
$\sigma(e^+e^-\to \mbox{\rm hadrons})$
and from the invariant-mass distribution of the final hadrons in
$\tau$ decays, which unfortunately provide slightly different results
\cite{DEHZ:03,HO:04}:
\bel{eq:QED_pred}
a_\mu^{\mathrm{th}} \, = \,\left\{ \bat
%%% (116\; 591 \; 961\;\pm\; 80) \,\cdot\, 10^{-11} \, ,
(116\; 591 \; 828\;\pm\; 73) \,\cdot\, 10^{-11}
&\qquad (e^+e^-\quad\mathrm{data})\, , \\
(116\; 592 \; 004\;\pm\; 68) \,\cdot\, 10^{-11}
&\qquad (\tau\quad\mathrm{data})\, .
\ea\right.
\ee
The quoted uncertainties include also the smaller
{\it light-by-light scattering} contributions (Fig.~\ref{fig:AnMagMom}.d).
The difference between the SM prediction and
the experimental value \eqn{eq:a_mu} corresponds to $2.7\,\sigma$  ($e^+e^-$)
\ or \ $0.9\,\sigma$ \ ($\tau$).
New precise $e^+e^-$ and $\tau$ data sets are needed to settle the
true value of $a_\mu^{\mathrm{th}}$.

\subsection{Quantum Chromodynamics}
\label{sec:QCDlagrangian}

\subsubsection{Quarks and colour}

%%%%%%%%%%%%%%%%%%%%%%%%%%%%%%%%%%%%%%%%%%%%%%
%%
%%   FIGURE ee --> qq
%%
\begin{figure}[htb]
\begin{center}
\includegraphics[width=5cm]{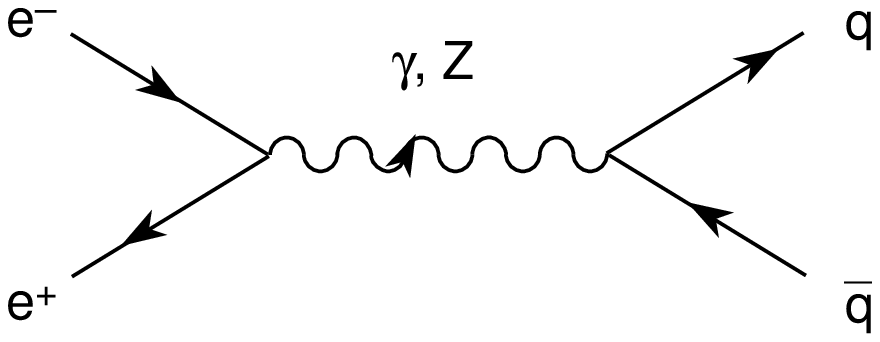}
\caption{Tree-level Feynman diagram for the \
$e^+e^-$ annihilation into hadrons.}
\label{fig:eeqqhad}
\end{center}
\end{figure}
%%%%%%%%%%%%%%%%%%%%%%%%%%%%%%%%%%%%%%%%%%%%%%%

The large number of known mesonic and baryonic states clearly signals the
existence of a deeper level of elementary constituents of matter:
{\it quarks}. Assuming that mesons are \ $M\equiv q\bar q$ \
states, while baryons have
three quark constituents, $B\equiv qqq$, one can nicely classify
the entire hadronic spectrum. However, in order to satisfy the
Fermi--Dirac statistics one needs to assume the existence of a
new quantum number, {\it colour},
such that each species of quark may have $N_C=3$
different colours: $q^\alpha$, $\alpha =1,2,3$ (red, green, blue).
Baryons and mesons are then described by
the colour--singlet combinations
\be\label{eq:m_b_wf}
B\, =\, {1\over\sqrt{6}}\,\epsilon^{\alpha\beta\gamma}\,
|q_\alpha q_\beta q_\gamma\rangle \, ,
\qquad\qquad
M\, =\, {1\over\sqrt{3}}\,\delta^{\alpha\beta} \,
|q_\alpha \bar q_\beta \rangle \, .
\ee
In order to avoid the existence of non-observed extra states
with non-zero colour,
one needs to further postulate that all asymptotic states
are colourless, i.e. singlets under rotations in colour space.
This assumption is known as the {\it confinement hypothesis},
because it implies the non-observability of free quarks:
since quarks carry colour they are confined within
colour--singlet bound states.

A direct test of the colour quantum number can be obtained from the
ratio
\be\label{eq:R_ee}
R_{e^+e^-} \;\equiv\;
{\sigma(e^+e^-\to \mbox{\rm hadrons})\over\sigma(e^+e^-\to\mu^+\mu^-)} \, .
\ee
The hadronic production occurs through
$e^+e^-\to\gamma^*, Z^*\to q\bar q\to \mbox{\rm hadrons}$.
Since quarks are assumed
to be confined, the probability to hadronize is just one; therefore,
summing over all possible quarks in the final state,
we can estimate the inclusive cross-section into hadrons.
The electroweak production factors which are common with the
$e^+e^-\to\gamma^*, Z^*\to\mu^+\mu^-$ process cancel in the ratio
\eqn{eq:R_ee}.
At energies well below the $Z$ peak, the cross-section is
dominated by the $\gamma$--exchange amplitude;
the ratio $R_{e^+e^-}$ is then given by the sum of the
quark electric charges squared:
\be\label{eq:R_ee_res}
R_{e^+e^-} \,\approx N_C \; \sum_{f=1}^{N_f} Q_f^2 \; = \;
\left\{
\begin{array}{cc}
\frac{2}{3} N_C = 2\, , \qquad & (N_f=3 \; :\; u,d,s)  \\[5pt]
\frac{10}{9} N_C = \frac{10}{3}\, , \qquad & (N_f=4 \; :\; u,d,s,c)  \\[5pt]
\frac{11}{9} N_C = \frac{11}{3} \, ,\qquad & (N_f=5 \; :\; u,d,s,c,b)
\ea\right. .
\ee
%

%%%%%%%%%%%%%%%%%%%%%%%%%%%%%%%%%%%%%%%%%%%%%%
%%
%%   FIGURE Ree
%%
\begin{figure}[bth]
\begin{center}
\includegraphics[width=12cm]{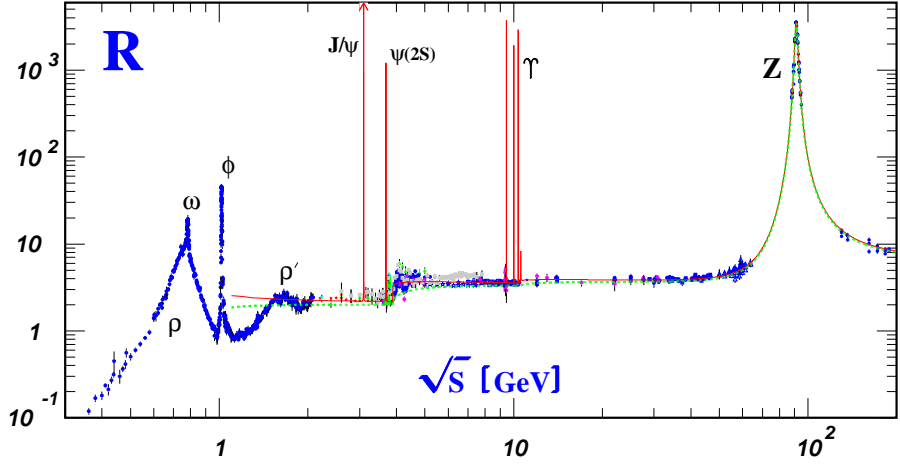}
\caption{World data on the ratio $R_{e^+e^-}$ \protect\cite{PDG:04}.
The broken lines show the naive quark model approximation with $N_C=3$.
The solid curve is the 3-loop perturbative QCD prediction.}
\label{fig:Ree}
\end{center}
\end{figure}
%%%%%%%%%%%%%%%%%%%%%%%%%%%%%%%%%%%%%%%%%%%%%%%

The measured ratio is shown in Fig. \ref{fig:Ree}. Although
the simple
formula \eqn{eq:R_ee_res} cannot explain the complicated
structure around the different quark thresholds,
it gives the right average
value of the cross-section (away from thresholds),
provided that $N_C$ is taken to be three.
The agreement is better at larger energies.
Notice that strong
interactions have not been taken into account;
only the confinement hypothesis has been used.

Electromagnetic interactions are associated with the fermion electric
charges, while the quark flavours (up, down, strange, charm, bottom, top)
are related to electroweak phenomena. The strong forces are
flavour conserving and flavour independent.
On the other side, the carriers of the electroweak interaction
($\gamma$, $Z$, $W^\pm$) do not couple to the quark colour. Thus, it seems
natural to take colour as the charge associated with the strong forces
and try to build a quantum field theory based on it \cite{FGL:73}.

\subsubsection{Non-abelian gauge symmetry}

Let us denote $q^\alpha_f$ a quark field of colour $\alpha$ and flavour $f$.
To simplify the equations, let us adopt a vector notation
in colour space:
$q_f^T \,\equiv\, (q^1_f\, ,\, q^2_f\, ,\, q^3_f) $.
The free Lagrangian
\bel{eq:L_free}
\cL_0\, =\, \sum_f\;\bar q_f \, \left( i\gamma^\mu\partial_\mu - m_f\right) q_f
\ee
is invariant under arbitrary global $SU(3)_C$ transformations in colour
space,
\bel{eq:q_transf}
q^\alpha_f \;\longrightarrow\;
(q^\alpha_f)'\, =\, U^\alpha_{\phantom{\alpha}\beta}\; q^\beta_f \; ,
\qquad\qquad
U\, U^\dagger\, =\, U^\dagger U\, =\, 1 \; , \qquad\qquad
\det U\, = 1\, \; .
\ee
The $SU(3)_C$ matrices can be written in the form
\bel{eq:U_def}
U\, =\, \exp\left\{ i\, {\lambda^a\over 2}\,\theta_a\right\} \; ,
\ee
where $\frac{1}{2}\,\lambda^a$ ($a=1,2,\ldots,8$) denote the
generators of the fundamental representation of the
$SU(3)_C$ algebra,
and $\theta_a$ are arbitrary parameters. The matrices $\lambda^a$
are traceless and
satisfy the commutation relations
\bel{eq:commutation}
\left[\, {\lambda^a\over 2}\, ,\, {\lambda^b\over 2}\,\right]\, =\,
i\, f^{abc} \; {\lambda^c\over 2} \; ,
% \left[\lambda^a,\lambda^b\right] =
% 2 i f^{abc} \, \lambda^c \, ,
\ee
with $f^{abc}$ the $SU(3)_C$
structure constants, which are real and totally antisymmetric.
Some useful properties of $SU(3)$ matrices are collected in Appendix~B.

As in the QED case, we can now require the Lagrangian to be also invariant
under {\it local} $SU(3)_C$ transformations, $\theta_a = \theta_a(x)$. To
satisfy this requirement, we need to change the quark derivatives by covariant
objects. Since we have now eight independent gauge parameters, eight different
gauge bosons $G^\mu_a(x)$, the so-called {\it gluons}, are needed:
\bel{eq:D_cov}
D^\mu q_f \,\equiv\, \left[ \partial^\mu - i g_s\, {\lambda^a\over 2}\,
G^\mu_a(x)\right]
  \, q_f \, \equiv\, \left[ \partial^\mu - i g_s\, G^\mu(x)\right] \, q_f \, .
\ee
Notice that we have introduced the compact matrix notation
\bel{eq:G_matrix}
  [G^\mu(x)]_{\alpha\beta}\,\equiv\, 
\left({\lambda^a\over 2}\right)_{\!\alpha\beta}\, G^\mu_a(x) \, .
\ee
%
%%%%%%%%%%%%%%%%%%%%%%%%%%%%%%%%%%%%%%%%%%%%%%
%%
%%   FIGURE QCD vertices
%%
\begin{figure}[tbh]
\begin{center}
\mbox{}\vskip .3cm
\includegraphics[width=12cm]{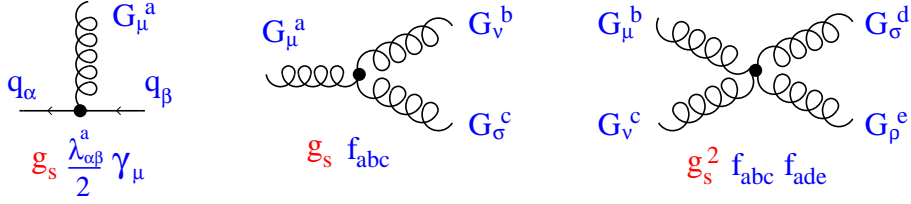}
\caption{Interaction vertices of the QCD Lagrangian.}
\label{fig:Lqcd}
\end{center}
\end{figure}
%%%%%%%%%%%%%%%%%%%%%%%%%%%%%%%%%%%%%%%%%%%%%%%
%
We want $D^\mu q_f$ to transform in exactly the same way as the
colour--vector $q_f$;
this fixes the transformation properties of the gauge fields:
\bel{eq:G_trans}
D^\mu \;\longrightarrow\; (D^\mu)'\, =\, U\, D^\mu\, U^\dagger \, ,
\qquad\qquad
G^\mu \;\longrightarrow\; (G^\mu)'\, =\, U\, G^\mu\, U^\dagger
  -{i\over g_s} \, (\partial^\mu U) \, U^\dagger \, .
\ee
Under an infinitesimal $SU(3)_C$ transformation,
\beqn\label{eq:inf_transf}
q^\alpha_f &\longrightarrow & (q^\alpha_f)'\, =\,
q^\alpha_f\, +\, i \,\left({\lambda^a\over 2}\right)_{\!\alpha\beta}
\,\delta\theta_a\; q^\beta_f \, ,
\no\\
G^\mu_a &\longrightarrow & (G^\mu_a)'\, =\,
G^\mu_a\, +\, {1\over g_s}\,\partial^\mu(\delta\theta_a)
\, \,- f^{abc} \,\delta\theta_b \, G^\mu_c \, .
\eeqn
The gauge transformation of the gluon fields is more complicated than the one
obtained in QED for the photon.
The non-commutativity of the $SU(3)_C$ matrices gives rise to an additional term
involving the gluon fields themselves.
For constant $\delta\theta_a$, the transformation rule for the gauge fields
is expressed in terms of the structure constants $f^{abc}$; thus,
the gluon fields belong to the adjoint representation of the colour group
(see Appendix~B).
Note also that there is a unique $SU(3)_C$
coupling $g_s$. In QED it was possible to assign arbitrary electromagnetic
charges to the
different fermions. Since the commutation relation \eqn{eq:commutation} is
non-linear, this freedom does not exist for $SU(3)_C$.

To build a gauge--invariant kinetic term for the gluon fields, we introduce
the corresponding field strengths:
\beqn\label{eq:G_tensor}
G^{\mu\nu}(x) & \equiv & {i\over g_s}\, [D^\mu, D^\nu] \, = \,
  \partial^\mu G^\nu - \partial^\nu G^\mu - i g_s\, [G^\mu, G^\nu] \, \equiv \,
  {\lambda^a\over 2}\, G^{\mu\nu}_a(x) \, ,
\no\\
G^{\mu\nu}_a(x) & = & \partial^\mu G^\nu_a - \partial^\nu G^\mu_a
  + g_s\, f^{abc}\, G^\mu_b\, G^\nu_c \, .
\eeqn
Under a gauge transformation,
\bel{eq:G_tensor_transf}
G^{\mu\nu}\;\longrightarrow\; (G^{\mu\nu})'\, =\, U\, G^{\mu\nu}\, U^\dagger \, ,
\ee
and the colour trace \
Tr$(G^{\mu\nu}G_{\mu\nu}) = \frac{1}{2}\, G^{\mu\nu}_aG_{\mu\nu}^a$ \
remains invariant.

Taking the proper normalization for the gluon kinetic term, we finally have
the $SU(3)_C$ invariant Lagrangian of Quantum Chromodynamics (QCD):
\bel{eq:L_QCD}
\cL_{\rms QCD} \,\equiv\, -{1\over 4}\, G^{\mu\nu}_aG_{\mu\nu}^a
\, +\, \sum_f\;\bar q_f \, \left( i\gamma^\mu D_\mu - m_f\right)\, q_f \, .
\ee
It is worthwhile to decompose the Lagrangian into its different pieces:
\beqn\label{eq:L_QCD_pieces}
\cL_{\rms QCD} & = &
 -\, {1\over 4}\, (\partial^\mu G^\nu_a - \partial^\nu G^\mu_a)\,
  (\partial_\mu G_\nu^a - \partial_\nu G_\mu^a)
\, +\, \sum_f\;\bar q^\alpha_f \, \left( i\gamma^\mu\partial_\mu - m_f\right)\,
q^\alpha_f
\qquad\no\\ && \mbox{}
 +\, g_s\, G^\mu_a\,\sum_f\;\bar q^\alpha_f\, \gamma_\mu\,
 \left({\lambda^a\over 2}\right)_{\!\alpha\beta}\, q^\beta_f
\\ && \mbox{}
 -\, {g_s\over 2}\, f^{abc}\, (\partial^\mu G^\nu_a - \partial^\nu G^\mu_a) \,
  G_\mu^b\, G_\nu^c \, - \,
{g_s^2\over 4} \, f^{abc} f_{ade} \, G^\mu_b\, G^\nu_c\, G_\mu^d\, G_\nu^e \, .
\no
\eeqn
The first line contains the correct kinetic terms for the different fields,
which
give rise to the corresponding propagators. The colour interaction between
quarks and gluons is given by the second line; it involves the $SU(3)_C$ 
matrices
$\lambda^a$. Finally, owing to the non-abelian character of the colour group,
the $G^{\mu\nu}_aG_{\mu\nu}^a$ term generates the cubic and quartic gluon
self-interactions shown in the last line; 
the strength of these interactions is given by the same coupling $g_s$ which
appears in the fermionic piece of the Lagrangian.

%%%%%%%%%%%%%%%  FIGURE %%%%%%%%%%%%%%%%%%%%%%%%%
\begin{figure}[t]\centering
\begin{minipage}[t]{.45\linewidth}\centering
\includegraphics[width=7cm,clip]{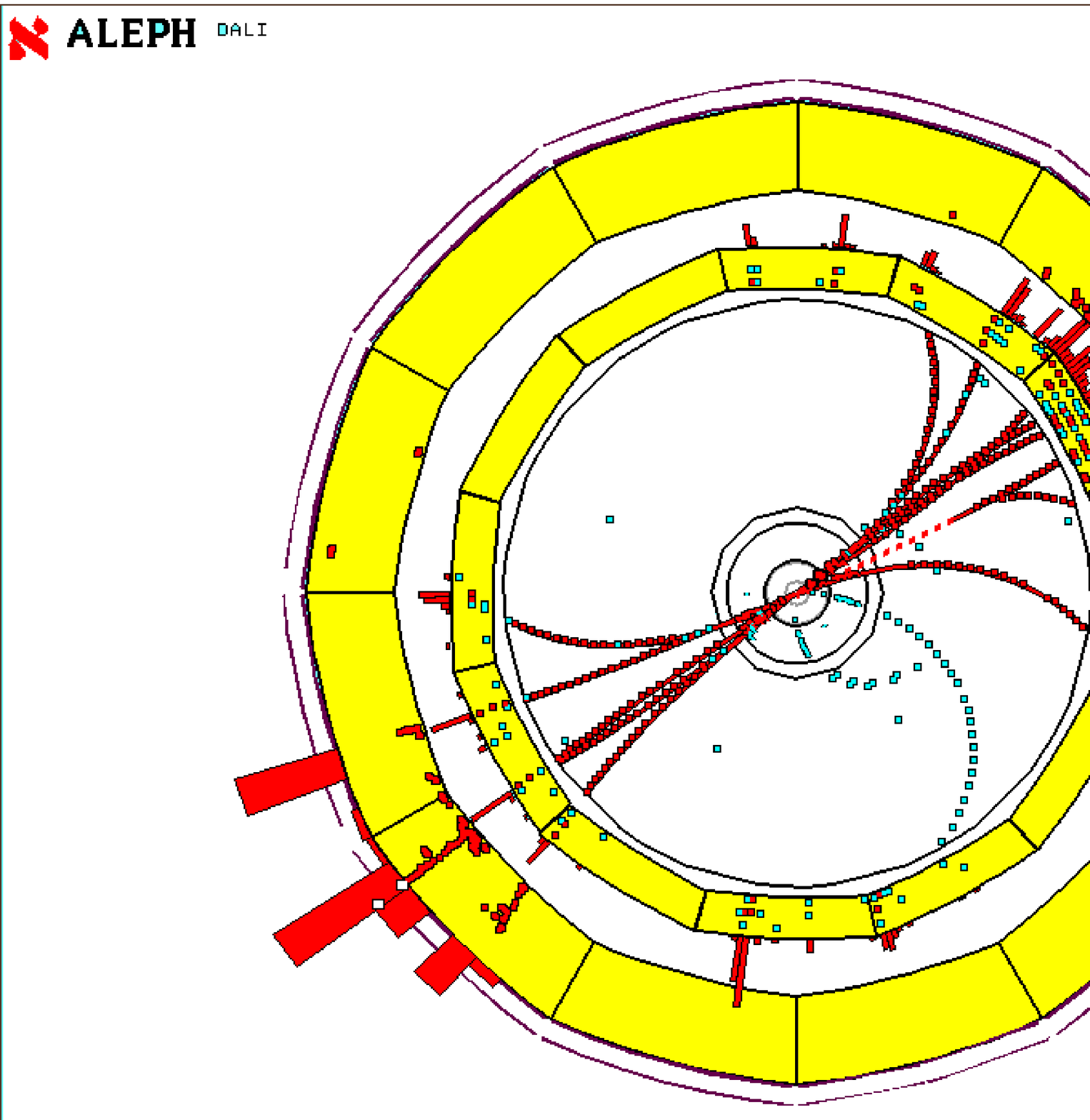} %%%2jetmod.eps
\end{minipage}
\hskip .75cm
\begin{minipage}[t]{.45\linewidth}\centering
\includegraphics[width=7cm,clip]{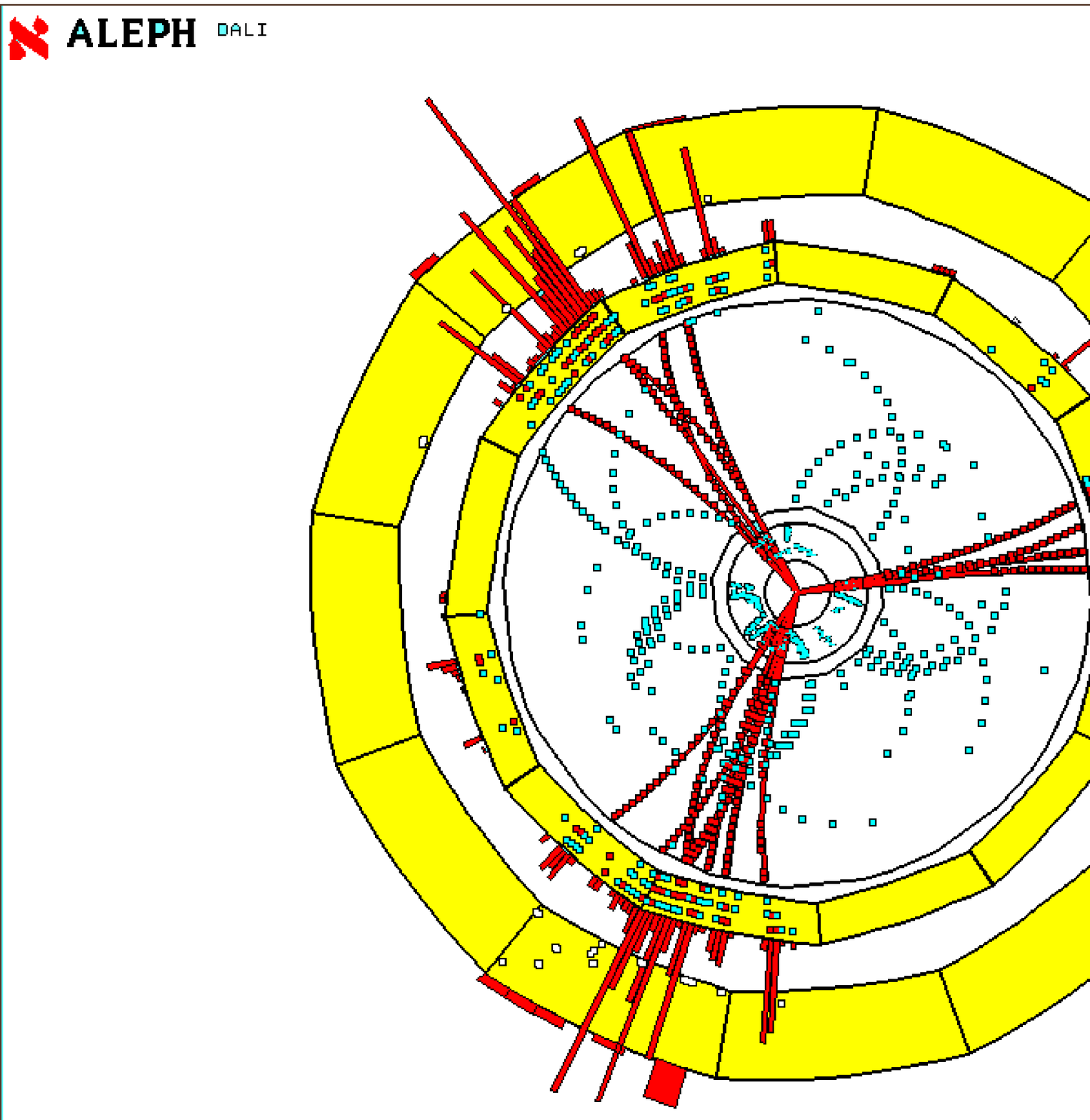} %%%3jetmod.eps
\end{minipage}
\caption{Two- and three-jet events from the hadronic $Z$ boson decays \
$Z\to q\bar q$ \ and \ $Z\to  q\bar q G$ \ (ALEPH).}
\label{fig:ThreeJets}
\end{figure}
%%%%%%%%%%%%%%%%%%%%%%%%%%%%%%%%%%%%%%%%%%%%%

In spite of the rich physics contained in it, the Lagrangian \eqn{eq:L_QCD}
looks very simple because of its colour symmetry properties.
All interactions are given in terms of a single universal coupling $g_s$,
which is called the {\it strong coupling constant}.
The existence of self-interactions among the gauge fields is a new feature
that was not present in QED; it seems then reasonable to expect that
these gauge self-interactions could explain properties
like asymptotic freedom (strong interactions become weaker at short distances)
and confinement (the strong forces increase at large distances),
which do not appear in QED \cite{PI:00,SE:04}.

Without any detailed calculation, one can already extract qualitative physical
consequences from $\cL_{\rms QCD}$.
Quarks can emit gluons. At lowest order
in $g_s$, the dominant process will be the emission of a single gauge boson.
Thus, the hadronic decay of the $Z$ should result in some $Z\to q\bar q G$ 
events, in addition to the dominant $Z\to q\bar q$ decays.
%in Section~\ref{sec:introduction}.
Figure~\ref{fig:ThreeJets} clearly shows that 3-jet events, with the required
kinematics, indeed appear in the LEP data. Similar events show up in
$e^+e^-$ annihilation into hadrons, away from the $Z$ peak.
The ratio between 3-jet and 2-jet events provides a simple estimate of
the strength of the strong interaction at LEP energies ($s=M_Z^2$):
$\alpha_s\equiv g_s^2/(4\pi) \sim 0.12$.

%%%%%%%%%% UNIFICATION %%%%%%%%%%

\setcounter{equation}{0}
\section{ELECTROWEAK \ UNIFICATION}
\label{sec:EWmodel}

\subsection{Experimental facts}

Low-energy experiments have provided a large amount of information
about the dynamics underlying flavour-changing processes. The
detailed analysis of the energy and angular distributions in $\beta$
decays, such as \ $\mu^-\to e^-\bar\nu_e\,\nu_\mu$ \ or \
$n\to p\, e^-\bar\nu_e\,$, made clear that only the left-handed
(right-handed) fermion (antifermion)
chiralities participate in those weak transitions. Moreover, the
strength of the interaction appears to be universal. This is
further corroborated through the study of other processes
like \ $\pi^-\to e^-\bar\nu_e$ \ or \ $\pi^-\to \mu^-\bar\nu_\mu\,$,
which moreover show that neutrinos have left-handed chiralities
while anti-neutrinos are right-handed.

From neutrino scattering data, we learnt the existence of different
neutrino types ($\nu_e\not=\nu_\mu$) and that there are separately conserved
lepton quantum numbers which distinguish neutrinos from antineutrinos.
Thus, we observe the transitions \
$\bar\nu_e\, p\to e^+ n\,$, \ $\nu_e\, n\to e^- p\,$, \
$\bar\nu_\mu\, p\to \mu^+ n$ \ or \ $\nu_\mu\, n\to \mu^- p\,$,
but we do not see processes like \
$\nu_e\, p\not\to e^+ n\,$, \ $\bar\nu_e\, n\not\to e^- p\,$, \
$\bar\nu_\mu\, p\not\to e^+ n$ \ or \ $\nu_\mu\, n\not\to e^- p\,$.

Together with theoretical considerations related to unitarity
(a proper high-energy behaviour) and the absence of flavour-changing
neutral-current transitions ($\mu^-\not\to e^-e^-e^+$),
the low-energy information was good enough to
determine the structure of the modern electroweak theory \cite{PI:94}.
The intermediate vector bosons $W^\pm$ and $Z$ were theoretically
introduced and their masses correctly estimated, before their
experimental discovery.
Nowadays, we have accumulated huge numbers of $W^\pm$ and $Z$
decay events, which bring a much direct experimental evidence
of their dynamical properties.

\subsubsection{Charged currents}

%%%%%%%%%%%%%%%  FIGURE %%%%%%%%%%%%%%%%%%%%%%%%%
\begin{figure}[tbh]\centering
\begin{minipage}[t]{.4\linewidth}\centering
\includegraphics[width=6cm]{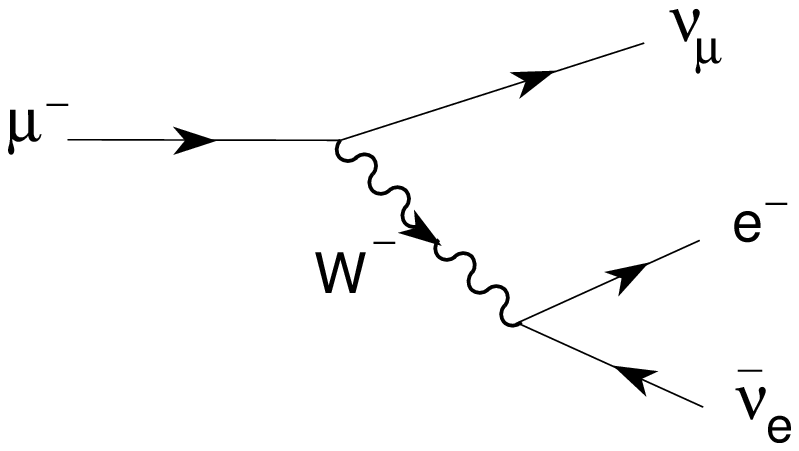}
\end{minipage}
\hskip .5cm
\begin{minipage}[t]{.4\linewidth}\centering
\includegraphics[width=5cm]{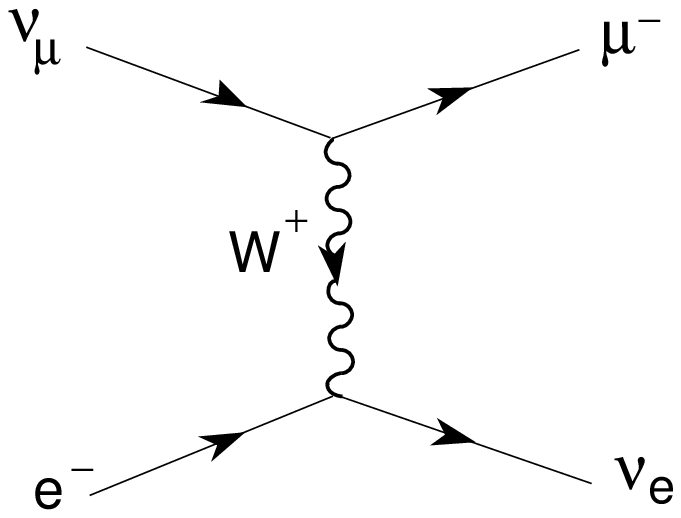}
\end{minipage}
\caption{Tree-level Feynman diagrams for \ $\mu^-\to e^-\bar\nu_e\,\nu_\mu$ \
and \ $\nu_\mu\, e^-\to\mu^-\nu_e$.}
\label{fig:ChargedCurrents}
\end{figure}
%%%%%%%%%%%%%%%%%%%%%%%%%%%%%%%%%%%%%%%%%%%%%

\noindent
The interaction of quarks and leptons with the $W^\pm$ bosons
exhibits the following features:

\bi

\item Only left-handed fermions and right-handed antifermions couple to the
$W^\pm$. Therefore, there is a 100\% breaking of
parity $\cP$ (left $\leftrightarrow$ right) and charge conjugation $\cC$
(particle $\leftrightarrow$ antiparticle).
However, the combined transformation \ $\cC\cP$ \ is still a good
symmetry.

\item The $W^\pm$ bosons couple to the fermionic doublets in \eqn{eq:families},
where the electric charges of the two fermion partners differ in one unit.
The decay channels of the $W^-$ are then:
\be
W^-\,\to\, e^-\bar\nu_e\, ,\, \mu^-\bar\nu_\mu\, ,\,\tau^-\bar\nu_\tau\, ,\,
d\,'\,\bar u\, ,\, s\,'\,\bar c\, .
\ee
Owing to the very high mass of the top quark,
$m_t = 178~\mathrm{GeV} > M_W = 80.4~\mathrm{GeV}$,
its on-shell production through \ $W^-\to  b\,'\, \bar t$ \ is
kinematically forbidden.

\item All fermion doublets couple to the $W^\pm$ bosons with the same
universal strength.

\item The doublet partners of the up, charm and top quarks appear to be
mixtures of the three charge $-\frac{1}{3}$ quarks:
\be
\left(\ba d\,'\\ s\,'\\ b\,'\ea\right) \, = \, \bV\;
\left(\ba d\\ s\\ b\ea\right)\; ,
\qquad\qquad
\bV\,\bV^\dagger\, =\,\bV^\dagger\, \bV\, =\, 1\, .
\ee
Thus, the weak eigenstates $d\,'\, ,\, s\,'\, ,\, b\,'\,$ are different than
the mass eigenstates $d\, ,\, s\, ,\, b\,$. They are related through the
$3\times 3$ unitary matrix $\bV$,
which characterizes flavour-mixing phenomena.

\item The experimental evidence of neutrino oscillations shows that $\nu_e$,
$\nu_\mu$ and $\nu_\tau$ are also mixtures of mass eigenstates. However,
the neutrino masses are tiny: \
$m^2_{\nu_3} - m^2_{\nu_2}\sim 3\cdot 10^{-3}\, \mathrm{eV}^2\,$, \
$m^2_{\nu_2} - m^2_{\nu_1}\sim 8\cdot 10^{-5}\, \mathrm{eV}^2\,$ \cite{GO:04}.

\ei

\subsubsection{Neutral currents}

%%%%%%%%%%%%%%%  FIGURE %%%%%%%%%%%%%%%%%%%%%%%%%
\begin{figure}[t]\centering
\begin{minipage}[t]{.4\linewidth}\centering
\includegraphics[width=5.5cm]{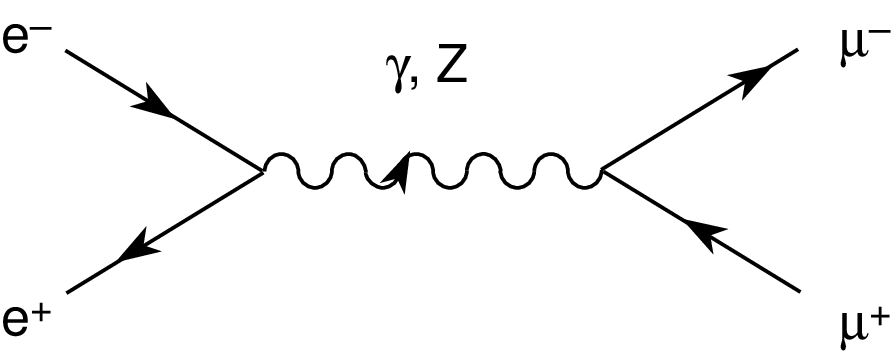}
\end{minipage}
\hskip 1cm
\begin{minipage}[t]{.4\linewidth}\centering
\includegraphics[width=5.5cm]{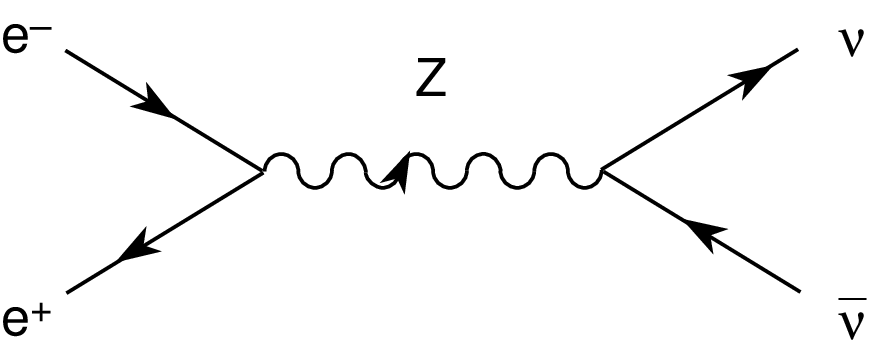}
\end{minipage}
\caption{Tree-level Feynman diagrams for \ $e^+ e^-\to\mu^+\mu^-$ \
and \ $e^+ e^-\to\nu\,\bar\nu$.}
\label{fig:NeutralCurrents}
\end{figure}
%%%%%%%%%%%%%%%%%%%%%%%%%%%%%%%%%%%%%%%%%%%%%

\noindent
The neutral carriers of the electromagnetic and weak interactions
have fermionic couplings with the following properties:

\bi

\item All interacting vertices are flavour conserving. Both the $\gamma$
and the $Z$ couple to a fermion and its own antifermion, i.e. \
$\gamma\, f\,\bar f$ \ and \ $Z\, f\,\bar f$. Transitions of the type \
$\mu\not\to e\gamma$ \ or \ $Z\not\to e^\pm\mu^\mp$ \ have never been observed.

\item The interactions depend on the fermion electric charge $Q_f$.
Fermions with the same $Q_f$ have exactly the same universal couplings.
Neutrinos do not have electromagnetic interactions ($Q_\nu = 0$),
but they have a non-zero coupling to the $Z$ boson.

\item Photons have the same interaction for both fermion chiralities, but
the $Z$ couplings are different for left-handed and right-handed fermions.
The neutrino coupling to the $Z$ involves only left-handed chiralities.

\item There are three different light neutrino species.

\ei

\subsection{The \ $\mathbf{SU(2)_L\otimes U(1)_Y}$ \ theory}
\label{sec:model}

Using gauge invariance, we have been able to determine
the right QED and QCD Lagrangians.
To describe weak interactions, we need a more elaborated
structure, with several fermionic flavours and
different properties for left- and right-handed fields.
Moreover, the left-handed fermions should appear in doublets,
and we would like to have massive gauge bosons $W^\pm$ and
$Z$ in addition to the photon.
The simplest group with doublet representations is $SU(2)$.
We want to include also the electromagnetic interactions;
thus we need an additional $U(1)$ group. The obvious
symmetry group to consider is then
\bel{eq:group}
G\,\equiv\, SU(2)_L\otimes U(1)_Y \, ,
\ee
where $L$ refers to left-handed fields.
We do not specify, for the moment, the meaning of the
subindex $Y$ since, as we will see, the naive identification
with electromagnetism does not work.

For simplicity, let us consider a single family of quarks,
and introduce the notation
\bel{eq:psi_def}
\psi_1(x)\, =\,\left(\ba u \\ d \ea\right)_L \, , \qquad\quad
\psi_2(x) \, =\, u_R\, , \qquad\quad\psi_3(x) \, = \, d_R \, .
\ee
Our discussion will also be valid for the
lepton sector, with the identification
\bel{eq:psi_def_l}
\psi_1(x)\, =\,\left(\ba \nu_e \\ e^- \ea\right)_L\, , \qquad\quad
\psi_2(x) \, =\, \nu^{}_{eR} \, , \qquad\quad\psi_3(x) \, = \, e^-_R\, .
\ee

As in the QED and QCD cases, let us consider the free Lagrangian
\bel{eq:free_l}
\cL_0\; =\; i\,\bar u(x)\,\gamma^\mu\,\partial_\mu u(x)\, +\,
i\,\bar d(x)\,\gamma^\mu\,\partial_\mu d(x)
\; =\;\sum_{j=1}^3 \;i\,
\overline{\psi}_j(x)\,\gamma^\mu\,\partial_\mu\psi_j(x)\, .
\ee
$\cL_0$ is invariant under global $G$ transformations in flavour space:
\beqn\label{eq:G_transf}
\psi_1(x)&\toG & \psi'_1(x)\,\equiv
\,\exp{\left\{iy_1\beta\right\}}\; U_L\;\psi_1(x) \, ,
\no\\
\psi_2(x)& \toG & \psi'_2(x)\,\equiv\,
\exp{\left\{iy_2\beta\right\}}\;\psi_2(x) \, ,
\\
\psi_3(x)& \toG & \psi'_3(x)\,\equiv\,
\exp{\left\{iy_3\beta\right\}}\;\psi_3(x) \, ,
\no
\eeqn
where the $SU(2)_L$ transformation
\bel{eq:U_L}
U_L^{\phantom{\dagger}}
\,\equiv\,\exp{\left\{i\,\frac{\sigma_i}{2}\,\alpha^i\right\}}
\qquad\qquad\qquad (i=1,2,3)
\ee
only acts on the doublet field $\psi_1$.
The parameters $y_i$ are called hypercharges, since
the $U(1)_Y$ phase transformation is analogous to the QED one. The
matrix transformation $U_L$ is non-abelian like in QCD.
Notice that we have not included a mass term in \eqn{eq:free_l}
because it would mix the
left- and right-handed fields [see Eq.~\eqn{eq:DiracL_LR}],
spoiling therefore our
symmetry considerations.
% (helicity is not a good quantum number for massive fermions).

We can now require the Lagrangian to be also invariant under
local $SU(2)_L\otimes U(1)_Y$ gauge transformations,
i.e. with $\alpha^i=\alpha^i(x)$ and $\beta=\beta(x)$.
In order to satisfy this symmetry requirement, we need to
change the fermion derivatives by covariant objects.
Since we have now 4 gauge parameters, $\alpha^i(x)$
and $\beta(x)$, 4 different gauge bosons are needed:
\beqn\label{eq:cov_der}
D_\mu\psi_1(x)&\equiv &\left[\partial_\mu
-i\,g\, \widetilde W_\mu(x)- i\,\gp\, y_1\, B_\mu(x)\right]
\,\psi_1(x)\, ,
% \qquad\qquad
% \widetilde W_\mu(x)\,\equiv\,{\sigma_i\over 2}\,W^i_\mu(x)\, ,
\no\\
D_\mu\psi_2(x)&\equiv &\left[\partial_\mu
- i\,\gp\, y_2\, B_\mu(x)\right]\,\psi_2(x)\, ,
\\[5pt]
D_\mu\psi_3(x)&\equiv &\left[\partial_\mu
- i\,\gp\, y_3\, B_\mu(x)\right]\,\psi_3(x)\, .
\no
\eeqn
where
\bel{eq:Wmatrix}
\widetilde W_\mu(x)\,\equiv\,{\sigma_i\over 2}\,W^i_\mu(x)
\ee
denotes a $SU(2)_L$ matrix field.
Thus, we have the correct number of gauge fields to describe
the $W^\pm$, $Z$ and $\gamma$.

We want $D_\mu\psi_j(x)$ to
transform in exactly the same way as the $\psi_j(x)$ fields;
this fixes the transformation properties of the gauge fields:
\beqn\label{eq:B_transf}
B_\mu(x)&\toG &B'_\mu(x)\,\equiv\, B_\mu(x) +{1\over \gp}\,
\partial_\mu\beta(x) ,
\\ \label{eq:W-transf}
\widetilde W_\mu&\toG &
\widetilde W'_\mu\,\equiv\,
U_L^{\phantom{\dagger}}(x)\, \widetilde W_\mu\, U_L^\dagger(x)
- {i\over g}\, \partial_\mu U_L^{\phantom{\dagger}}(x)\, U_L^\dagger(x) ,
\eeqn
where \
$U_L(x)\equiv\exp{\left\{i\,\frac{\sigma_i}{2}\,\alpha^i(x)\right\}}$.
The transformation of $B_\mu$ is identical to the
one obtained in QED for the photon, while
the $SU(2)_L$ $W^i_\mu$ fields transform in a way analogous to the gluon fields of QCD.
Note that the $\psi_j$ couplings to $B_\mu$ are completely free
as in QED, i.e. the hypercharges $y_j$ can be arbitrary parameters.
Since the $SU(2)_L$ commutation relation is non-linear,
this freedom does not exist for the $W^i_\mu$:
there  is only a unique $SU(2)_L$ coupling $g$.

The Lagrangian
\bel{eq:lagrangian}
\cL\; =\; \sum_{j=1}^3 \;i\,
\overline{\psi}_j(x)\,\gamma^\mu\, D_\mu\psi_j(x)\, ,
\ee
is invariant under local $G$ transformations.
In order to build the gauge-invariant kinetic term for the
gauge fields, we introduce the corresponding field strengths:
\be\label{eq:b_mn}
B_{\mu\nu}  \;\equiv\;  \partial_\mu B_\nu - \partial_\nu B_\mu\, ,
\ee
\bel{eq:W_mn}
\widetilde W_{\mu\nu} \;\equiv\; {i\over g}\,\left[
\left(\partial_\mu -i\,g\, \widetilde W_\mu\right)\, ,\,
\left(\partial_\nu -i\,g\, \widetilde W_\nu\right)\right]
\, =\,\partial_\mu \widetilde W_\nu -\partial_\nu \widetilde W_\mu
-i g\,\left[ W_\mu , W_\nu \right]\, ,
\ee
\bel{eq:Wi_mn}
\widetilde W_{\mu\nu} \;\equiv\; {\sigma_i\over 2}\, W^i_{\mu\nu}\, ,
\qquad\qquad
W^i_{\mu\nu}\, =\,\partial_\mu W^i_\nu - \partial_\nu W^i_\mu
+ g\,\epsilon^{ijk}\, W^j_\mu\, W^k_\nu\, .
\ee
$B_{\mu\nu}$ remains invariant under $G$ transformations,
while $\widetilde W_{\mu\nu}$ transforms covariantly:
\bel{eq:W_mn_transf}
B_{\mu\nu}\,\toG\, B_{\mu\nu}\, ,
\qquad\qquad
\widetilde W_{\mu\nu}\,\toG\, U_L^{\phantom{\dagger}}\,
\widetilde W_{\mu\nu}\, U_L^\dagger
\, .
\ee
Therefore, the properly normalized kinetic Lagrangian is given by
\bel{eq:kinetic}
\cL_{\rms Kin} \, = \,
-{1\over 4}\, B_{\mu\nu}\, B^{\mu\nu} - {1\over 2}\,
\mbox{\rm Tr}\left[\widetilde W_{\mu\nu}\, \widetilde W^{\mu\nu}\right]
\, = \,
-{1\over 4}\, B_{\mu\nu}\, B^{\mu\nu} - {1\over 4}\,
W_{\mu\nu}^i\, W^{\mu\nu}_i\, .
\ee
Since the field strengths
$W^i_{\mu\nu}$ contain a quadratic piece, the Lagrangian
$\cL_{\rms Kin}$ gives rise to cubic and
quartic self-interactions among the gauge fields.
The strength of these interactions
is given by the same $SU(2)_L$ coupling $g$  which appears in the
fermionic piece of the Lagrangian.

The gauge symmetry forbids to write a mass term for the
gauge bosons. Fermionic masses are also not possible, because
they would communicate the left- and right-handed fields,
which have different transformation properties,
and therefore would produce an explicit
breaking of the gauge symmetry.
Thus, the $SU(2)_L\otimes U(1)_Y$ Lagrangian in
Eqs.~\eqn{eq:lagrangian} and \eqn{eq:kinetic}
only contains massless fields.

\subsection{Charged-current interaction}
\label{subsec:cc}

%%%%%%%%%%%%%%%  FIGURE %%%%%%%%%%%%%%%%%%%%%%%%%
\begin{figure}[tbh]\centering
\begin{minipage}[t]{.4\linewidth}\centering
\includegraphics[width=4cm]{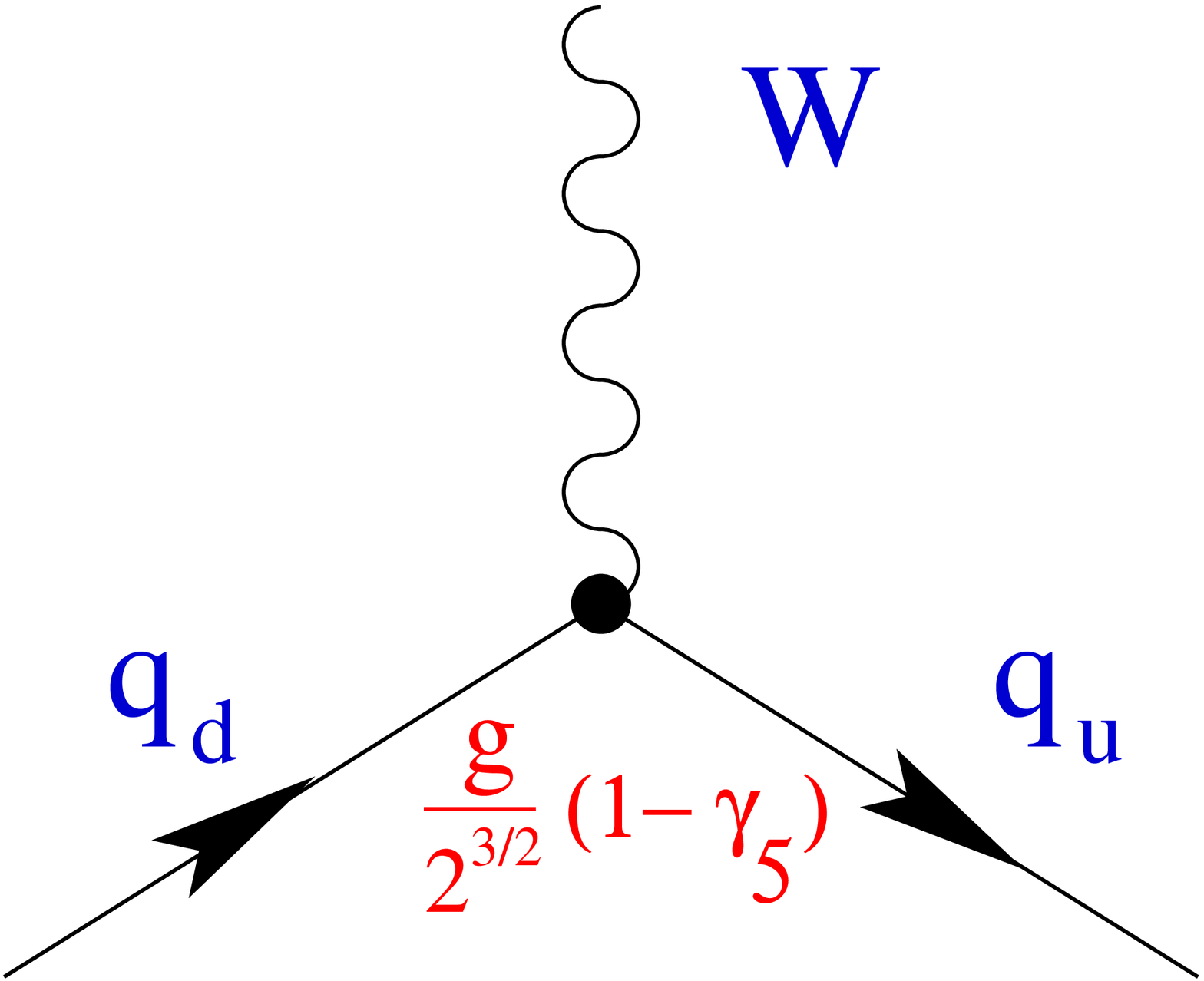}
\end{minipage}
\hskip 1cm
\begin{minipage}[t]{.4\linewidth}\centering
\includegraphics[width=4cm]{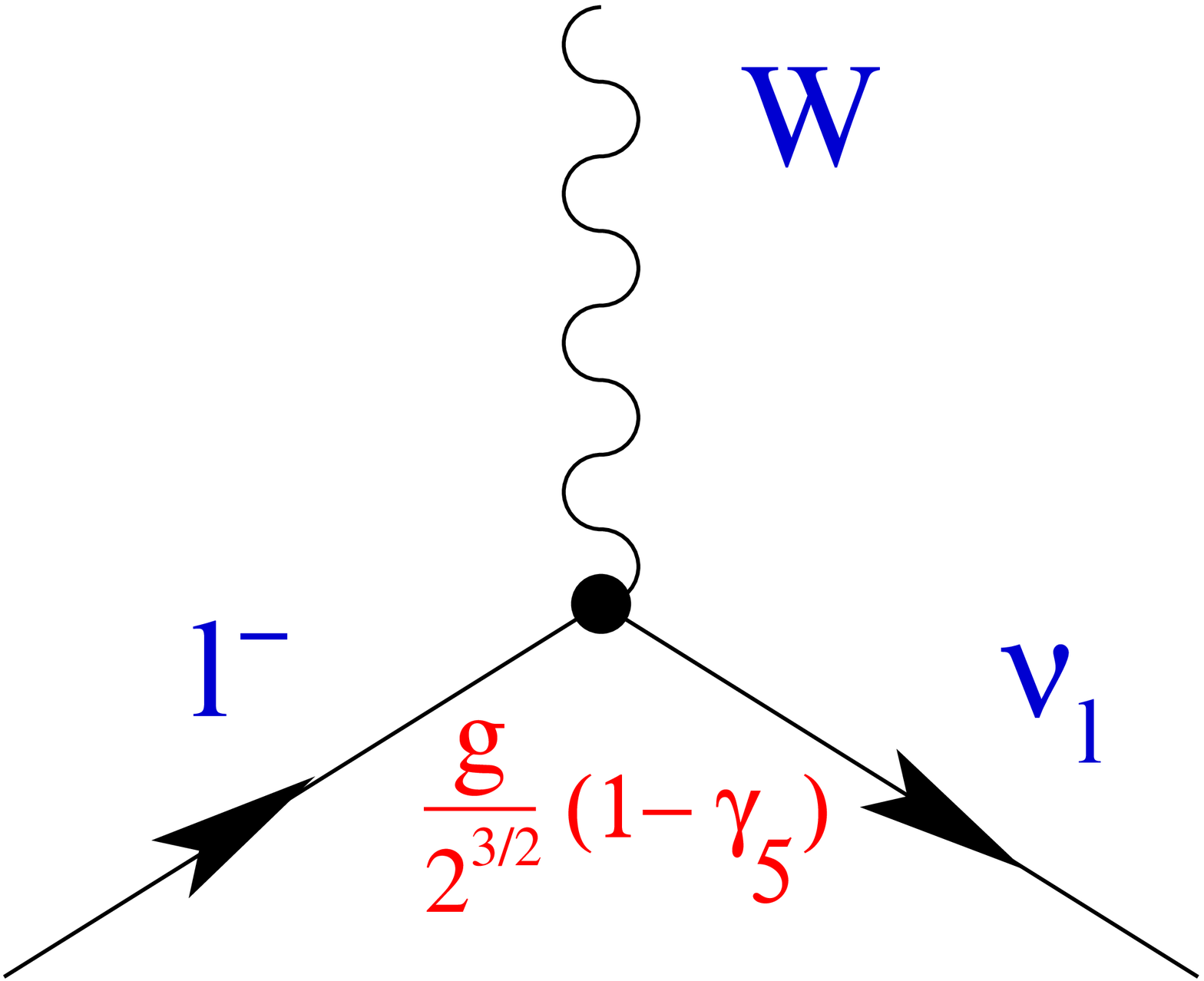}
\end{minipage}
\caption{Charged-current interaction vertices.}
\label{fig:L_CC}
\end{figure}
%%%%%%%%%%%%%%%%%%%%%%%%%%%%%%%%%%%%%%%%%%%%%

The Lagrangian \eqn{eq:lagrangian} contains interactions of the
fermion fields with the gauge bosons,
\bel{eq:int}
\cL\quad\longrightarrow\quad
g\,\overline{\psi}_1\gamma^\mu\widetilde  W_\mu
\psi_1\, + \,\gp\,B_\mu\,
\sum_{j=1}^3\, y_j\,\overline{\psi}_j\gamma^\mu\psi_j \, .
\ee
The term containing the $SU(2)_L$ matrix
\bel{eq:W_matrix}
\widetilde W_\mu\, =\, {\sigma^i\over 2}\, W_\mu^i\, = \, {1\over\sqrt{2}}
\,\left(\bat
\sqrt{2}\, W^3_\mu & W_\mu^\dagger \\[5pt] W_\mu & -\sqrt{2}\, W^3_\mu
\ea\right)
\ee
gives rise to charged-current interactions with the boson field \
$W_\mu\equiv (W_\mu^1+i\, W_\mu^2)/\sqrt{2}$ \ and its complex-conjugate \
$W^\dagger_\mu\equiv (W_\mu^1-i\, W_\mu^2)/\sqrt{2}$.
For a single family of quarks and leptons,
\bel{eq:W_interactions}
\cL_{\rms CC}\, = \, {g\over 2\sqrt{2}}\,\left\{
W^\dagger_\mu\,\left[
\bar u\gamma^\mu(1-\gamma_5) d\, +\, \bar\nu_e\gamma^\mu(1-\gamma_5) e
\right]\, + \, \mbox{\rm h.c.}\right\}\, .
\ee
The universality of the quark and lepton interactions is now a direct
consequence of the assumed gauge symmetry.
Note, however, that \eqn{eq:W_interactions} cannot
describe the observed dynamics, because the gauge bosons are massless
and, therefore, give rise to long-range forces.

\subsection{Neutral-current interaction}
\label{subsec:nc_int}

%%%%%%%%%%%%%%%  FIGURE %%%%%%%%%%%%%%%%%%%%%%%%%
\begin{figure}[tbh]\centering
\begin{minipage}[t]{.4\linewidth}\centering
\includegraphics[width=4cm]{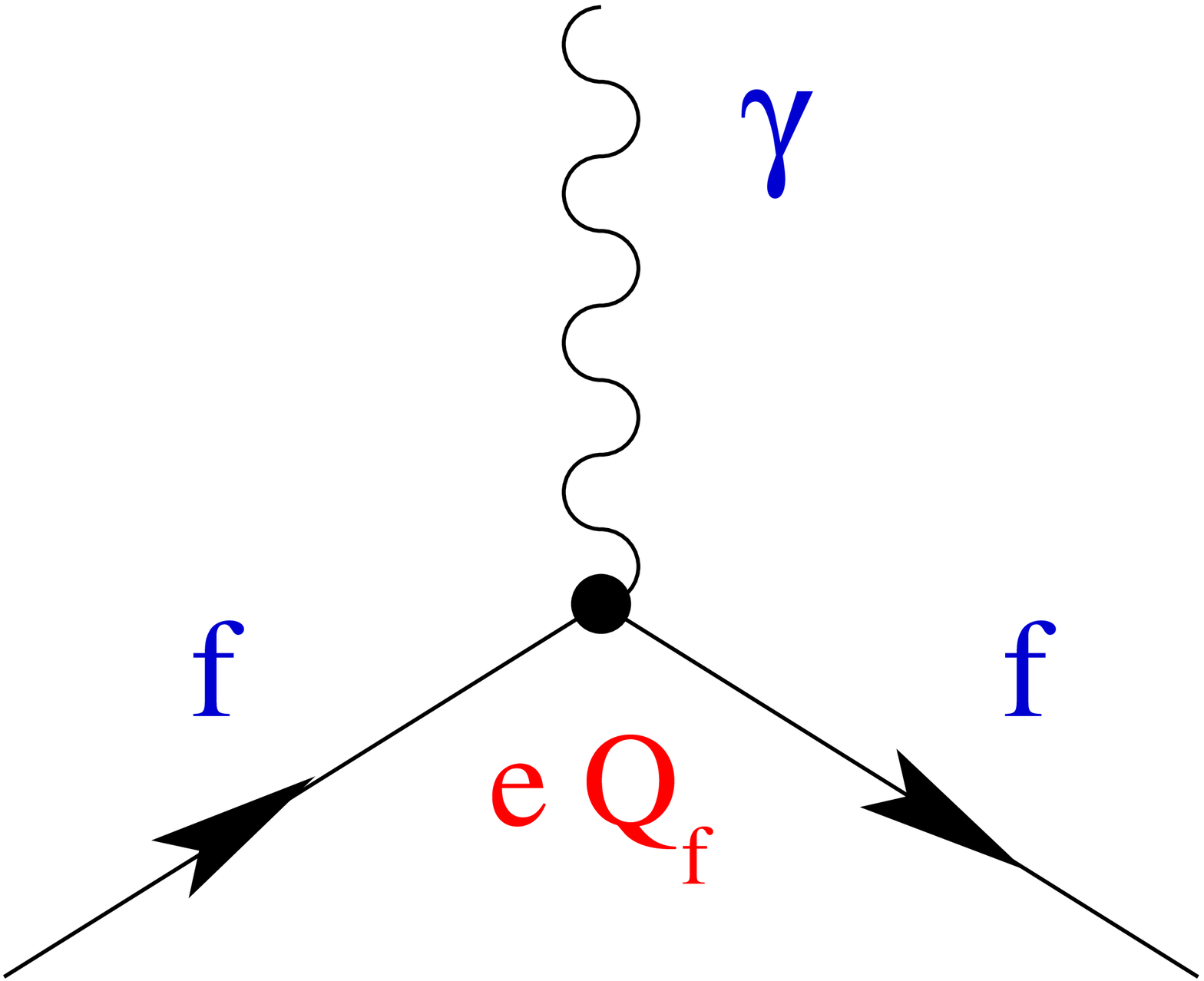}
\end{minipage}
\hskip 1cm
\begin{minipage}[t]{.4\linewidth}\centering
\includegraphics[width=4cm]{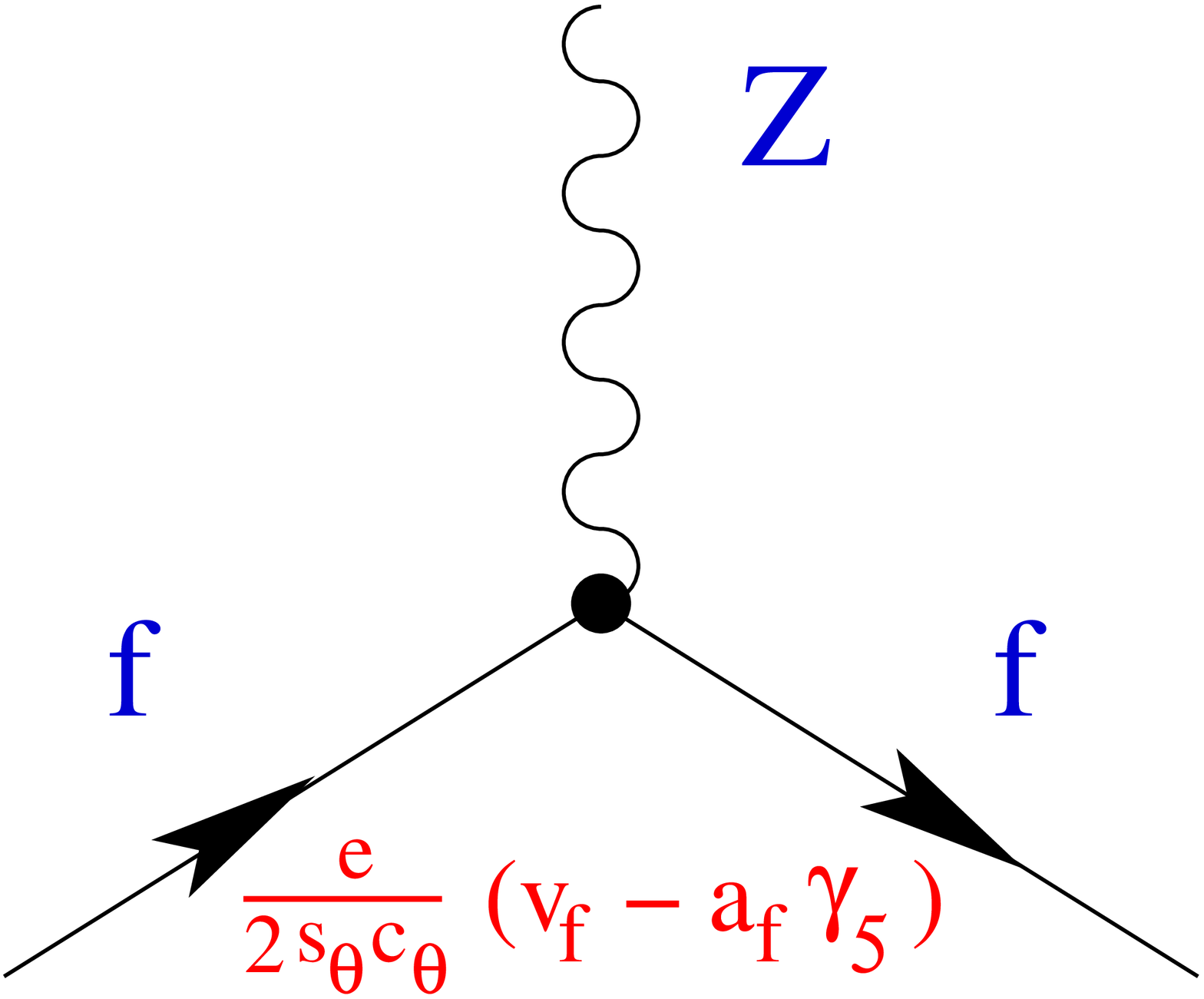}
\end{minipage}
\caption{Neutral-current interaction vertices.}
\label{fig:L_NC}
\end{figure}
%%%%%%%%%%%%%%%%%%%%%%%%%%%%%%%%%%%%%%%%%%%%%

Eq.~\eqn{eq:int} contains also interactions with
the neutral gauge fields
$W^3_\mu$ and $B_\mu$. We would like to identify these bosons with
the $Z$ and the $\gamma$. However, since the photon has the same
interaction with both fermion chiralities, the singlet gauge
boson $B_\mu$ cannot be equal to the electromagnetic field.
That would require \ $y_1 = y_2 = y_3$ \ and \ $\gp y_j = e\, Q_j$, which
cannot be simultaneously true.

Since both fields are neutral, we can try with
an arbitrary combination of them:
\bel{eq:Z_g_mixing}
\left(\ba W_\mu^3 \\ B_\mu\ea\right) \,\equiv\,
\left(\bat
\cos{\theta_W} & \sin{\theta_W} \\ -\sin{\theta_W} & \cos{\theta_W}
\ea\right) \, \left(\ba Z_\mu \\ A_\mu\ea\right)\, .
\ee
The physical $Z$ boson has a mass different from zero, which is
forbidden by the local gauge symmetry. We will see in the next section
how it is possible to generate non-zero boson masses, through the
SSB mechanism.
For the moment, we will just assume that something breaks the symmetry,
generating the $Z$ mass, and that the neutral mass eigenstates are a mixture
of the triplet and singlet $SU(2)_L$ fields.
In terms of the fields $Z$ and $\gamma$,
the neutral-current Lagrangian is given by
\be\label{eq:L_NC}
\cL_{\rms NC}\, =\,
\sum_j\,\overline{\psi}_j\,\gamma^\mu\left\{ A_\mu\,
\left[ g\, {\sigma_3\over 2}\,\sin{\theta_W} + \gp\,y_j\,\cos{\theta_W}\right]
\, +\,
Z_\mu\,
\left[ g\, {\sigma_3\over 2}\,\cos{\theta_W} - \gp\,y_j\,\sin{\theta_W}\right]
\right\}\,\psi_j\, .
\ee
In order to get QED from the $A_\mu$ piece, one needs to impose the
conditions:
\bel{eq:unification}
g\,\sin{\theta_W} \, = \, \gp\,\cos{\theta_W}\, = \, e\, ,
\qquad\qquad\qquad
 Y \, = \, Q-T_3\, ,
\ee
where \ $T_3\equiv\sigma_3/2$ \ and \
$Q$ \ denotes the electromagnetic charge operator
\bel{eq:Q}
Q_1\,\equiv\,\left(\bat Q_{u/\nu}& 0 \\ 0 & Q_{d/e}\ea\right)\, ,
\qquad\qquad Q_2\, =\,Q_{u/\nu}\, ,\qquad\qquad Q_3\, =\,Q_{d/e}
\, .
\ee
The first equality relates the $SU(2)_L$ and $U(1)_Y$ couplings
to the electromagnetic coupling, providing the wanted unification
of the electroweak interactions. The second identity, fixes
the fermion hypercharges in terms of their electric charge and
weak isospin quantum numbers:
$$  %%\bel{eq:hypercharges}
\begin{array}{lccccc}
\mbox{Quarks:}\quad &
y_1\, =\, Q_u-{1\over 2}\, =\, Q_d+{1\over 2}\, =\, {1\over 6}\, ,
&\quad & y_2\, =\, Q_u\, =\, {2\over 3}\, ,
&\quad & y_3\, =\, Q_d\, =\, -{1\over 3}\, ,
\\[10pt]
\mbox{Leptons:}\quad &
y_1\, =\, Q_\nu-{1\over 2}\, =\, Q_e+{1\over 2}\, =\, -{1\over 2}\, ,
&\quad & y_2\, =\, Q_\nu\, =\, 0\, ,
&\quad & y_3\, =\, Q_e\, =\, -1\, .
\end{array}
$$  %%\ee
A hypothetical
right-handed neutrino would have both electric charge
and weak hypercharge equal to zero. Since it would not couple
either to the $W^\pm$ bosons, such a particle would not have any
kind of interaction (sterile neutrino).
For aesthetical reasons, we will then not consider right-handed
neutrinos any longer.

Using the relations \eqn{eq:unification},
the neutral-current Lagrangian can be written as
\bel{eq:L_NCb}
\cL_{\rms NC}\, = \,
\cL_{\rms QED}\, + \,
\cL_{\rms NC}^Z\, ,
\ee
where
\bel{eq:L_QED}
\cL_{\rms QED}\, = \,e\,A_\mu\,\sum_j\,
\overline{\psi}_j\gamma^\mu Q_j\psi_j
\,\equiv\, e\, A_\mu\, J^\mu_{\rms em}
\ee
is the usual QED Lagrangian and
\bel{eq:Z_NC}
\cL_{\rms NC}^Z \, =\,
{e\over 2 \sin{\theta_W}\cos{\theta_W}} \,
J^\mu_Z\, Z_\mu
\ee
contains the interaction of the $Z$-boson with the neutral fermionic current
\bel{eq:J_NC}
J^\mu_Z \,\equiv\, \sum_j\,\overline{\psi}_j\gamma^\mu
\left(\sigma_3-2\sin^2{\theta_W}Q_j\right)\psi_j
\, = \, J^\mu_3 - 2 \sin^2{\theta_W}\, J^\mu_{\rms em}\, .
\ee
In terms of the more usual fermion fields,
$\cL_{\rms NC}^Z$ has the form
\bel{eq:Z_Lagrangian}
\cL_{\rms NC}^Z\, = \,
{e\over 2\sin{\theta_W}\cos{\theta_W}} \,Z_\mu\,\sum_f\,
\bar f \gamma^\mu (v_f-a_f\gamma_5) \, f\, ,
\ee
where \
$a_f =  T_3^f$ \ and \
$v_f = T_3^f \left( 1 - 4 |Q_f| \sin^2{\theta_W}\right)$.
Table~\ref{tab:nc_couplings} shows the neutral-current
couplings of the different fermions.

%%%%%%%%%%%%%%% Table %%%%%%%%%%%%%%%%%
\begin{table}[t]
\begin{center}
{\renewcommand{\arraystretch}{1.5}
\caption{Neutral-current couplings.\label{tab:nc_couplings}}
\vspace{0.2cm}
\begin{tabular}{|c||c|c||c|c|}
\hline
& $u$ & $d$ & $\;\nu_e\;$ & $e$
\\ \hline\hline
$\, 2\, v_f\, $ & $1 -{8\over 3} \sin^2{\theta_W}$ &
$-1 +{4\over 3} \sin^2{\theta_W}$ & $\,\, 1\,\,$ &
$-1 +4 \sin^2{\theta_W}$
\\
$2\, a_f$ & $1$ & $-1$ & $1$ & $-1$
\\ \hline
\end{tabular}}
\end{center}
\end{table}
%%%%%%%%%%%% End Table %%%%%%%%%%%%%%%%

\subsection{Gauge self-interactions}
\label{subsec:self-interactions}

%%%%%%%%%%%%%%%  FIGURE %%%%%%%%%%%%%%%%%%%%%%%%%
\begin{figure}[tbh]\centering
\includegraphics[width=15.5cm]{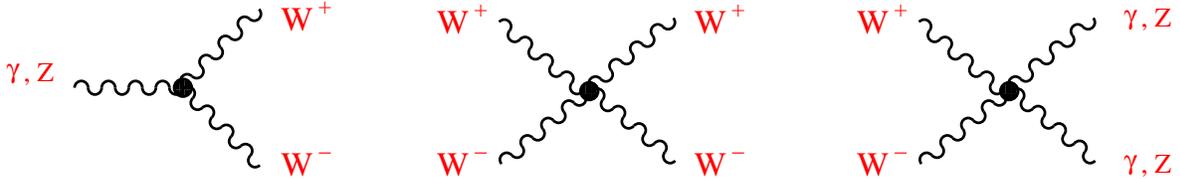}
\caption{Gauge boson self-interaction vertices.}
\label{fig:L_GSI}
\end{figure}
%%%%%%%%%%%%%%%%%%%%%%%%%%%%%%%%%%%%%%%%%%%%%

In addition to the usual kinetic terms, the Lagrangian
\eqn{eq:kinetic} generates cubic and quartic self-interactions
among the gauge bosons:
\beqn\label{eq:cubic}
\cL_3 &\!\!\!\! = &\!\!\!\!
-i e \cot{\theta_W}\left\{
\left(\partial^\mu W^\nu -\partial^\nu W^\mu\right)
 W^\dagger_\mu Z_\nu -
\left(\partial^\mu W^{\nu\dagger} -\partial^\nu W^{\mu\dagger}\right)
 W_\mu Z_\nu +
W_\mu W^\dagger_\nu\left(\partial^\mu Z^\nu -\partial^\nu Z^\mu\right)
\right\}
\no\\[10pt] &&\!\!\!\!\mbox{}
-i e \left\{
\left(\partial^\mu W^\nu -\partial^\nu W^\mu\right)
 W^\dagger_\mu A_\nu -
\left(\partial^\mu W^{\nu\dagger} -\partial^\nu W^{\mu\dagger}\right)
 W_\mu A_\nu +
W_\mu W^\dagger_\nu\left(\partial^\mu A^\nu -\partial^\nu A^\mu\right)
\right\}  ;
\no\\[10pt] &&\\[10pt] %%%\label{eq:quartic}
\cL_4 &\!\!\!\! = &\!\!\!\!\mbox{}
-{e^2\over 2\sin^2{\theta_W}}\left\{
\left(W^\dagger_\mu W^\mu\right)^2 - W^\dagger_\mu W^{\mu\dagger}
W_\nu W^\nu \right\}
- e^2 \cot^2{\theta_W}\,\left\{
W_\mu^\dagger W^\mu Z_\nu Z^\nu - W^\dagger_\mu Z^\mu W_\nu Z^\nu
\right\}
\no\\[10pt] &\!\!\!\! &\!\!\!\!\mbox{}
- e^2 \cot{\theta_W}\left\{
2 W_\mu^\dagger W^\mu Z_\nu A^\nu - W^\dagger_\mu Z^\mu W_\nu A^\nu
- W^\dagger_\mu A^\mu W_\nu Z^\nu
\right\}
\no\\[10pt] &\!\!\!\! &\!\!\!\!\mbox{}
- e^2\,\left\{
W_\mu^\dagger W^\mu A_\nu A^\nu - W^\dagger_\mu A^\mu W_\nu A^\nu
\right\} .
\no\eeqn
Notice that there are always at least a pair of charged $W$ bosons.
The $SU(2)_L$ algebra does not generate any
neutral vertex with only photons and $Z$ bosons.

%%%%%%%%%% SSB %%%%%%%%%%

\setcounter{equation}{0}
\section{SPONTANEOUS \ SYMMETRY \ BREAKING}
\label{sec:ssb}

%%%%%%%%%%%%%%%  FIGURE %%%%%%%%%%%%%%%%%%%%%%%%%
\begin{figure}[tbh]\centering
\begin{minipage}[t]{.47\linewidth}%\centering
\includegraphics[width=7.25cm,clip]{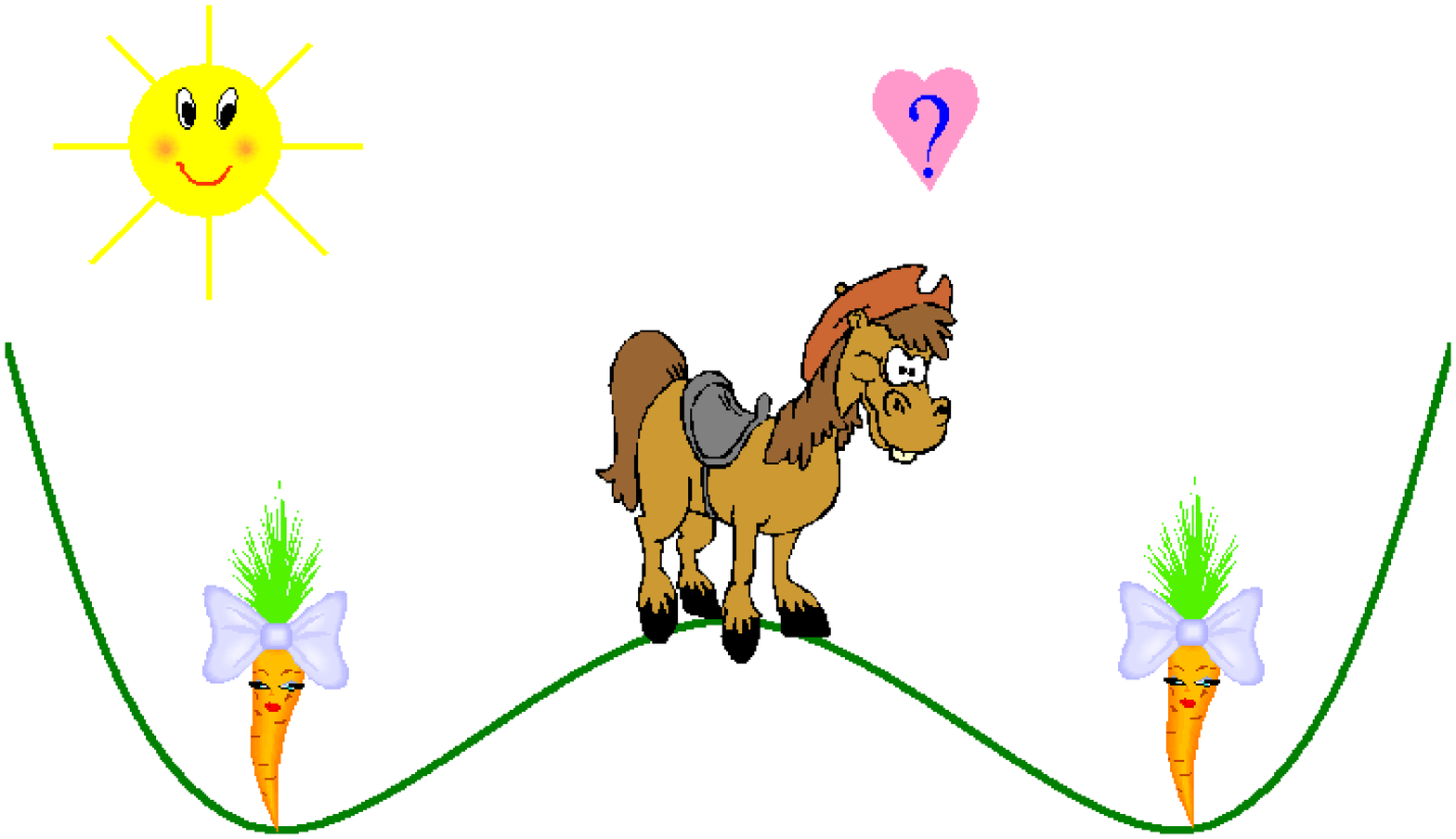}
\end{minipage}
\hfill
\begin{minipage}[t]{.47\linewidth}%\centering
\includegraphics[width=7.25cm,clip]{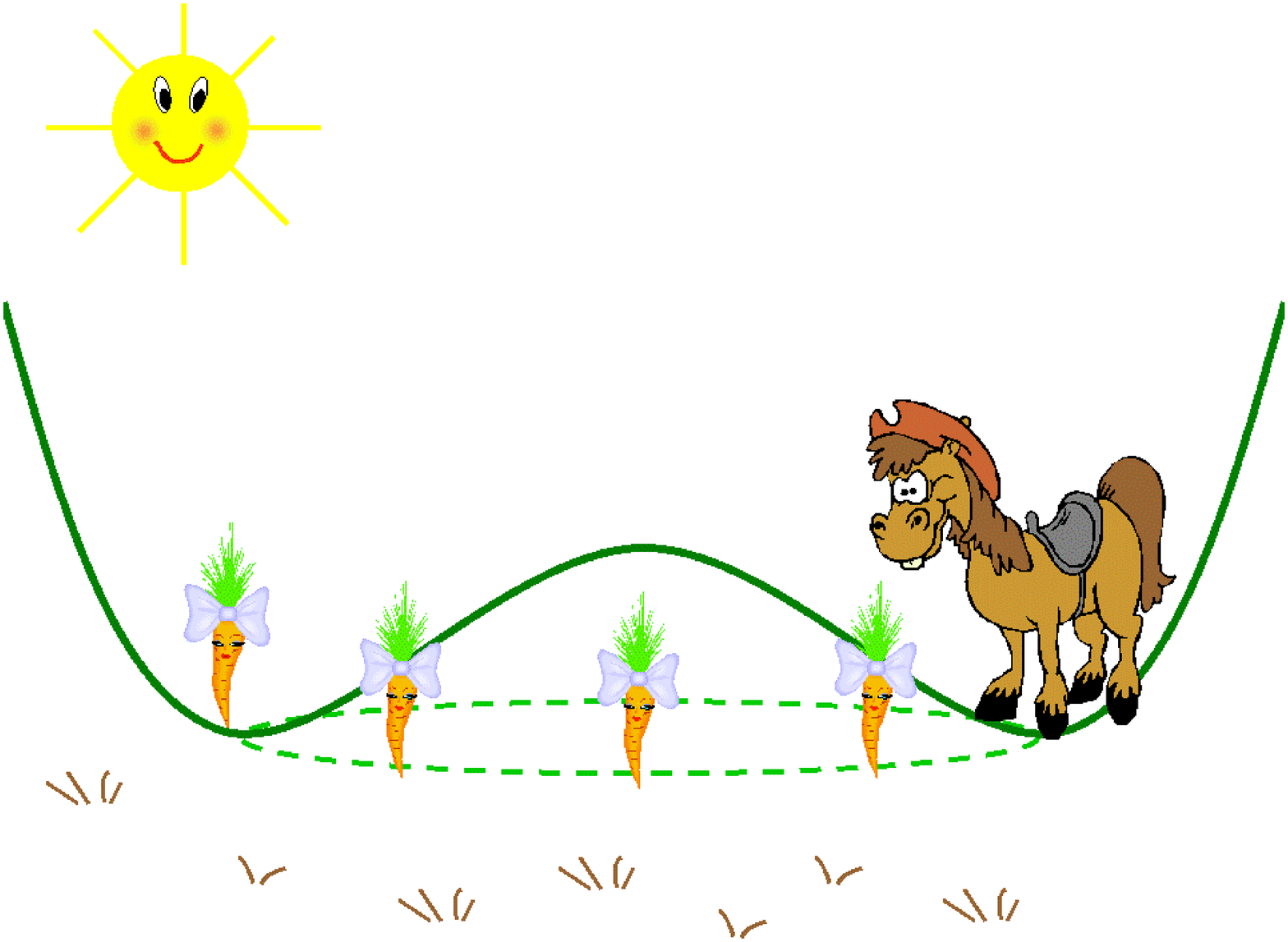}
\end{minipage}
\caption{Although Nicol\'as likes the symmetric food configuration,
he must break the symmetry deciding which carrot is more appealing. In 3 dimensions,
there is a continuous valley where Nicol\'as can move from one carrot to the
next without effort.}
\label{fig:Nicolas}
\end{figure}
%%%%%%%%%%%%%%%%%%%%%%%%%%%%%%%%%%%%%%%%%%%%%

So far, we have been able to derive charged- and neutral-current
interactions of the type needed to describe weak decays;
we have nicely incorporated QED into the
same theoretical framework; and, moreover, we have got additional
self-interactions of the gauge bosons, which are generated by the
non-abelian structure of the $SU(2)_L$ group.
Gauge symmetry also guarantees that we have a well-defined
renormalizable Lagrangian.
However, this Lagrangian has very little to do with
reality. Our gauge bosons are massless particles; while this is
fine for the photon field, the physical $W^\pm$ and
$Z$ bosons should be quite heavy objects.

In order to generate masses, we need to break the gauge symmetry
in some way; however, we also need a fully symmetric Lagrangian
to preserve renormalizability.
A possible solution to this dilemma, is based on the fact that
it is possible to get non-symmetric results from an invariant
Lagrangian.

Let us consider a Lagrangian, which:

\begin{enumerate}
\item Is invariant under a group $G$ of transformations.

\item Has a degenerate set of states with minimal energy,
which transform under $G$ as the members of a given multiplet.
\end{enumerate}

\noindent
If one arbitrarily selects one of those states as the ground
state of the system, one says that the symmetry becomes
spontaneously broken.

A well-known physical example is provided by a ferromagnet: although
the Hamiltonian is invariant under rotations, the ground state
has the spins aligned into some arbitrary direction. Moreover,
any higher-energy state, built from the ground state by a finite
number of excitations, would share its anisotropy.
In a Quantum Field Theory, the ground state is the vacuum.
Thus, the SSB mechanism will appear in those cases where
one has a symmetric Lagrangian, but a non-symmetric vacuum.

The horse in Fig.~\ref{fig:Nicolas} illustrates in a very simple way
the phenomenon of SSB. Although the left and right carrots are identical,
Nicol\'as must take a decision if he wants to get food. The important
thing is not whether he goes left or right, which are equivalent options,
but the fact that the symmetry gets broken. In two dimensions (discrete
left-right symmetry), after eating the first carrot Nicol\'as would
need to make some effort in order to climb the hill and reach the carrot on the
other side. However, in three dimensions (continuous rotation symmetry)
there is a marvellous flat circular valley along which Nicol\'as can move
from one carrot to the next without any effort.

The existence of flat directions connecting the degenerate states of minimal
energy is a general property of the SSB of continuous symmetries.
In a Quantum Field Theory it implies the existence of massless degrees of freedom.

\subsection{Goldstone theorem}
\label{subsec:goldstone}

%%%%%%%%%%%%%%%  FIGURE %%%%%%%%%%%%%%%%%%%%%%%%%
\begin{figure}[tbh]\centering
\begin{minipage}[c]{.4\linewidth}\centering
\includegraphics[width=5.3cm]{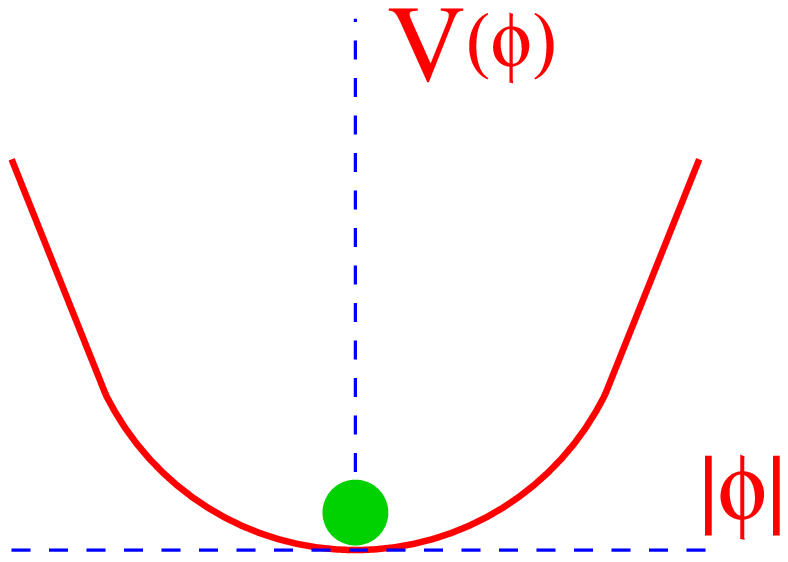}
\vskip .5cm\mbox{}
\end{minipage}
\hskip 1cm
\begin{minipage}[c]{.4\linewidth}\centering
\includegraphics[width=6cm]{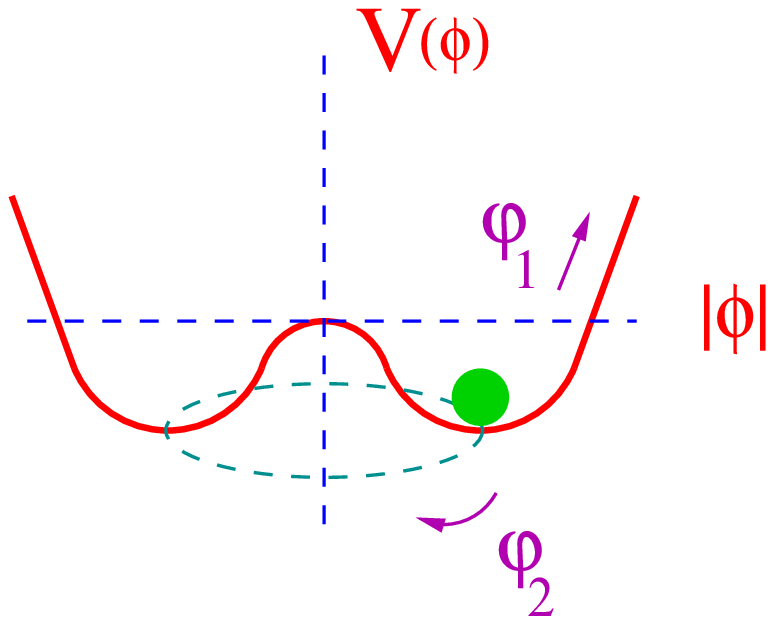}
\end{minipage}
\vskip -.3cm
\caption{Shape of the scalar potential for \ $\mu^2>0$ \ (left) \
and \ $\mu^2<0$ \ (right). In the second case there is a continuous
set of degenerate vacua, corresponding to different phases \ $\theta$,
connected through a massless field excitation \ $\varphi_2$.}
\label{fig:HiggsPot}
\end{figure}
%%%%%%%%%%%%%%%%%%%%%%%%%%%%%%%%%%%%%%%%%%%%%

Let us consider a complex scalar field $\phi(x)$, with
Lagrangian
\bel{eq:L_phi}
\cL\, = \, \partial_\mu\phi^\dagger \partial^\mu\phi - V(\phi)\, ,
\qquad\qquad
V(\phi)\, = \, \mu^2 \phi^\dagger\phi + h
\left(\phi^\dagger\phi\right)^2\, .
\ee
$\cL$ is invariant under global phase transformations
of the scalar field
\bel{eq:phi_transf}
\phi(x)\,\longrightarrow\,\phi'(x)\,\equiv\, 
\exp{\left\{i\theta\right\}}\,\phi(x) \, .
\ee

In order to have a ground state the potential should be bounded
from below, i.e. $h>0$. For the quadratic piece there are two
possibilities:
\vskip .1cm
\begin{enumerate}

\item \mbox{\boldmath $\mu^2>0$:} \
The potential has only the trivial minimum $\phi=0$.
It describes a massive scalar particle with mass $\mu$
and quartic coupling $h$.
\vskip .25cm

\item \mbox{\boldmath $\mu^2<0$:} \
The minimum is obtained for those field configurations
satisfying
\bel{eq:minimum}
|\phi_0|\, = \, \sqrt{{-\mu^2\over 2 h}} \,\equiv\, {v\over\sqrt{2}}
\, > \, 0 \, ,
\qquad\qquad\qquad V(\phi_0)\, =\, -{h\over 4} v^4\,  .
\ee
Owing to the $U(1)$ phase-invariance of the Lagrangian,
there is an infinite number of degenerate states of
minimum energy,
$\phi_0(x) = {v\over\sqrt{2}}\, \exp{\left\{i\theta\right\}}$.
By choosing a particular solution, $\theta=0$ for example, as the
ground state, the symmetry gets spontaneously broken.
If we parametrize the excitations over the ground state as
\bel{eq:perturbations}
\phi(x)\, \equiv \, {1\over\sqrt{2}}\,\left[
v + \varphi_1(x) + i\, \varphi_2(x)\right]\, ,
\ee
where $\varphi_1$ and $\varphi_2$ are real fields,
the potential takes the form
\bel{eq:pot}
V(\phi)\, = \, V(\phi_0) -\mu^2\varphi_1^2 +
h\, v \,\varphi_1 \left(\varphi_1^2+\varphi_2^2\right) +
{h\over 4} \left(\varphi_1^2+\varphi_2^2\right)^2\, .
\ee
Thus, $\varphi_1$ describes a massive state of mass
$m_{\varphi_1}^2 = -2\mu^2$, while $\varphi_2$ is massless.

\end{enumerate}
\vskip .1cm

The first possibility ($\mu^2>0$) is just the usual situation
with a single ground state.
The other case, with SSB, is more interesting.
The appearance of a massless particle when $\mu^2<0$
is easy to understand: the field $\varphi_2$ describes excitations
around a flat direction in the potential, i.e. into states
with the same energy as the chosen ground state.
Since those excitations do not cost any energy, they obviously
correspond to a massless state.

The fact that there are massless excitations associated with the
SSB mechanism is a completely general result, known
as the Goldstone theorem \cite{goldstone}:
if a Lagrangian is invariant under a continuous symmetry group
$G$, but the vacuum is only invariant under a subgroup $H\subset G$,
then there must exist as many massless spin--0 particles
(Goldstone bosons) as broken generators
(i.e. generators of $G$ which do not belong to $H$).

\subsection{The Higgs--Kibble mechanism}
\label{subsec:Higgs-Kibble}

At first sight, the Goldstone theorem has very little to do with
our mass problem; in fact, it makes it worse since we want
massive states and not massless ones.
However, something very interesting happens when there is a local
gauge symmetry \cite{HI:66}.

Let us consider \cite{WE:67}
an $SU(2)_L$ doublet of complex scalar fields
\bel{eq:scalar_multiplet}
\phi(x)\,\equiv\,\left(\ba \phi^{(+)}(x)\\ \phi^{(0)}(x)\ea\right)\, .
\ee
The gauged scalar Lagrangian of the Goldstone model (\ref{eq:L_phi}),
\bel{eq:LS}
\cL_S \, = \, \left(D_\mu\phi\right)^\dagger D^\mu\phi
-\mu^2\phi^\dagger\phi - h \left( \phi^\dagger\phi\right)^2\, ,
\qquad\qquad\qquad (h>0\, ,\, \mu^2<0)\, ,
\ee
\bel{eq:DS}
D^\mu\phi \, = \,
\left[\partial^\mu
-i\, g\,\widetilde W^\mu
- i\,\gp\, y_\phi\, B^\mu\right]\,\phi\, ,
\qquad\qquad\qquad
y_\phi\, =\, Q_\phi - T_3 \, =\, \frac{1}{2}\, ,
\ee
is invariant under local $SU(2)_L\otimes U(1)_Y$ transformations.
The value of the scalar hypercharge is fixed by the requirement
of having the correct couplings between $\phi(x)$ and
$A^\mu(x)$; i.e. that the photon does not couple to $\phi^{(0)}$,
and one has the right electric charge for $\phi^{(+)}$.

The potential is very similar to the one considered before.
There is a infinite set of degenerate states with minimum
energy, satisfying
\bel{eq:vev}
\big|\langle 0|\phi^{(0)}|0\rangle\big|
\, = \, \sqrt{{-\mu^2\over 2 h}} \,\equiv\, {v\over\sqrt{2}} \, .
\ee
Note that we have made explicit the association of the classical
ground state with the quantum vacuum. Since the electric charge
is a conserved quantity, only the neutral scalar field can acquire
a vacuum expectation value.
Once we choose a particular ground state,
the $SU(2)_L\otimes U(1)_Y$ symmetry gets spontaneously broken to
the electromagnetic subgroup $U(1)_{\rms QED}$,
which by construction still remains a true symmetry of the vacuum.
According to Goldstone theorem 3 massless states should then appear.

Now, let us parametrize the scalar doublet in the general form
\bel{eq:parametrization}
\phi(x)\, = \, \exp{\left\{
i\, {\sigma_i\over 2}\,\theta^i(x)\right\}}
\; {1\over\sqrt{2}}\,
\left(\ba 0 \\ v + H(x) \ea\right)\, ,
\ee
with 4 real fields $\theta^i(x)$ and $H(x)$.
The crucial point to be realized is that the local $SU(2)_L$
invariance of the Lagrangian allows us to rotate away any
dependence on $\theta^i(x)$.
These 3 fields are precisely the would-be massless Goldstone bosons
associated with the SSB  mechanism.
The additional ingredient of gauge symmetry makes these
massless excitations unphysical.

The covariant derivative \eqn{eq:DS} couples the scalar
multiplet to the $SU(2)_L\otimes U(1)_Y$ gauge bosons.
If one takes the physical (unitary) gauge \ $\theta^i(x)=0\,$,
the kinetic piece of the scalar Lagrangian \eqn{eq:LS}
takes the form:
\bel{eq:unitary_gauge}
\left(D_\mu\phi\right)^\dagger D^\mu\phi\quad\toTheta\quad
{1\over 2}\, \partial_\mu H \partial^\mu H
+ (v+H)^2 \, \left\{ {g^2\over 4}\, W_\mu^\dagger W^\mu
+ {g^2\over 8\cos^2{\theta_W}}\, Z_\mu Z^\mu \right\}\, .
\ee
The vacuum expectation value of the neutral scalar has
generated a quadratic term for the $W^\pm$ and the $Z$,
i.e. those gauge bosons have acquired masses:
\bel{eq:boson_masses}
M_Z\,\cos{\theta_W}\, = \, M_W \, = \,  \frac{1}{2}\, v\, g \, .
\ee

Therefore, we have found a clever way of giving masses to the
intermediate carriers of the weak force. We just add $\cL_S$
to our $SU(2)_L\otimes U(1)_Y$ model.
The total Lagrangian is invariant under gauge transformations,
which guarantees \cite{TH:71}
the renormalizability of the associated
Quantum Field Theory. However, SSB occurs. The 3 broken generators
give rise to 3 massless Goldstone bosons which,
owing to the underlying local gauge symmetry,
can be eliminated from the Lagrangian.
Going to the unitary gauge, we discover that the $W^\pm$ and the
$Z$ (but not the $\gamma$, because $U(1)_{\rms QED}$ is
an unbroken symmetry) have acquired masses, which are moreover related
as indicated in Eq.~\eqn{eq:boson_masses}.
Notice that (\ref{eq:Z_g_mixing}) has now the meaning of writing
the gauge fields in terms of the physical boson fields with
definite mass.

It is instructive to count the number of degrees of freedom (d.o.f.).
Before the SSB mechanism, the Lagrangian contains massless
$W^\pm$ and $Z$ bosons, i.e. $3\times 2 = 6$ d.o.f., due to the 2
possible polarizations of a massless spin--1 field, and 4
real scalar fields. After SSB, the 3 Goldstone modes are ``eaten''
by the weak gauge bosons, which become massive and, therefore,
acquire one additional longitudinal polarization.
We have then $3\times 3=9$ d.o.f. in the gauge sector,
plus the remaining
scalar particle $H$, which is called the Higgs boson. The total
number of d.o.f. remains of course the same.

\subsection{Predictions}
\label{subsec:predictions}

We have now all the needed ingredients to describe the electroweak
interaction within a well-defined Quantum Field Theory.
Our theoretical framework implies the existence of massive
intermediate gauge bosons, $W^\pm$ and $Z$. Moreover,
the Higgs-Kibble mechanism has produced a precise
prediction\footnote{
%%%%%%%%%%%%% Footnote %%%%%%%%%%%
Note, however, that the relation
$M_Z\cos{\theta_W} =  M_W$ has a more general validity.
It is a direct consequence of the symmetry properties of
$\cL_S$ and does not depend on its detailed dynamics.}
%%%%%%%%%%% End Footnote %%%%%%%%%
for
the $W^\pm$ and $Z$ masses, relating them to the vacuum expectation
value of the scalar field through Eq.~\eqn{eq:boson_masses}.
Thus, $M_Z$ is predicted to be bigger than $M_W$ in agreement
with the measured masses \cite{LEPEWWG:04}:
\bel{eq:WZmass}
M_Z = 91.1875\pm 0.0021\:\mathrm{GeV}\, ,
\qquad\qquad
M_W = 80.425\pm 0.034\:\mathrm{GeV}\, .
\ee
From these experimental numbers, one obtains the electroweak
mixing angle
\bel{eq:thetaW}
\sin^2{\theta_W}\, =\, 1 - {M_W^2\over M_Z^2} \, =\,  0.222\, .
\ee

We can easily get and independent estimate of $\sin^2{\theta_W}$ from
the decay $\mu^-\to e^-\bar\nu_e\,\nu_\mu$. The momentum
transfer \ $q^2 = (p_\mu - p_{\nu_\mu})^2 = (p_e + p_{\nu_e})^2
\lsim m_\mu^2$ \ is much smaller than $M_W^2$. Therefore,
the $W$ propagator in Fig.~\ref{fig:ChargedCurrents} shrinks
to a point and can be well approximated through a local
four-fermion interaction, i.e.
\bel{eq:4fermion}
{g^2\over M_W^2 - q^2}\,\approx\, {g^2\over M_W^2}
\, =\, {4\pi\alpha\over\sin^2{\theta_W} M_W^2}
\,\equiv\, 4\sqrt{2}\, G_F\, .
\ee
The measured muon lifetime,
$\tau_\mu = (2.19703\pm 0.00004)\cdot 10^{-6}$~s \cite{PDG:04}, provides
a very precise determination of the Fermi coupling constant $G_F$:
\bel{eq:muonlifetime}
{1\over \tau_\mu}\, =\, \Gamma_\mu\, =\,
{G_F^2 m_\mu^5\over 192\,\pi^3}\, f(m_e^2/m_\mu^2)
\,\left( 1 +\delta_{\mathrm{RC}}\right)\, ,
\qquad
f(x)\,\equiv\, 1-8x+8x^3-x^4-12x^2\log{x}\, .
\ee
Taking into account the radiative corrections $\delta_{\mathrm{RC}}$,
which are known to $O(\alpha^2)$ \cite{muRC}, one gets
\cite{PDG:04}:
\bel{eq:G_F}
G_F\, =\, (1.16637\pm 0.00001)\cdot 10^{-5}~\mathrm{GeV}^{-2}\, .
\ee
The measured values of
$\alpha^{-1} = 137.03599911\, (46)$, $M_W$ and $G_F$ imply
\bel{eq:thetaW2}
\sin^2{\theta_W}\, =\, 0.215\, ,
\ee
in very good agreement with \eqn{eq:thetaW}. We will see later
than the small difference between these two numbers
can be understood in terms of higher-order quantum corrections.
The Fermi coupling gives also a direct determination of
the electroweak scale, i.e the scalar vacuum expectation value:
\bel{eq:v}
v \, =\, \left(\sqrt{2}\, G_F\right)^{-1/2}\, =\, 246\:\mbox{\rm GeV}\, .
\ee

\subsection{The Higgs boson}
\label{subsec:Higgs}

%%%%%%%%%%%%%%%%%%%%%%%%%%%%%%%%%%%%%%%%%%%%%%
%%
%%   FIGURE Higgs-Gauge couplings
%%
\begin{figure}[htb]
\begin{center}
\includegraphics[width=10cm]{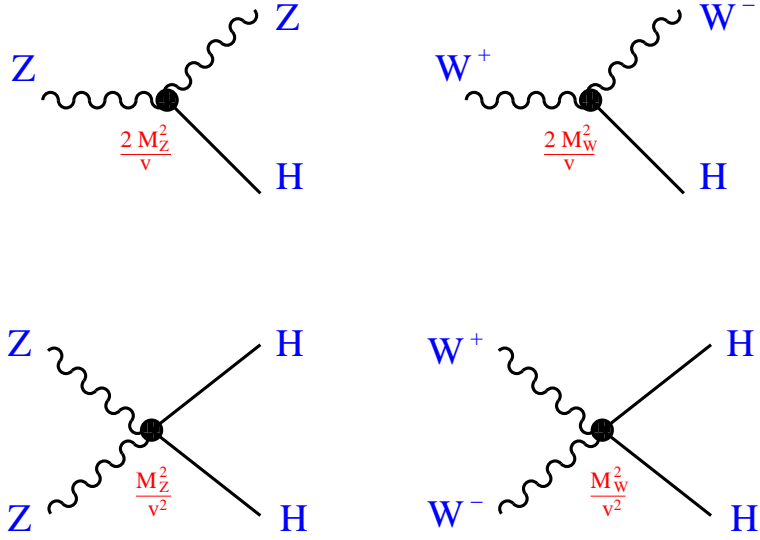}
\caption{Higgs couplings to the gauge bosons.}
\label{fig:HiggsWZcoup}
\end{center}
\end{figure}
%%%%%%%%%%%%%%%%%%%%%%%%%%%%%%%%%%%%%%%%%%%%%%%

%In addition to generate the needed gauge-boson masses, through the
%SSB mechanism,
The scalar Lagrangian \eqn{eq:LS}
has introduced a new scalar particle into the model: the Higgs $H$.
In terms of the physical fields (unitary gauge), $\cL_S$ takes the
form
\bel{eq:H_lag}
\cL_S\, = \, {1\over 4}\, h\, v^4\, + \,\cL_H \, + \, \cL_{HG^2} \, ,
\ee
where
\beqn\label{eq:H_int}
&&\cL_H \, = \, {1\over 2}\, \partial_\mu H \partial^\mu H -
{1\over 2}\, M_H^2\, H^2
- {M_H^2\over 2 v}\, H^3 - {M_H^2\over 8 v^2}\, H^4\, ,
\\[5pt] \label{eq:HGG}
\cL_{HG^2} &\! = &\! M_W^2\, W_\mu^\dagger W^\mu\,
\left\{ 1+ {2\over v}\, H
+ {H^2\over v^2}\right\}\, + \,
{1\over 2}\, M_Z^2\, Z_\mu Z^\mu \,\left\{ 1+{2\over v}\, H
+ {H^2\over v^2}\right\}\, ,
\eeqn
and the Higgs mass is given by
\bel{H_mass}
M_H\, = \, \sqrt{-2\mu^2}\, =\, \sqrt{2 h}\, v \, .
\ee
The Higgs interactions have a very characteristic form:
they are always proportional to the mass (squared) of the coupled boson.
All Higgs couplings
are determined by $M_H$, $M_W$, $M_Z$ and the vacuum expectation value $v$.

So far the experimental searches for the Higgs have only provided
a lower bound on its mass, corresponding to the exclusion
of the kinematical range accessible at LEP and the Tevatron
\cite{PDG:04}:   %%%{LEPEWWG:04}
\bel{eq:MH_low}
M_H\, >\, 114.4\:\mathrm{GeV}\qquad (95\%\;\mathrm{C.L.})\, .
\ee

\subsection{Fermion masses}
\label{subsec:f_masses}

%%%%%%%%%%%%%%%%%%%%%%%%%%%%%%%%%%%%%%%%%%%%%%
%%
%%   FIGURE Higgs-Fermion couplings
%%
\begin{figure}[htb]
\begin{center}
\includegraphics[width=4cm]{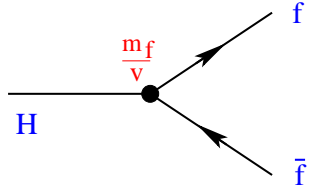}
\caption{Fermionic coupling of the Higgs boson.}
\label{fig:Hffcoup}
\end{center}
\end{figure}
%%%%%%%%%%%%%%%%%%%%%%%%%%%%%%%%%%%%%%%%%%%%%%%

A fermionic mass term \
$\cL_m = -m \,\overline{\psi}\psi = - m \left(\overline{\psi}_L\psi_R
+  \overline{\psi}_R\psi_L\right)$ \
is not allowed, because it breaks the gauge symmetry.
However, since we have introduced an additional scalar doublet into
the model, we can write the following gauge-invariant fermion-scalar coupling:
\bel{eq:yukawa}
\cL_Y\, =\, c_1\, \left(\bar u , \bar d\right)_L
\left(\ba \phi^{(+)}\\ \phi^{(0)}\ea\right)\, d_R \, + \,
c_2\,\left(\bar u , \bar d\right)_L
\left(\ba \phi^{(0)*}\\ -\phi^{(-)}\ea\right)\, u_R \, + \,
c_3\,\left(\bar \nu_e , \bar e\right)_L
\left(\ba \phi^{(+)}\\ \phi^{(0)}\ea\right)\, e_R
\, +\, \mbox{\rm h.c.}\, ,
\ee
where the second term involves the
$\cC$--conjugate scalar field $\phi^c\equiv i\,\sigma_2\,\phi^*$.
In the unitary gauge (after SSB),
this Yukawa-type Lagrangian takes the simpler form
\bel{eq:y_m}
\cL_Y\, =\, {1\over\sqrt{2}} \,(v+H)\,\left\{ c_1 \,\bar d d + c_2
\,\bar u u + c_3 \,\bar e e\right\}\, .
\ee
Therefore, the SSB mechanism generates also fermion masses:
\bel{eq:f_masses}
m_d\, = \, - c_1\, {v\over\sqrt{2}} \, , \qquad
m_u\, = \, - c_2 \, {v\over\sqrt{2}} \, , \qquad
m_e\, = \, - c_3\, {v\over\sqrt{2}} \, .
\ee

Since we do not know the parameters $c_i$, the values of the fermion masses
are arbitrary. Note, however, that all Yukawa couplings are fixed in terms
of the masses:
\bel{eq:y_mass}
\cL_Y\, =\, -\left( 1 + {H\over v}\right)
 \,\left\{ m_d \,\bar d d + m_u \,\bar u u
+ m_e \,\bar e e\right\}\, .
\ee
%

%%%%%%%%%% EW PHENOMENOLOGY %%%%%%%%%%

\setcounter{equation}{0}
\section{ELECTROWEAK \ PHENOMENOLOGY}
\label{sec:phenom}

In the gauge and scalar sectors, the SM Lagrangian contains only
4 parameters: $g$, $\gp$, $\mu^2$ and $h$. One could trade them
by $\alpha$, $\theta_W$, $M_W$ and $M_H$.
Alternatively, we can choose as free parameters \cite{PDG:04,LEPEWWG:04}:
\beqn\label{eq:SM_inputs}
G_F & = & (1.166\, 37 \pm 0.000\, 01) \cdot 10^{-5}\:\mathrm{GeV}^{-2}\, ,
\no\\
\alpha^{-1} & = & 137.035\, 999\, 11\pm 0.000\, 000\, 46\, ,
\\
M_Z & = & (91.1875\pm 0.0021)\,\mathrm{GeV}
\no\eeqn
and the Higgs mass $M_H$. This has the advantage of using the three
most precise experimental determinations to fix the interaction.
The relations
\bel{eq:A_def}
\sin^2{\theta_W}\,  =\,  1 - {M_W^2\over M_Z^2}\, ,
\qquad\qquad\qquad
M_W^2 \sin^2{\theta_W}\, =\, {\pi\alpha\over\sqrt{2}\, G_F}\,
%%% \equiv\, A\, =\,  37.2805\;\mathrm{GeV}^2\, ,
\ee
determine then \ $\sin^2{\theta_W} = 0.212$ \ and \ $M_W= 80.94\;\mathrm{GeV}$.
The predicted $M_W$ is in
good agreement with the measured value in \eqn{eq:WZmass}.
% \cite{PDG:04},
% $M_W = (80.425\pm 0.038)$ GeV.

%%%%%%%%%%%%%%%  FIGURE %%%%%%%%%%%%%%%%%%%%%%%%%
\begin{figure}[tbh]\centering
\begin{minipage}[c]{.4\linewidth}\centering
\includegraphics[width=4cm]{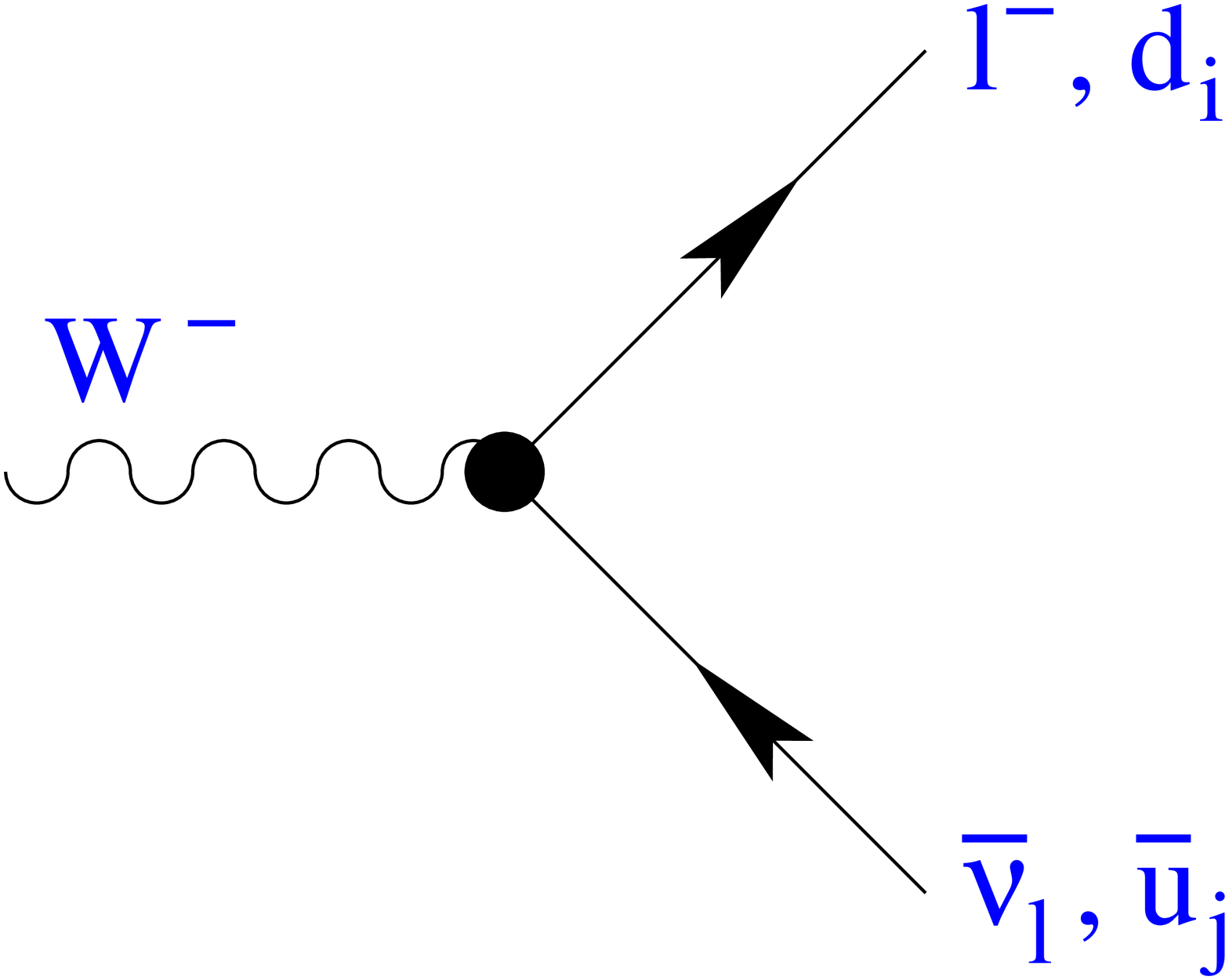}
\end{minipage}
\hskip 1cm
\begin{minipage}[c]{.4\linewidth}\centering
\includegraphics[width=3.4cm]{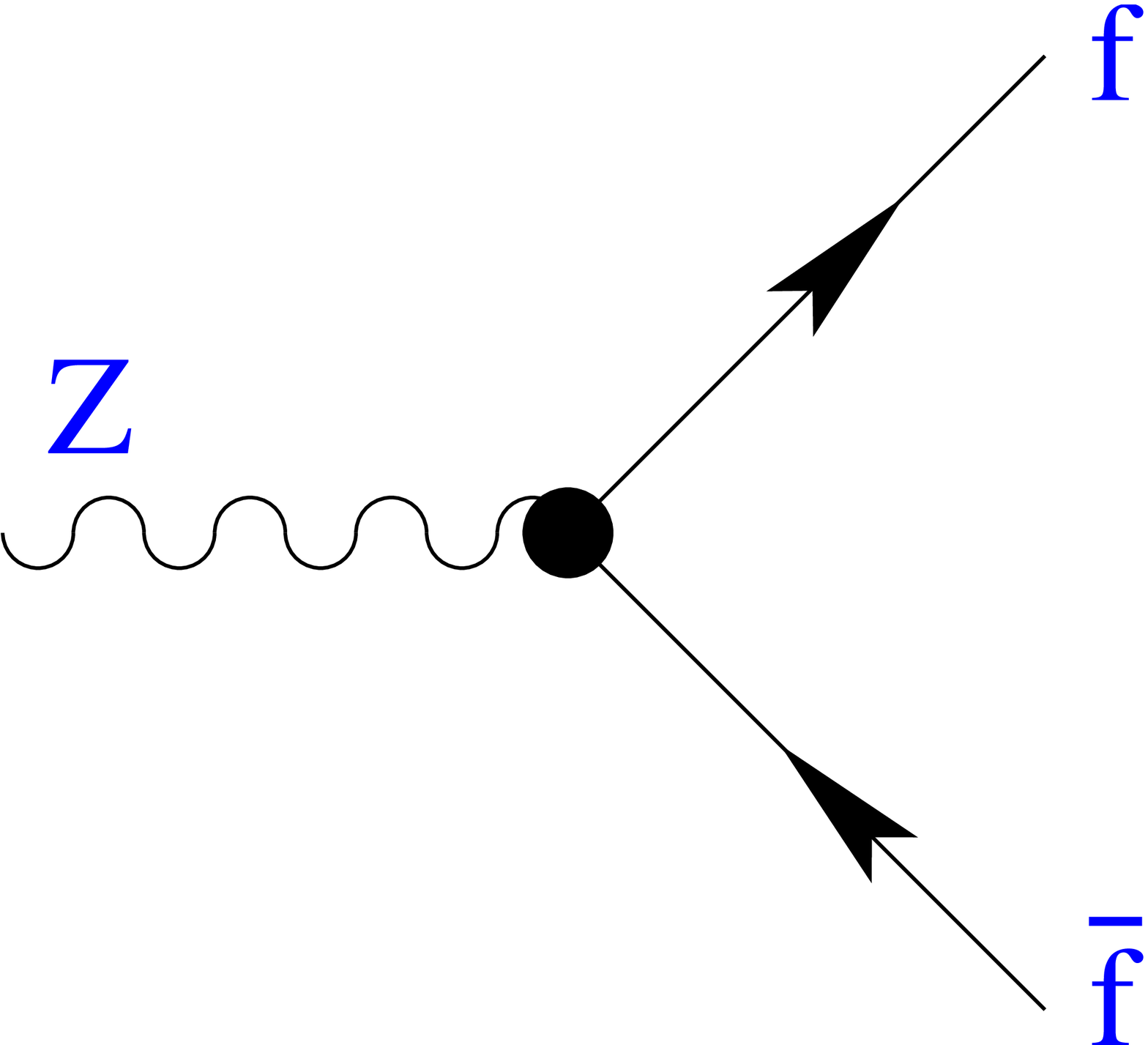}
\end{minipage}
%\vskip -.3cm
\caption{Tree-level Feynman diagrams contributing to the $W^\pm$ and $Z$ decays.}
\label{fig:WZdecay}
\end{figure}
%%%%%%%%%%%%%%%%%%%%%%%%%%%%%%%%%%%%%%%%%%%%%

At tree level, the decay widths of the weak gauge bosons can be
easily computed. The $W$ partial widths,
\bel{eq:W_width}
\Gamma\left(W^-\to \bar\nu_l l^-\right)\,
=\, {G_F M_W^3\over 6\pi\sqrt{2}} \, ,
%%% =\, 0.2320\;\mathrm{GeV}\, ,
\qquad\qquad\qquad
\Gamma\left(W^-\to \bar u_i d_j\right)\,
=\, N_C\; |\mathbf{V}_{\! ij}|^2\; {G_F M_W^3\over 6\pi\sqrt{2}}
%%%\, =\, 0.6960\, |V_{ij}|^2\:\mathrm{GeV}
\, ,
\ee
are equal for all leptonic decay modes (up to small kinematical mass
corrections from phase space). The quark modes involve also the
colour quantum number $N_C=3$ and the mixing factor \
$d\,'_i = \mathbf{V}_{\! ij}\, d_j$ \ relating weak and mass eigenstates.
The $Z$ partial widths are different for each decay mode, since
its couplings depend on the fermion charge:
\bel{eq:Z_width}
\Gamma\left( Z\to \bar f f\right)\, =\, N_f\,
{G_F M_Z^3\over 6\pi\sqrt{2}} \, \left(|v_f|^2 + |a_f|^2\right)
%%% \, =\, 0.3318\, N_f\, \left(|v_f|^2 + |a_f|^2\right)
%%% \, \,\mathrm{GeV}
\, ,
\ee
where $N_l=1$ and $N_q=N_C$.
Summing over all possible final fermion pairs, one predicts
the total widths \
$\Gamma_W=2.09$ GeV \ and \ $\Gamma_Z=2.48$ GeV, in excellent agreement
with the experimental values
$\Gamma_W=(2.133\pm 0.069)$ GeV \
and \ $\Gamma_Z=(2.4952\pm 0.0023)$ GeV \cite{LEPEWWG:04}.

The universality of the $W$ couplings implies
\bel{eq:W_lep}
\mathrm{Br}(W^-\to\bar\nu_l\, l^-)\, = \,
{1\over 3 + 2 N_C}\, = \, 11.1 \% \, ,
\ee
where we have taken into account that the decay into the top quark
is kinematically forbidden.
Similarly, the leptonic decay widths of the $Z$ are predicted
to be \
$\Gamma_l\equiv\,\Gamma(Z\to l^+l^-) = 84.85\;\mathrm{MeV}$.
As shown in Table~\ref{tab:leptonic_modes},
these predictions are in good agreement with the
measured leptonic widths, confirming the universality of the
$W$ and $Z$ leptonic couplings.
There is however an excess of the branching ratio $W\to\tau\,\bar\nu_\tau$ \
with respect to \ $W\to e\,\bar\nu_e$ \ and \ $W\to\mu\,\bar\nu_\mu\,$, which
represents a $2.8\,\sigma$ effect \cite{LEPEWWG:04}.

The universality of the leptonic $W$ couplings can be also tested
indirectly, through weak decays mediated by charged-current
interactions.
Comparing the measured decay widths of leptonic or semileptonic decays
which only differ by the lepton flavour, one can test experimentally
that the $W$ interaction is indeed the same, i.e. that \
$g_e = g_\mu = g_\tau \equiv g\, $.
As shown in Table~\ref{tab:univtm}, the present data \cite{PDG:04,LEPEWWG:04}
verify the universality of the leptonic charged-current couplings
to the 0.2\% level.

%%%%%%%%%%%%%  Table %%%%%%%%%%%%%%%%%%%%%%%%
\begin{table}[htb]
\begin{center}
{\renewcommand{\arraystretch}{1.3}
\caption{Measured values of \
$\mbox{\rm Br} (W^-\to\bar\nu_l\; l^-)$ \ and \ $\Gamma(Z\to l^+l^-)$
%%% and the leptonic forward-backward asymmetries
\protect\cite{LEPEWWG:04}.
The average of the three leptonic modes is shown in the last column
(for a massless charged lepton $l$).
\label{tab:leptonic_modes}}
\vspace{0.2cm}
\begin{tabular}{|c||c|c|c||c|}
\hline
& $e$ & $\mu$ & $\tau$ & $l$
\\ \hline\hline
Br($W^-\to\bar\nu_l l^-$) \, (\%) & $10.66\pm 0.17$
& $10.60\pm 0.15$ & $11.41\pm 0.22$ & $10.84\pm 0.09$
\\
$\Gamma(Z\to l^+l^-)$ \, (MeV) & $83.92\pm 0.12$
& $83.99\pm 0.18$ & $84.08\pm 0.22$ & $83.984\pm 0.086$
% \\
% $\cA_{\mbox{\rms FB}}^{0,l}$ \, (\%) & $1.56\pm 0.34$
% & $1.41\pm 0.21$ & $2.28\pm 0.26$ & $1.70\pm 0.16$
\\ \hline
\end{tabular}}
\end{center}
\end{table}
%%%%%%%%%%%%% End Table %%%%%%%%%%%%%%%%%%%%%

%%%%%%%%%%%%%%%%%%%%  TABLES  %%%%%%%%%%%%%%%%%
\begin{table}[htb]
\caption{Experimental determinations of the ratios \ $g_l/g_{l'}$}
\begin{center}
\renewcommand{\arraystretch}{1.3}
\begin{tabular}{|c|ccc|}
\hline
& $\Gamma_{\tau\to\nu_\tau\mu\,\bar\nu_\mu}/
\Gamma_{\tau\to\nu_\tau e\,\bar\nu_e}$ &
$\Gamma_{\pi\to\mu\,\bar\nu_\mu} /\Gamma_{\pi\to e\,\bar\nu_e}$ &
$\Gamma_{W\to\mu\,\bar\nu_\mu} /\Gamma_{W\to e\,\bar\nu_e}$
\\ \hline
$|g_\mu/g_e|$
& $0.9999\pm 0.0020$ & $1.0017\pm 0.0015$ & $0.997\pm 0.010$
\\ \hline
\end{tabular}
\end{center}
%\label{tab:univtm}
%\end{table}
%%%%%%%%%%%%%%%%%%%%%%%%%%%%%%%%%%%%%%%%%%%%%%%%
%\begin{table}
%\caption{Present constraints on $|g_\tau/g_\mu|$}
\begin{center}
\renewcommand{\arraystretch}{1.3}
\begin{tabular}{|c|cccc|}
\hline
& $\Gamma_{\tau\to\nu_\tau e\,\bar\nu_e}/\Gamma_{\mu\to\nu_\mu e\,\bar\nu_e}$ &
$\Gamma_{\tau\to\nu_\tau\pi}/\Gamma_{\pi\to\mu\,\bar\nu_\mu}$ &
$\Gamma_{\tau\to\nu_\tau K}/\Gamma_{K\to\mu\,\bar\nu_\mu}$ &
$\Gamma_{W\to\tau\,\bar\nu_\tau}/\Gamma_{W\to\mu\,\bar\nu_\mu}$
%$\Gamma_{W\to\tau\,\bar\nu_\tau/\mu\,\bar\nu_\mu}$
\\ \hline
$|g_\tau/g_\mu|$ & $1.0004\pm 0.0023$ & $0.9999\pm 0.0036$ &
$0.979\pm 0.017$ & $1.037\pm 0.014$
%%% $1.008\pm 0.014$
\\ \hline
\end{tabular}
\end{center}
%\label{tab:univtm}
%\end{table}
%%%%%%%%%%%%%%%%%%%%%%%%%%%%%%%%%%%%%%%%%%%%%%%%
%\begin{table}
%\caption{Present constraints on $|g_\tau/g_e|$}
\begin{center}
\renewcommand{\arraystretch}{1.3}
\begin{tabular}{|c|cc|}
\hline
& $\Gamma_{\tau\to\nu_\tau \mu\,\bar\nu_\mu}/\Gamma_{\mu\to\nu_\mu e\,\bar\nu_e}$ &
$\Gamma_{W\to\tau\,\bar\nu_\tau}/\Gamma_{W\to e\,\bar\nu_e}$
\\ \hline
$|g_\tau/g_e|$ & $1.0002\pm 0.0022$ & $1.034\pm 0.014$
%%% $1.001\pm 0.014$
\\ \hline
\end{tabular}
\end{center}
\label{tab:univtm}
\end{table}
%%%%%%%%%%%%%%%%%%%%%%%%%%%%%%%%%%%%%%%%%%%%%%%%

Another interesting quantity is the $Z$ decay width into invisible modes,
\bel{eq:Z_inv}
 {\Gamma_{\rms inv}\over\Gamma_l}\,\equiv\,
{N_\nu\;\Gamma(Z\to\bar\nu\,\nu)\over\Gamma_l}\, = \,
{2\, N_\nu\over (1 - 4\, \sin^2{\theta_W})^2 + 1}\; ,
%%% {N_\nu\over 2\, (|v_l|^2 + |a_l|^2)} \, = \, 5.865 \, ,
\ee
which is usually normalized to the charged leptonic width.
% and the ratio \
% $R_Z\equiv\Gamma(Z\to\mathrm{hadrons})/\Gamma_l = 20.29 \,$.
The comparison with the measured value,
$\Gamma_{\rms inv}/\Gamma_l = 5.942 \pm 0.016$  \cite{PDG:04,LEPEWWG:04},
%%% and \ $R_Z = 20.767\pm 0.025\,$,
provides a very strong experimental evidence of the existence of
three different light neutrinos.

\subsection{Fermion-pair production at the $Z$ peak}
\label{subsec:Zpeak}

%%%%%%%%%%%%%%%  FIGURE %%%%%%%%%%%%%%%%%%%%%%%%%
\begin{figure}[tbh]\centering
\includegraphics[width=10cm]{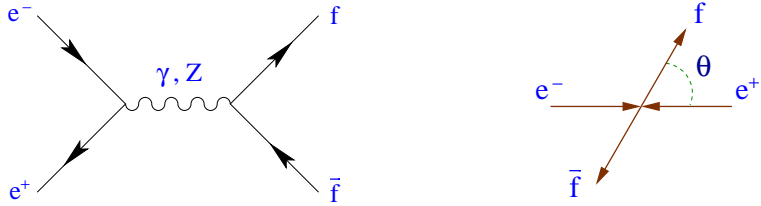}
\caption{Tree-level contributions to \ $e^+e^-\to\bar f f \,$ \
and kinematical configuration in the centre-of-mass system.}
\label{fig:eeZff}
\end{figure}
%%%%%%%%%%%%%%%%%%%%%%%%%%%%%%%%%%%%%%%%%%%%%

Additional information can be obtained from the study of the
process \ $e^+e^-\to\gamma,Z\to\bar f f \,$.
For unpolarized $e^+$ and $e^-$ beams, the differential %%%production
cross-section can be written, at lowest order, as
\bel{eq:dif_cross}
{d\sigma\over d\Omega}\, = \, {\alpha^2\over 8 s} \, N_f \,
         \left\{ A \, (1 + \cos^2{\theta}) \, + B\,  \cos{\theta}\,
     - \, h_f \left[ C \, (1 + \cos^2{\theta}) \, +\, D \cos{\theta}
         \right] \right\} ,
\ee
where \ $h_f=\pm 1$ \ denotes the sign of the helicity of
the produced fermion $f$ and $\theta$ is the scattering angle
between $e^-$ and $f$ in the centre-of-mass system. Here,
\beqn\label{eq:A}
A & = & 1 + 2\, v_e v_f \,\mathrm{Re}(\chi)
 + \left(v_e^2 + a_e^2\right) \left(v_f^2 + a_f^2\right) |\chi|^2\,,
\no\\ \label{eq:B}
B & = & 4\, a_e a_f \,\mathrm{Re}(\chi) + 8\, v_e a_e v_f a_f\,  |\chi|^2\, ,
\no\\ \label{eq:C}
C & = & 2\, v_e a_f \,\mathrm{Re}(\chi) + 2\, \left(v_e^2 + a_e^2\right) v_f
      a_f\, |\chi|^2\, ,
\no\\ \label{eq:D}
D & = & 4\, a_e v_f \,\mathrm{Re}(\chi) + 4\, v_e a_e \left(v_f^2 +
      a_f^2\right) |\chi|^2\,  ,
\eeqn
and  $\chi$  contains the $Z$  propagator
\bel{eq:Z_propagator}
\chi \, = \, {G_F M_Z^2 \over 2 \sqrt{2} \pi \alpha }
     \; {s \over s - M_Z^2 + i s \Gamma_Z  / M_Z } \, .
\ee

The coefficients $A$, $B$, $C$ and $D$ can be experimentally determined,
by measuring the total cross-section, the forward-backward asymmetry,
the polarization asymmetry and the forward-backward polarization
asymmetry, respectively:
$$
\sigma(s)\, =\,
{4 \pi \alpha^2 \over 3 s } \, N_f \, A \, ,
\qquad\qquad\qquad\qquad
\cA_{\rms FB}(s)\,\equiv\, {N_F - N_B \over N_F + N_B}
   \, =\,  {3 \over 8} {B \over A}\, ,
$$

\bel{eq:A_pol}
\cA_{\rms Pol}(s) \,\equiv\,
{\sigma^{(h_f =+1)}
- \sigma^{(h_f =-1)} \over \sigma^{(h_f =+1)} + \sigma^{(h_f = -1)}}
\, =\,  - {C \over A} \, ,
\ee

$$
\cA_{\rms FB,Pol}(s)\, \equiv\,
{N_F^{(h_f =+1)} -
N_F^{(h_f = -1)} - N_B^{(h_f =+1)} + N_B^{(h_f = -1)} \over
 N_F^{(h_f =+1)} + N_F^{(h_f = -1)} + N_B^{(h_f =+1)} + N_B^{(h_f = -1)}}
\, =\,  -{3 \over 8} {D \over A}\, .
$$

\noindent
Here, $N_F$ and $N_B$ denote the number of $f$'s
emerging in the forward and backward hemispheres,
respectively, with respect to the electron direction.
The measurement of the final fermion polarization can be done for $f=\tau$,
by measuring the distribution of the final $\tau$ decay products.

For $s = M_Z^2$, the real part of the $Z$ propagator vanishes
and the photon-exchange terms can be neglected
in comparison with the $Z$-exchange contributions
($\Gamma_Z^2 / M_Z^2 << 1$). Eqs. \eqn{eq:A_pol}
become then,
$$
\sigma^{0,f} \,\equiv\,\sigma(M_Z^2)\, =\,
 {12 \pi  \over M_Z^2 } \; {\Gamma_e \Gamma_f\over\Gamma_Z^2}\, ,
\qquad\qquad\qquad
\cA_{\rms FB}^{0,f}\,\equiv\,\cA_{FB}(M_Z^2)\, =\, {3 \over 4}\,
\cP_e \cP_f \, ,
$$
\bel{eq:A_pol_Z}
\cA_{\rms Pol}^{0,f} \,\equiv\,
  \cA_{\rms Pol}(M_Z^2)\, =\, \cP_f \, ,
\qquad\qquad\qquad
\cA_{\rms FB,Pol}^{0,f} \,\equiv\,
\cA_{\rms FB,Pol}(M_Z^2)\, =\,  {3 \over 4}\, \cP_e  \, ,
\ee
where $\Gamma_f$ is the $Z$ partial decay width into the $\bar f f$ final state,
and
\bel{eq:P_f}
\cP_f \,\equiv \, - A_f \,\equiv \,
{ - 2\, v_f a_f \over v_f^2 + a_f^2}
\ee
is the average longitudinal polarization of the fermion $f$,
which only depends on the ratio of the vector and axial-vector couplings.

With polarized $e^+e^-$ beams, which have been available at SLC,
one can also study the left-right
asymmetry between the cross-sections for initial left- and right-handed
electrons, and the corresponding forward-backward left-right asymmetry:
\bel{eq:A_LR}
\cA_{\rms LR}^0\,\equiv\,
\cA_{\rms LR}(M_Z^2)
  \, = \, {\sigma_L(M_Z^2)
- \sigma_R(M_Z^2) \over \sigma_L(M_Z^2) + \sigma_R(M_Z^2)}
\, = \, - \cP_e \,  ,
\qquad\quad
\cA_{\rms FB,LR}^{0,f} \,\equiv\,
\cA_{\rms FB,LR}(M_Z^2)\, =\, - {3 \over 4}\, \cP_f \, .
\ee
At the $Z$ peak, $\cA_{\rms LR}^0$ measures
the average initial lepton polarization, $\cP_e$,
without any need for final particle identification,
while $\cA_{\rms FB,LR}^{0,f}$ provides a direct determination
of the final fermion polarization.

$\cP_f$ is a very sensitive function of $\sin^2{\theta_W}$.
Small higher-order corrections can produce large variations on the
predicted lepton polarization, because
$|v_l| =\frac{1}{2}\, |1-4\,\sin^2{\theta_W}|\ll 1$.
Therefore, $\cP_l$ provides an
interesting window to search for electroweak quantum effects.

\subsection{Higher-order corrections}
\label{subsec:loops}

%%%%%%%%%%%%%%%  FIGURE %%%%%%%%%%%%%%%%%%%%%%%%%
\begin{figure}[tbh]\centering
\begin{minipage}[c]{.3\linewidth}\centering
\includegraphics[width=5.5cm]{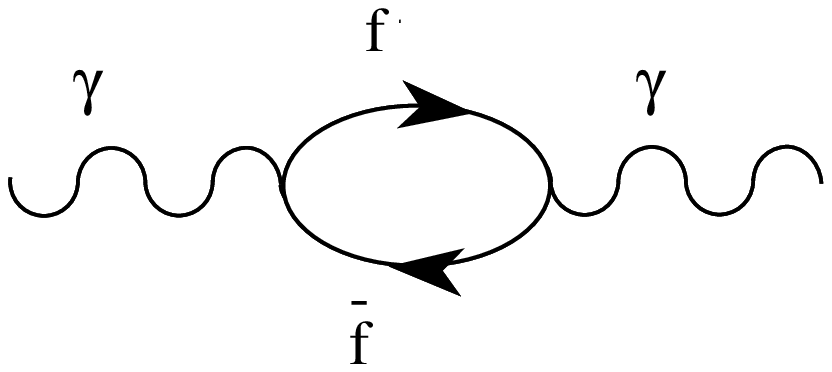}
\end{minipage}
\hskip 2cm
\begin{minipage}[c]{.5\linewidth}\centering
\includegraphics[width=7cm]{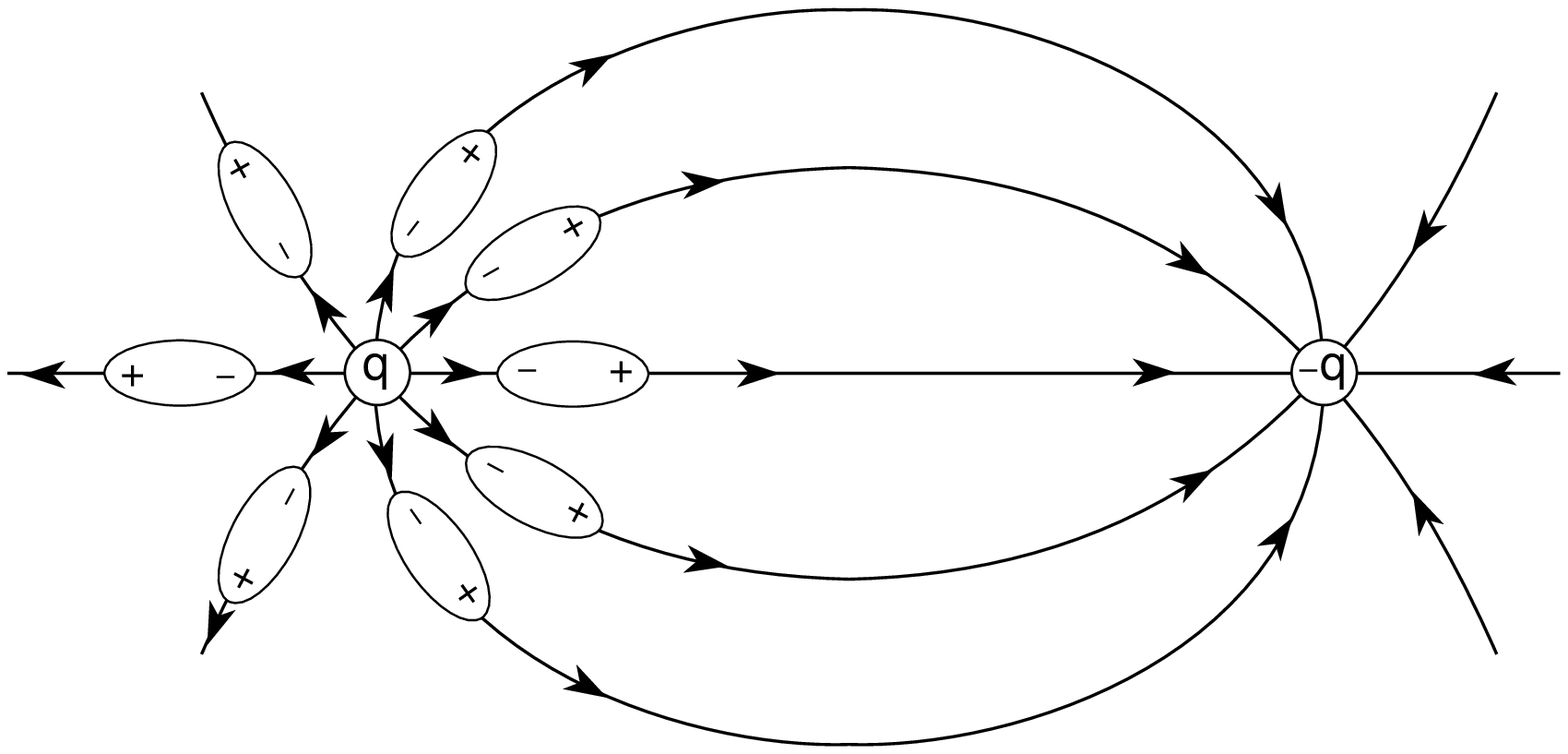}
\end{minipage}
\caption{The photon vacuum polarization (left)
generates a charge screening effect, making $\alpha(s)$ smaller
at larger distances.}
\label{fig:PhotonPolarization}
\end{figure}
%%%%%%%%%%%%%%%%%%%%%%%%%%%%%%%%%%%%%%%%%%%%%

Before trying to analyze the relevance of higher-order electroweak
contributions, it is instructive to consider the numerical impact
of the well-known QED and QCD corrections.
The photon propagator gets vacuum polarization corrections, induced by
virtual fermion-antifermion pairs.
This kind of QED loop corrections can be taken into account through
a redefinition of the QED coupling, which depends on the energy scale.
The resulting QED running coupling $\alpha(s)$ decreases at large
distances. This can be intuitively understood as the charge screening
generated by the virtual fermion pairs. The physical QED vacuum behaves
as a polarized dielectric medium.
The huge difference between the electron and $Z$ mass scales makes
this quantum correction relevant at LEP energies \cite{PDG:04,LEPEWWG:04}:
\begin{equation}\label{eq:alpha}
\alpha(m_e^2)^{-1}\; =\; 137.035\, 999\, 11\, (46)
\;\; > \;\;
\alpha(M_Z^2)^{-1}\; =\; 128.95\pm 0.05\; .
\end{equation}

The running effect generates an important change in Eq.~\eqn{eq:A_def}.
Since $G_F$ is measured at low energies, while $M_W$ is a
high-energy parameter, the relation between both quantities is
modified by vacuum-polarization contributions.
Changing $\alpha$ by $\alpha(M_Z^2)$,
%%% in Eqs.~\eqn{eq:A_def},
one gets the corrected predictions:
\bel{eq:new_pred}
\sin^2{\theta_W}\, = \, 0.231 \, ,
\qquad\qquad\qquad
M_W\, = \, 79.96\,\mathrm{GeV} \, .
\ee
The experimental value of $M_W$ is in the range between the two
results obtained with either $\alpha$ or $\alpha(M_Z^2)$,
showing its sensitivity to quantum corrections.
The effect is more spectacular in the leptonic asymmetries at the
$Z$ peak. The small variation of $\sin^2{\theta_W}$ from 0.212 to 0.231,
induces a large shift on the vector $Z$ coupling to charged leptons
from \ $v_l= -0.076$ \ to \ $-0.038\,$, changing the predicted average lepton
polarization $\cP_l$ by a factor of two.

So far, we have treated quarks and leptons on an equal footing.
However, quarks are strong-interacting particles.
The gluonic corrections to the decays
$Z\to\bar q q$ and $W^-\to\bar u_i d_j$ can be directly incorporated
into the formulae given before, by taking an ``effective'' number of
colours:
\bel{N_C_eff}
N_C \quad \Longrightarrow \quad
N_C\,\left\{ 1 + {\alpha_s\over\pi} + \ldots\right\}\,
\approx\, 3.115 \, ,
\ee
where we have used the value of $\alpha_s$ at $s=M_Z^2$, \
$\alpha_s(M_Z^2)= 0.1182\pm 0.0027\, $ \cite{PDG:04,BE:04}.

Note that the strong coupling also ``runs''. However,
the gluon self-interactions generate an anti-screening effect,
through gluon-loop corrections to the gluon propagator, which spread
out the QCD charge. Since this correction is larger than
the screening of the colour charge induced by virtual quark--antiquark
pairs, the net result is that the strong coupling decreases at short
distances. Thus, QCD has the required property of
asymptotic freedom: quarks behave as free particles when $Q^2\to\infty$
\cite{PI:00,SE:04}.

QCD corrections increase the probabilities of the $Z$ and the $W^\pm$
to decay into hadronic modes. Therefore, their leptonic branching fractions
become smaller. The effect can be easily estimated from Eq.~\eqn{eq:W_lep}.
The probability of the decay $W^-\to \bar\nu_e\, e^-$ gets reduced from
11.1\% to 10.8\%, improving the agreement with the measured value in
Table~\ref{tab:leptonic_modes}.

%%%%%%%%%%%%%%%  FIGURE %%%%%%%%%%%%%%%%%%%%%%%%%
\begin{figure}[tbh]\centering
\includegraphics[width=11cm]{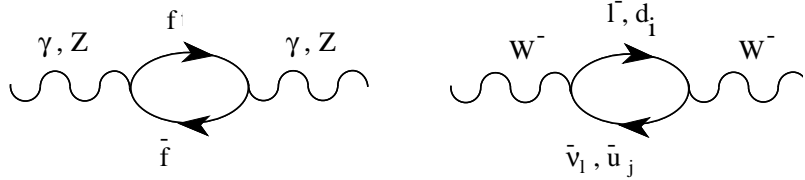}
\caption{Self-energy corrections to the gauge boson propagators.}
\label{fig:Oblique}
\end{figure}
%%%%%%%%%%%%%%%%%%%%%%%%%%%%%%%%%%%%%%%%%%%%%

Quantum corrections offer the possibility to be sensitive to
heavy particles, which cannot be kinematically accessed,
through their virtual loop effects.
In QED and QCD the vacuum polarization contribution of a heavy fermion pair
is suppressed by inverse powers of the fermion mass.
At low energies, the information on the heavy fermions is then lost.
This ``decoupling'' of the heavy fields happens in theories
with only vector couplings and an exact gauge symmetry \cite{AC:75},
where the effects generated by the heavy particles can always be
reabsorbed into a redefinition of the low-energy parameters.

The SM involves, however, a broken chiral gauge symmetry. This
has the very interesting implication of avoiding the decoupling
theorem \cite{AC:75}. The vacuum polarization contributions
induced by a heavy top generate corrections to
the $W^\pm$ and $Z$ propagators,
which increase quadratically with the top mass \cite{VE:77}.
Therefore, a heavy top does not decouple.
For instance, with $m_t = 178$ GeV, the leading quadratic correction
to the second relation in \eqn{eq:A_def} amounts to
a sizeable $3\% $ effect.
The quadratic mass contribution originates in the strong breaking
of weak isospin generated by the top and bottom quark masses, i.e.
the effect is actually proportional to $m_t^2-m_b^2$.

Owing to an accidental $SU(2)_C$ symmetry of the scalar sector
(the so-called custodial symmetry), the
virtual production of Higgs particles does not generate any
quadratic dependence on the Higgs mass at one loop \cite{VE:77}.
The dependence on $M_H$ is only logarithmic.
The numerical size of the corresponding correction in
Eq.~\eqn{eq:A_def} varies from a 0.1\% to a 1\% effect
for $M_H$ in the range from 100 to 1000 GeV.

%%%%%%%%%%%%%%%  FIGURE %%%%%%%%%%%%%%%%%%%%%%%%%
\begin{figure}[bth]\centering
\includegraphics[width=9cm]{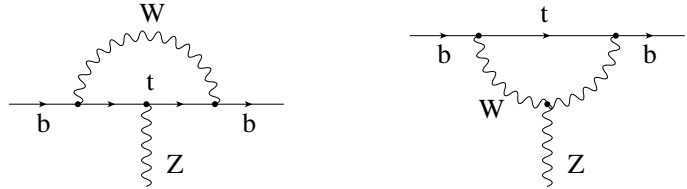}
\caption{One-loop corrections to the $Z\bar b b$ vertex,
involving a virtual top.}
\label{fig:Zbb}
\end{figure}
%%%%%%%%%%%%%%%%%%%%%%%%%%%%%%%%%%%%%%%%%%%%%

Higher-order corrections to the different electroweak couplings
are non-universal and usually smaller than the self-energy contributions.
There is one interesting exception, the $Z \bar bb$ vertex, which is sensitive
to the top quark mass \cite{BPS:88}.
The $Z\bar f f$ vertex gets one-loop corrections where a virtual
$W^\pm$ is exchanged between the two fermionic legs. Since, the $W^\pm$
coupling changes the fermion flavour, the decays
$Z\to \bar d d, \bar s s, \bar b b$ \ get contributions with a top quark
in the internal fermionic lines,
i.e. $Z\to\bar t t\to \bar d_i d_i$. Notice that this mechanism can also
induce the flavour-changing neutral-current decays
$Z\to \bar d_i d_j$ with $i\not=j$.
These amplitudes are suppressed by the small CKM mixing factors
$|\bV^{\phantom{*}}_{\! tj}\bV^*_{\! ti}|^2$.
However, for the $Z\to\bar b b$ vertex, there is no suppression
because $|\bV_{\! tb}|\approx 1$.

The explicit calculation
\cite{BPS:88,ABR:86} shows the presence of hard $m_t^2$ corrections to
the $Z\to\bar b b$ vertex. This effect can be easily understood
\cite{BPS:88}
in non-unitary gauges where the unphysical charged scalar $\phi^{(\pm)}$
is present. The fermionic couplings of the charged scalar
are proportional to the fermion masses; therefore, the exchange
of a virtual $\phi^{(\pm)}$ gives rise to a $m_t^2$ factor.
In the unitary gauge, the charged scalar has been ``eaten'' by the
$W^\pm$ field; thus, the effect comes now from the exchange of a
longitudinal $W^\pm$,
with terms proportional to $q^\mu q^\nu$ in the propagator that
generate fermion masses.
Since the $W^\pm$ couples only to left-handed fermions, the induced
correction is the same for the vector
and axial-vector $Z\bar b b$ couplings
and for $m_t = 178$~GeV amounts to a 1.7\% reduction of the
$Z\to \bar b b$ decay width \cite{BPS:88}.

The ``non-decoupling'' present in the
$Z\bar b b$ vertex is quite different from the one happening in
the boson self-energies.
The vertex correction does not have any dependence with the
Higgs mass. Moreover,
while any kind of new heavy particle
coupling to the gauge bosons would contribute to the $W$ and $Z$
self-energies, the possible new physics contributions to the
$Z\bar b b$ vertex are much more restricted and, in any case,
different.
Therefore, the independent experimental measurement of the two effects
is very valuable in order to disentangle possible
new physics contributions from the SM corrections.
In addition, since the ``non-decoupling'' vertex effect is related
to $W_L$-exchange, it is sensitive to the SSB mechanism.

\subsection{SM electroweak fit}
\label{subsec:EWfit}

%%%%%%%%%%%%%%%  FIGURE %%%%%%%%%%%%%%%%%%%%%%%%%
\begin{figure}[tbh]\centering
\begin{minipage}[c]{.45\linewidth}\centering
\includegraphics[width=7cm]{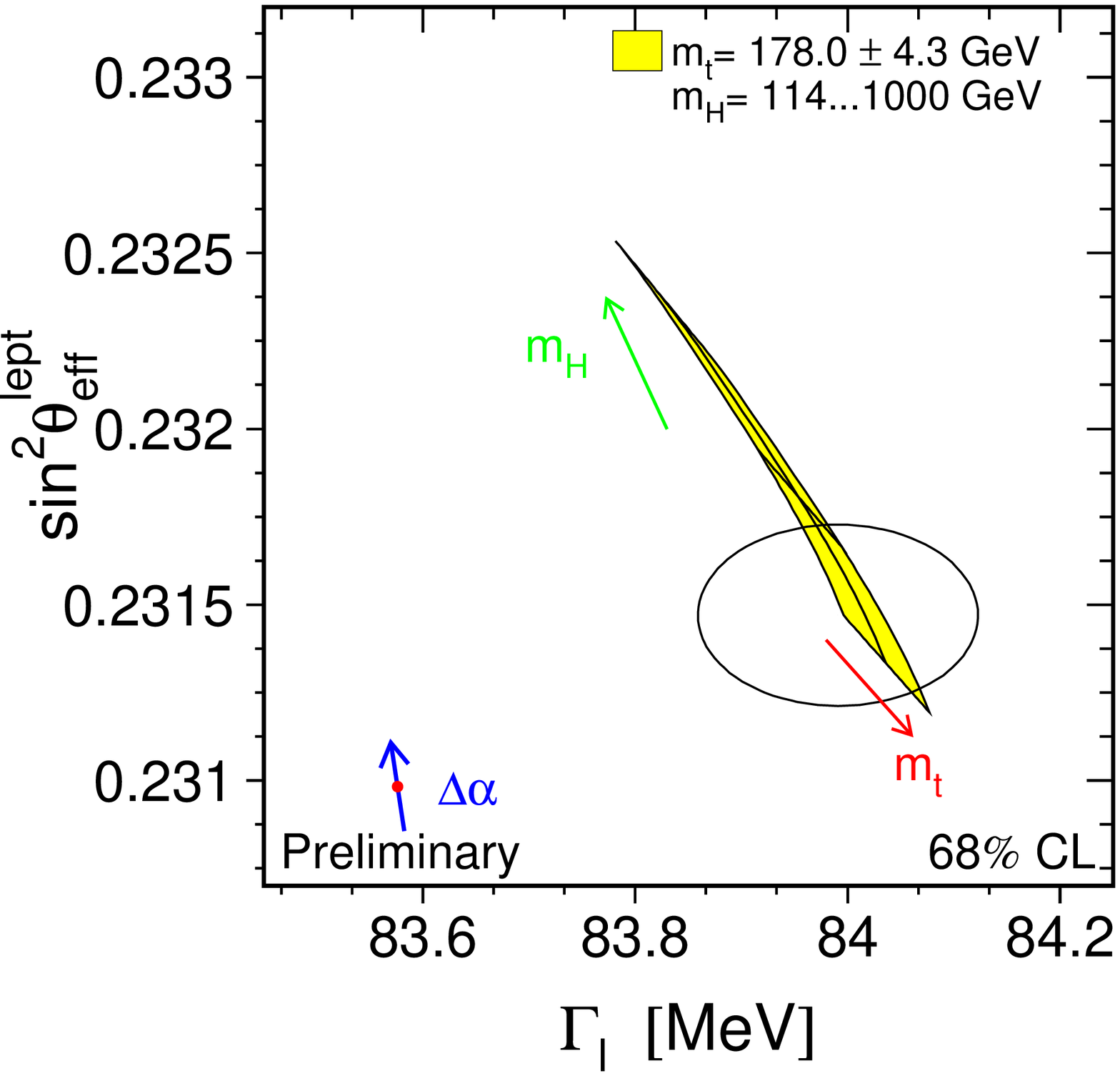}
\end{minipage}
\hskip 1cm
\begin{minipage}[c]{.45\linewidth}\centering
\includegraphics[width=7cm]{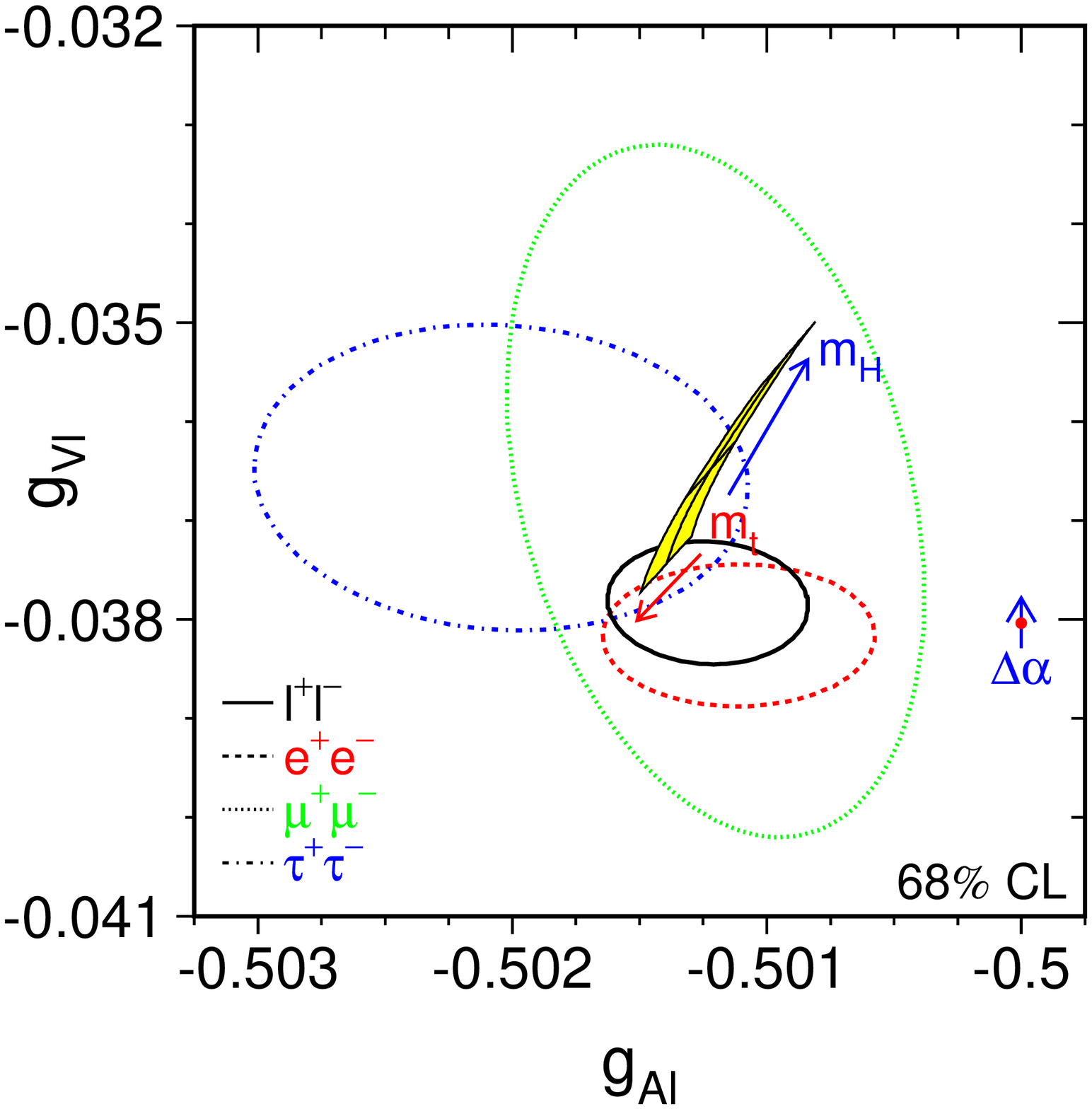}
\end{minipage}
\caption{Combined LEP and SLD measurements of
$\sweff$ and $\Gamma_l$ (left)
and the corresponding effective vector and axial-vector couplings
$v_l$ and $a_l$ (right). The shaded region shows the SM prediction.
The arrows point in the direction of increasing values of $m_t$ and $M_H$.
The point shows the predicted values if among the electroweak radiative
corrections only the photon vacuum polarization is included.
Its arrow indicates the variation induced by
the uncertainty in $\alpha(M_Z^2)$ \cite{LEPEWWG:04}.}
\label{fig:Zcouplings}
\end{figure}
%%%%%%%%%%%%%%%%%%%%%%%%%%%%%%%%%%%%%%%%%%%%%

The leptonic asymmetry measurements from LEP and SLD
can all be combined to determine the ratios $v_l/a_l$
of the vector and axial-vector couplings of the
three charged leptons,
or equivalently the effective electroweak mixing angle
\bel{eq:sW-eff}
\sweff\,\equiv\,
\frac{1}{4}\,\left(1 -\frac{v_l}{a_l}\right)\, .
\ee
The sum $(v_l^2 + a_l^2)$ is derived from
the leptonic decay widths of the $Z$, i.e. from
Eq.~\eqn{eq:Z_width}
corrected with a multiplicative factor \
$\left(1 + {3\over4}\, {\alpha\over\pi}\right)$
to account for final-state QED corrections.
The signs of $v_l$ and $a_l$ are fixed by requiring $a_e<0$.

%%%%%%%%%%%%%%%  FIGURE %%%%%%%%%%%%%%%%%%%%%%%%%
\begin{figure}[tbh]\centering
\begin{minipage}[t]{.46\linewidth}\centering
\includegraphics[width=6.25cm]{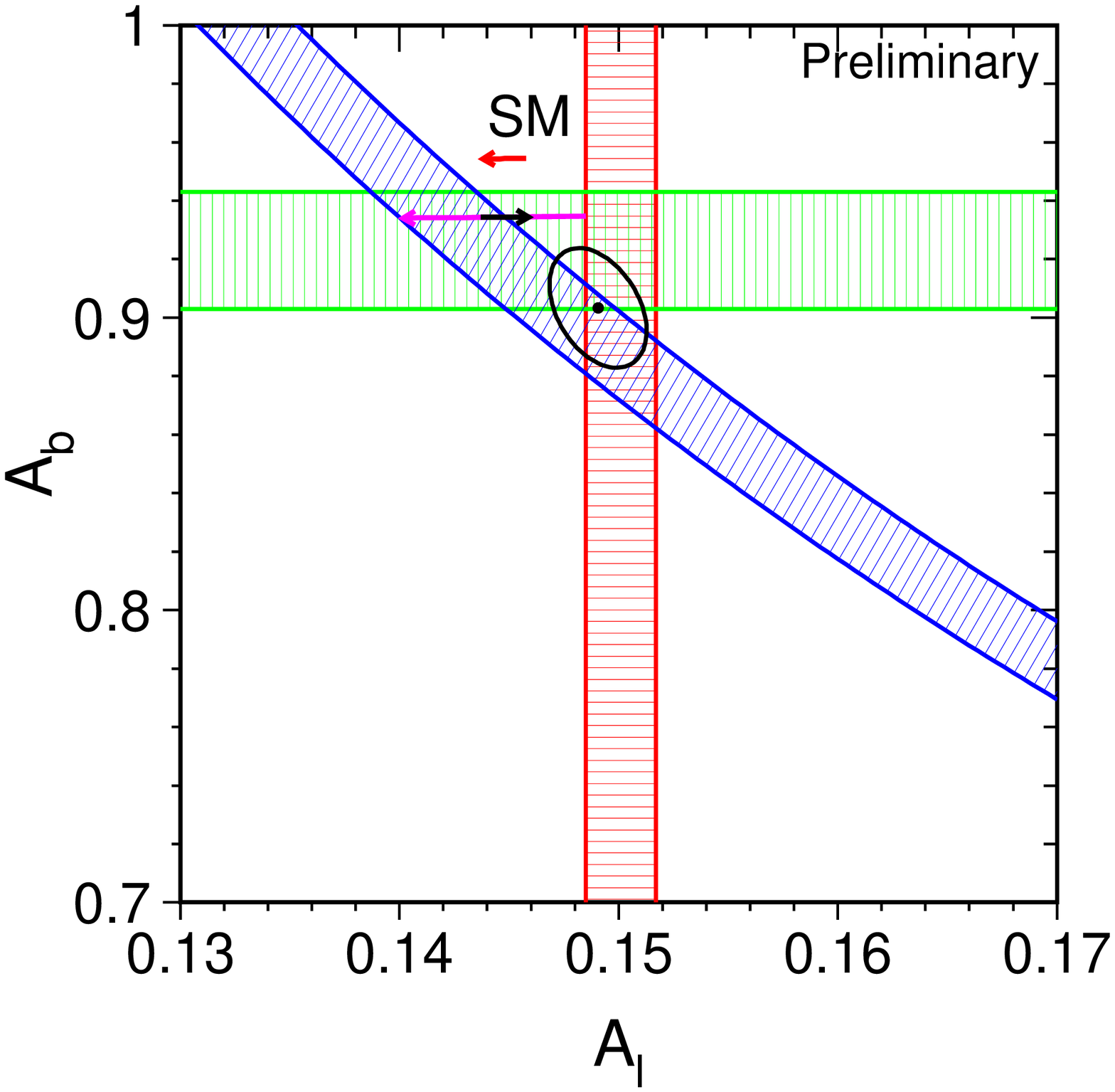}
\caption{Measurements of $A_l$ (LEP + SLD),
$A_b$ (SLD) and
$\cA_{\rms FB}^{0,b}$ (LEP, diagonal band), compared to
the SM expectations (arrows). Also shown is the
combined 68\% C.L. contour. The arrows pointing to
the left (right) show the variations of the SM prediction
with $M_H = 300\,{}^{+700}_{-186}\:\mathrm{GeV}$
($m_t = 178.0\pm 4.3\:\mathrm{GeV}$). The small
arrow oriented to the left shows the additional uncertainty
from $\alpha(M_Z^2)$ \cite{LEPEWWG:04}.}
\label{fig:Al_Ab}
\end{minipage}
\hfill
\begin{minipage}[t]{.46\linewidth}\centering
\includegraphics[width=6.5cm]{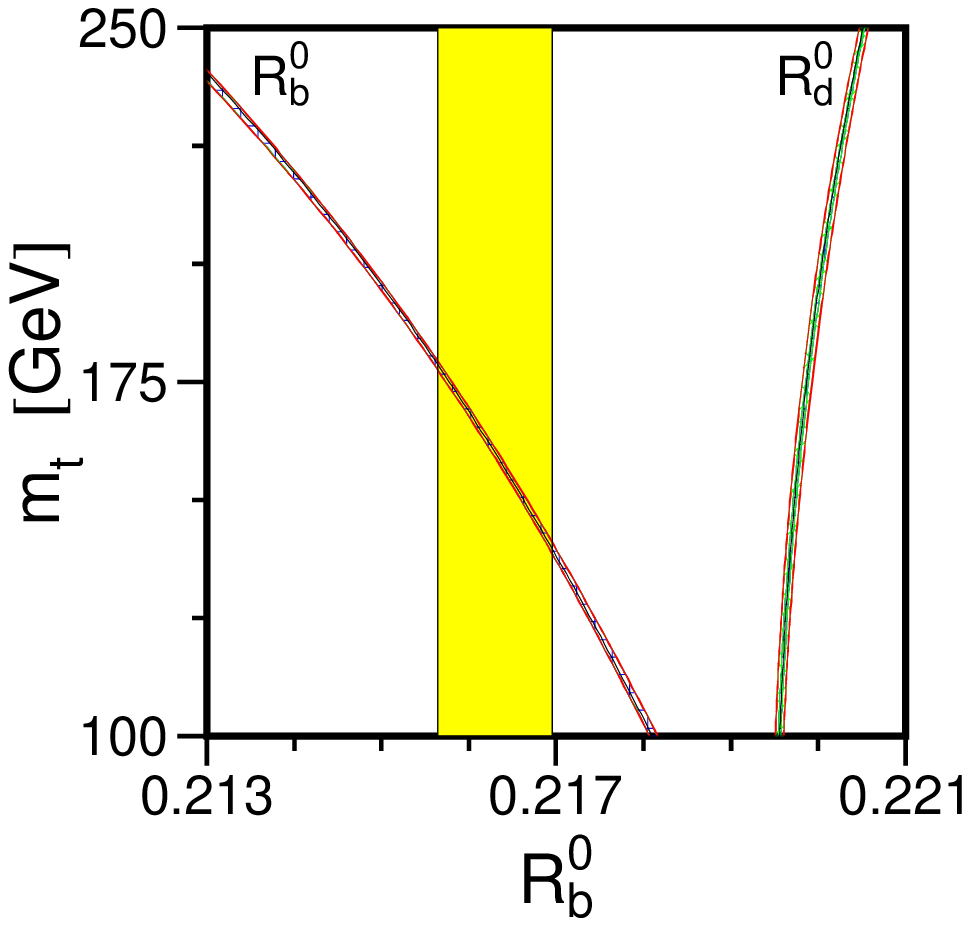}
\caption{The SM prediction of the ratios $R_b$ and $R_d$ \
[$R_q\equiv\Gamma(Z\to\bar q q)/\Gamma(Z\to\mathrm{hadrons})$],
as a function of the top mass.
The measured value of $R_b$ (vertical
band) provides a determination of $m_t$
\cite{LEPEWWG:04}.}
\label{fig:Rb}
\end{minipage}
\end{figure}
%%%%%%%%%%%%%%%%%%%%%%%%%%%%%%%%%%%%%%%%%%%%%

The resulting 68\% probability contours are shown in
Fig.~\ref{fig:Zcouplings}, which provides strong evidence
of the electroweak radiative corrections. The good agreement
with the SM predictions, obtained for low values of the
Higgs mass, is lost if only the QED vacuum
polarization contribution is taken into account, as indicated
by the point with an arrow.
The shaded region showing the SM prediction corresponds to
the input values
$m_t = 178.0\pm 4.3\;\mathrm{GeV}$,
$\alpha(M_Z^2)^{-1}=128.95\pm 0.05$
and
$M_H = 300\,{}^{+700}_{-186}\;\mathrm{GeV}$.
Notice that the uncertainty induced
by the input value of $\alpha(M_Z^2)$ is sizeable.
The measured couplings of the three charged leptons
confirm lepton universality in the neutral-current sector.
The solid contour combines the three measurements assuming
universality.

The neutrino couplings can also be determined from the invisible $Z$ decay
width, by assuming three identical neutrino generations
with left-handed couplings, and fixing the sign from neutrino scattering
data. Alternatively, one can use the SM prediction for $\Gamma_{\rms inv}$
to get a determination of the number of light neutrino flavours
\cite{LEPEWWG:04}:
\bel{eq:Nnu}
N_\nu = 2.9841\pm 0.0083\, .
\ee

Fig.~\ref{fig:Al_Ab} shows the measured values of
$A_l$ (LEP + SLD) and $A_b$ (SLD), together with
the joint constraint obtained from $\cA_{\rms FB}^{0,b}$
at LEP (diagonal band). The combined determination
of $A_b$ is $2.5\,\sigma$ below the SM prediction.
The discrepancy originates in the $A_b$ value
obtained from $A_l$ and $\cA_{\rms FB}^{0,b}$,
which is significantly lower than both the SM and
the direct measurement of $A_b$ at SLD.
Heavy quarks seem to prefer a high value of the
Higgs mass, while leptons favour a light Higgs.
The combined analysis prefers low values of $M_H$,
because of the influence of $A_l$.

The strong sensitivity to the top quark mass of the ratio
$R_b\equiv\Gamma(Z\to\bar b b)/\Gamma(Z\to\mathrm{hadrons})$
is shown in Fig.~\ref{fig:Rb}. Owing to the $|V_{td}|^2$
suppression, such a dependence is not present in the
analogous ratio $R_d$. Combined with all other
electroweak precision measurements at the $Z$ peak,
$R_b$ provides a determination of $m_t$, in good agreement
with the direct and most precise measurement at the Tevatron.
This is shown in Fig.~\ref{fig:MW_Mt_MH}, which compares
the information on $M_W$ and $m_t$ obtained at
LEP1 and SLD, with the direct measurements performed
at LEP2 and the Tevatron.
A similar comparison for $m_t$ and $M_H$ is also shown.
The lower bound on $M_H$ obtained from direct searches
excludes a large portion of the 68\% C.L. allowed domain
from precision measurements.
%
%%%%%%%%%%%%%%%  FIGURE %%%%%%%%%%%%%%%%%%%%%%%%%
\begin{figure}[hbt]\centering
\begin{minipage}[c]{.45\linewidth}\centering
\includegraphics[width=6.7cm]{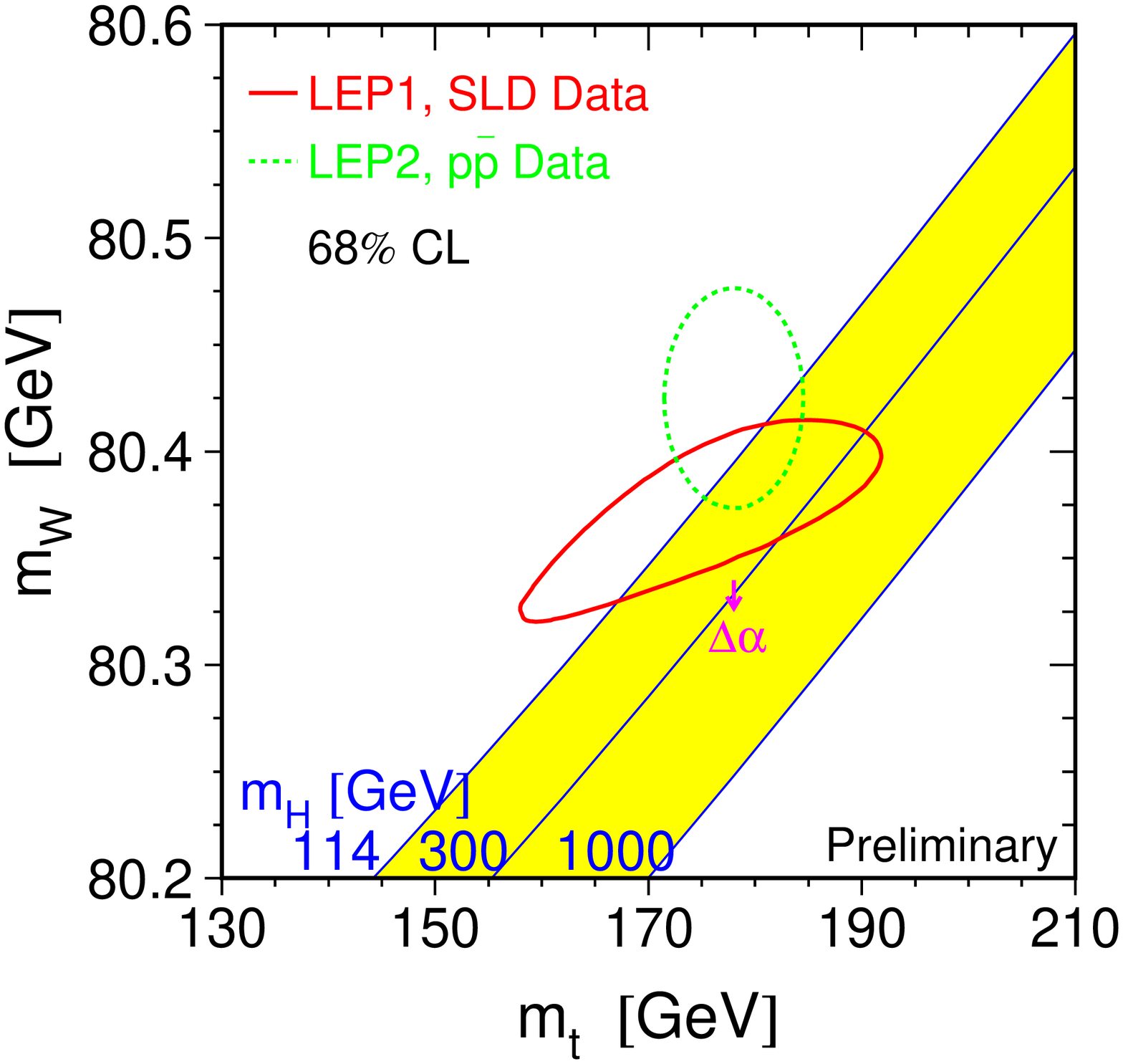}
\end{minipage}
\hskip 1cm
\begin{minipage}[c]{.45\linewidth}\centering
\includegraphics[width=6.7cm]{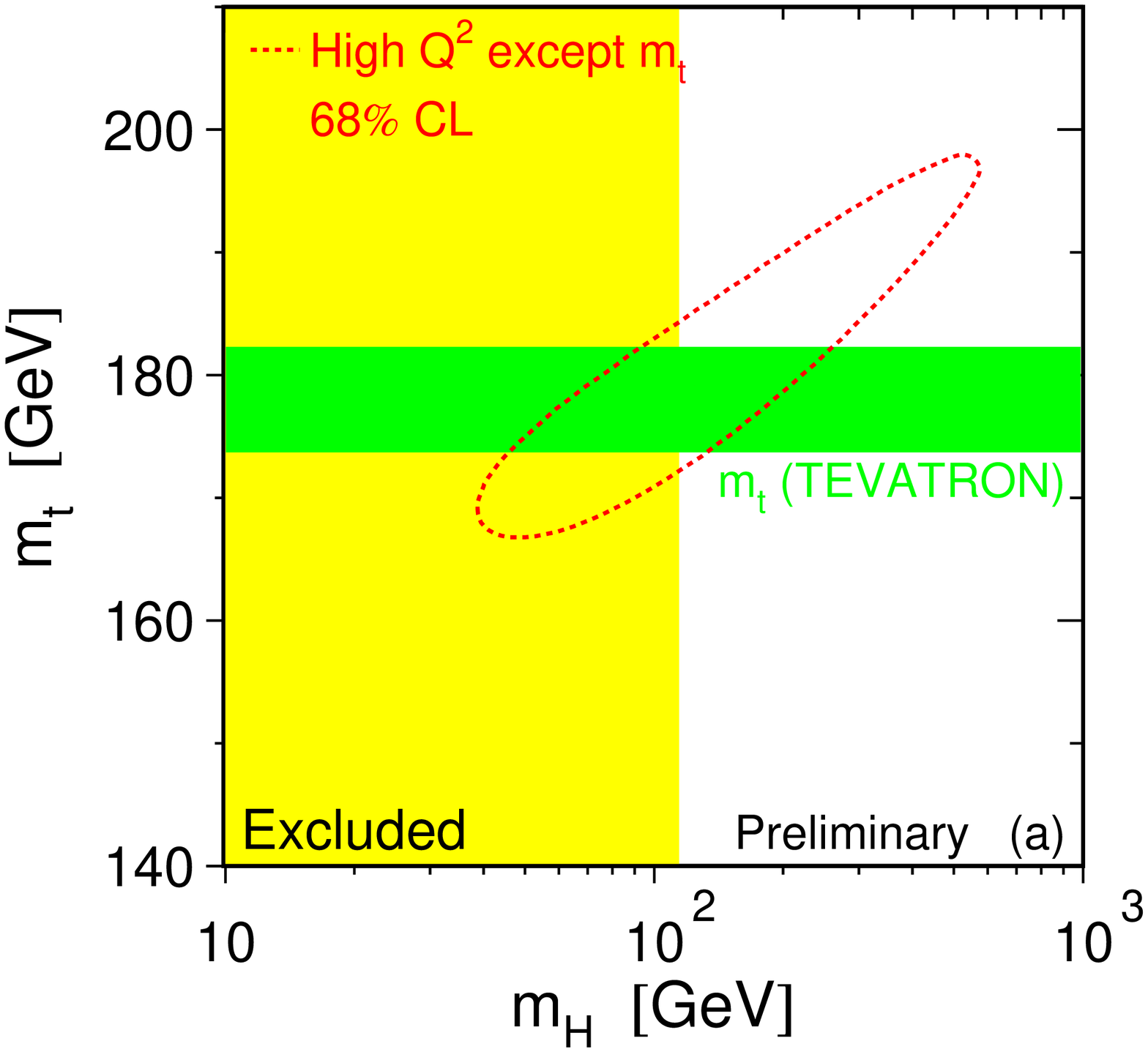}
\end{minipage}
\caption{Comparison (left ) of the direct measurements of $M_W$ and $m_t$
(LEP2 and Tevatron data) with the indirect determination through
electroweak radiative corrections (LEP1 and SLD). Also shown in the
SM relationship for the masses as function of $M_H$.
The figure on the right makes
the analogous comparison for $m_t$ and $M_H$ \cite{LEPEWWG:04}.}
\label{fig:MW_Mt_MH}
\end{figure}
%%%%%%%%%%%%%%%%%%%%%%%%%%%%%%%%%%%%%%%%%%%%%

%%%%%%%%%%%%%%%  FIGURE %%%%%%%%%%%%%%%%%%%%%%%%%
\begin{figure}[tbh]\centering
\begin{minipage}[t]{.45\linewidth}\centering
\includegraphics[width=7.25cm]{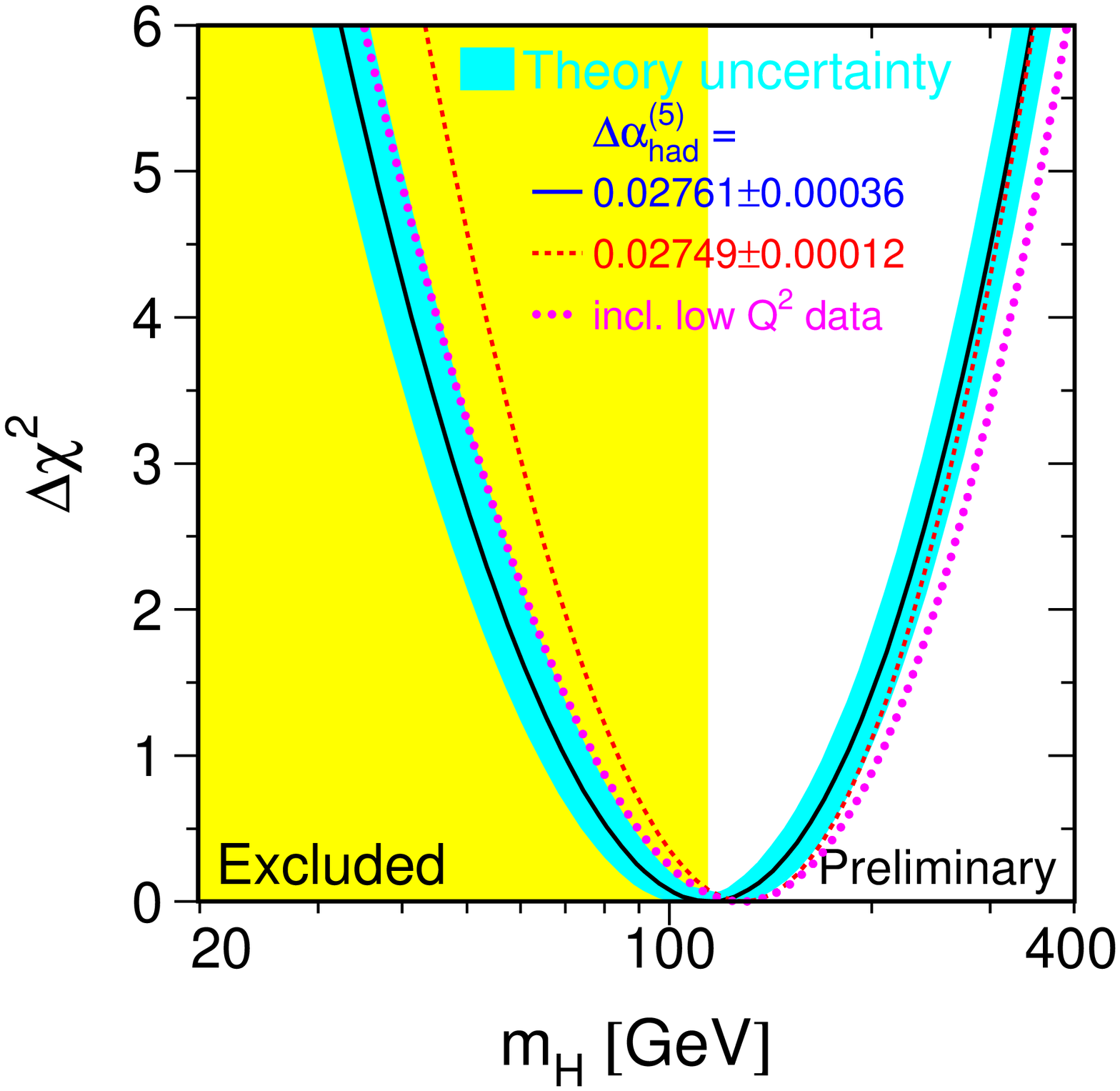}
\caption{$\Delta\chi^2 = \chi^2-\chi^2_{\rms min}$ versus $M_H$, from
the global fit to the electroweak data. The vertical band indicates
the 95\% exclusion limit from direct searches \cite{LEPEWWG:04}.}
\label{fig:MH_chi2}
\end{minipage}
\hfill
\begin{minipage}[t]{.45\linewidth}\centering
\includegraphics[width=6.5cm]{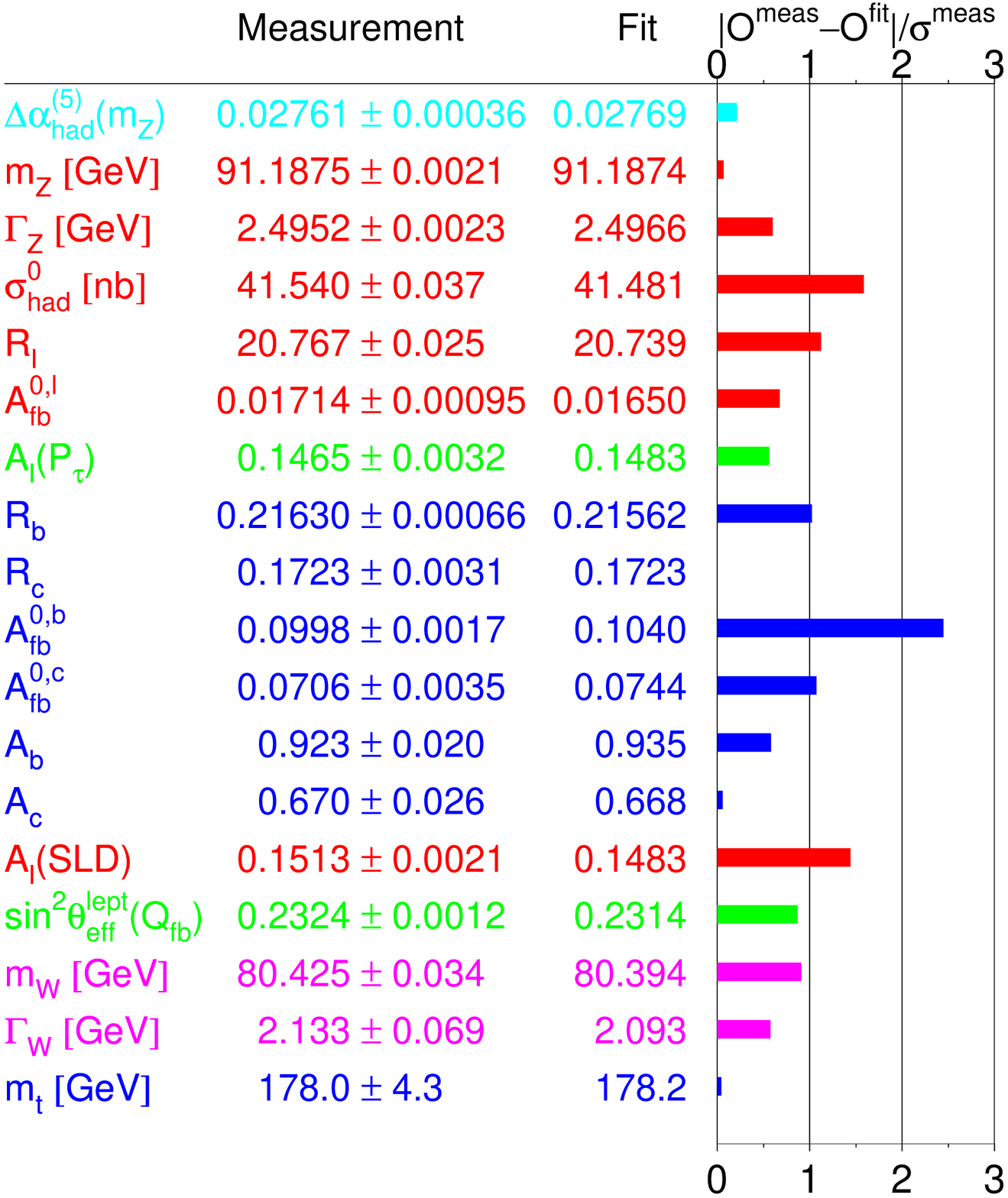}
\caption{Comparison between the measurements included in the
combined analysis of the SM and the results from the global
electroweak fit \cite{LEPEWWG:04}.}
\label{fig:pulls}
\end{minipage}
\end{figure}
%%%%%%%%%%%%%%%%%%%%%%%%%%%%%%%%%%%%%%%%%%%%%

Taking all direct and indirect data into account, one obtains the
best constraints on $M_H$. The global electroweak fit results in the
$\Delta\chi^2 = \chi^2-\chi^2_{\rms min}$ curve shown in
Fig.~\ref{fig:MH_chi2}. The lower limit on $M_H$
obtained from direct searches is close to the
point of minimum $\chi^2$. At 95\% C.L., one gets \cite{LEPEWWG:04}
\bel{eq:MH_limits}
114.4\;\mathrm{GeV}\; <\; M_H\; <\; 260\;\mathrm{GeV} .
\ee
The fit provides also a very accurate value of the strong coupling
constant,
$\alpha_s(M_Z^2) = 0.1186\pm 0.0027$,
in very good agreement with the world average value
$\alpha_s(M_Z^2) = 0.1182\pm 0.0027$ \cite{PDG:04,BE:04}.
The largest discrepancy between theory and experiment occurs
for $\cA_{\rms FB}^{0,b}$, with the fitted value being
$2.4\,\sigma$ larger than the measurement.
As shown in Fig.~\ref{fig:pulls}, a good agreement is obtained for
all other observables.

\subsection{Gauge self-interactions}

%%%%%%%%%%%%%%%  FIGURE %%%%%%%%%%%%%%%%%%%%%%%%%
\begin{figure}[tbh]\centering
\includegraphics[width=15cm]{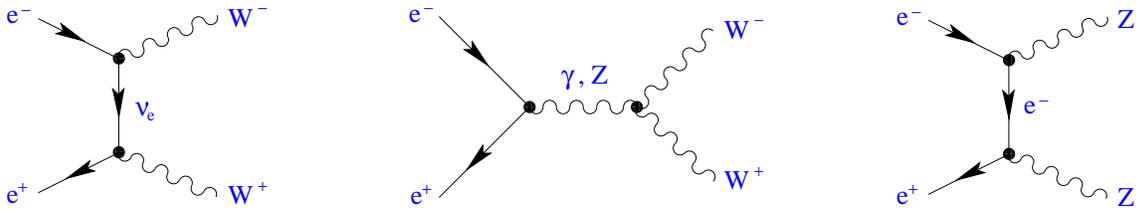}
\caption{Feynman diagrams contributing to \ $e^+e^-\!\to W^+W^-$
\ and \ $e^+e^-\!\to ZZ$.}
\label{fig:eeWW}
\end{figure}
%%%%%%%%%%%%%%%%%%%%%%%%%%%%%%%%%%%%%%%%%%%%%

At tree level, the $W$--pair production process \ $e^+e^-\to W^+W^-$ \
involves three different contributions (Fig.~\ref{fig:eeWW}),
corresponding to the exchange of $\nu_e$, $\gamma$ and $Z$.
The cross-section measured at LEP2 agrees very well with the
SM predictions. As shown in Fig.~\ref{fig:sigma_eeWW_ZZ},
the $\nu_e$--exchange contribution alone would lead to an unphysical
growing of the cross-section at large energies and, therefore,
would imply a violation of unitarity. Adding the $\gamma$--exchange
contribution softens this behaviour, but a clear disagreement
with the data persists. The $Z$--exchange mechanism, which
involves the $ZWW$ vertex, appears to be crucial in order
to explain the data.

%%%%%%%%%%%%%%%  FIGURE %%%%%%%%%%%%%%%%%%%%%%%%%
\begin{figure}[tbh]\centering
\begin{minipage}[c]{.45\linewidth}\centering
\includegraphics[width=7.25cm]{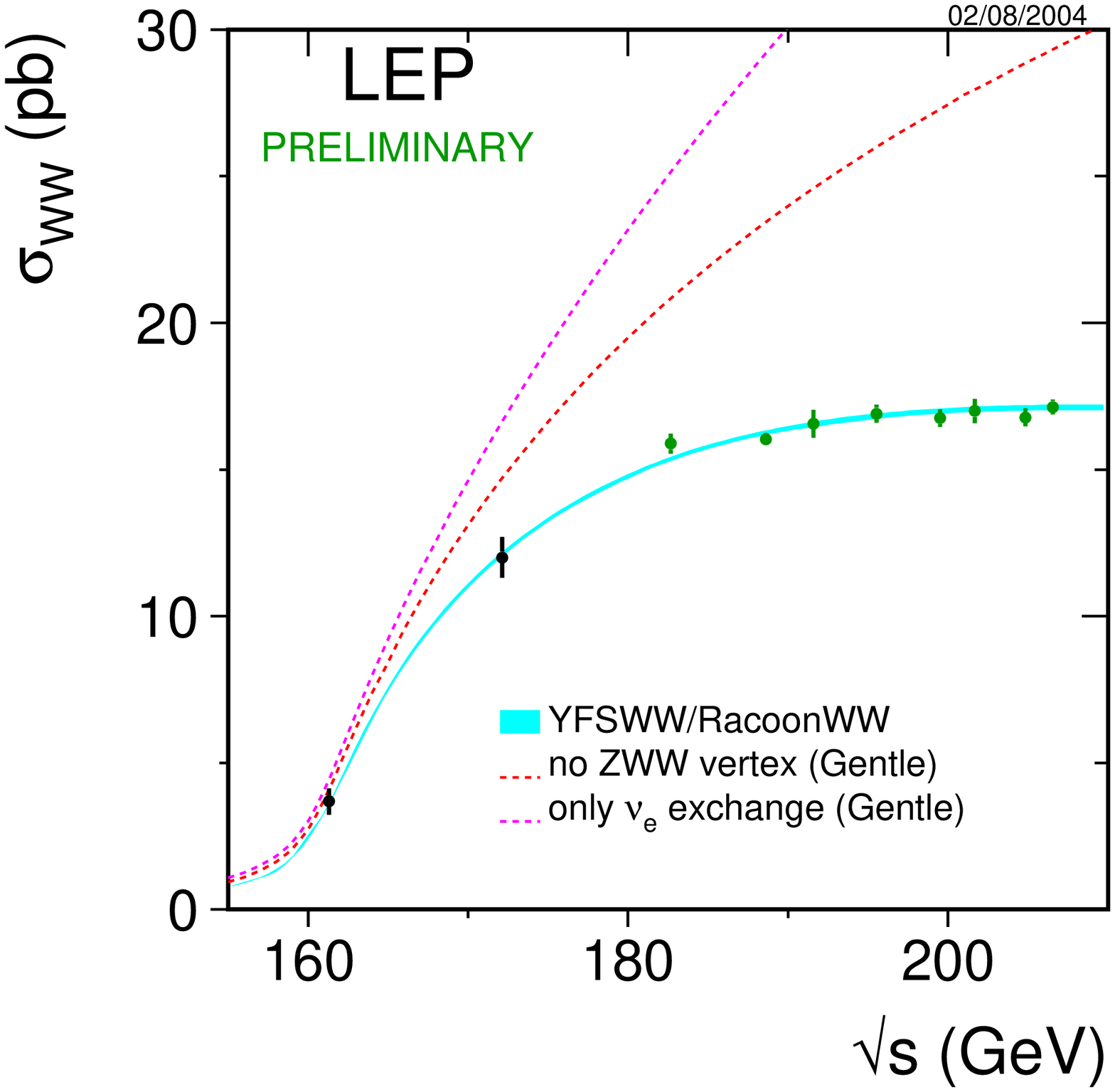}
\end{minipage}
\hfill
\begin{minipage}[c]{.45\linewidth}\centering
\includegraphics[width=7.25cm]{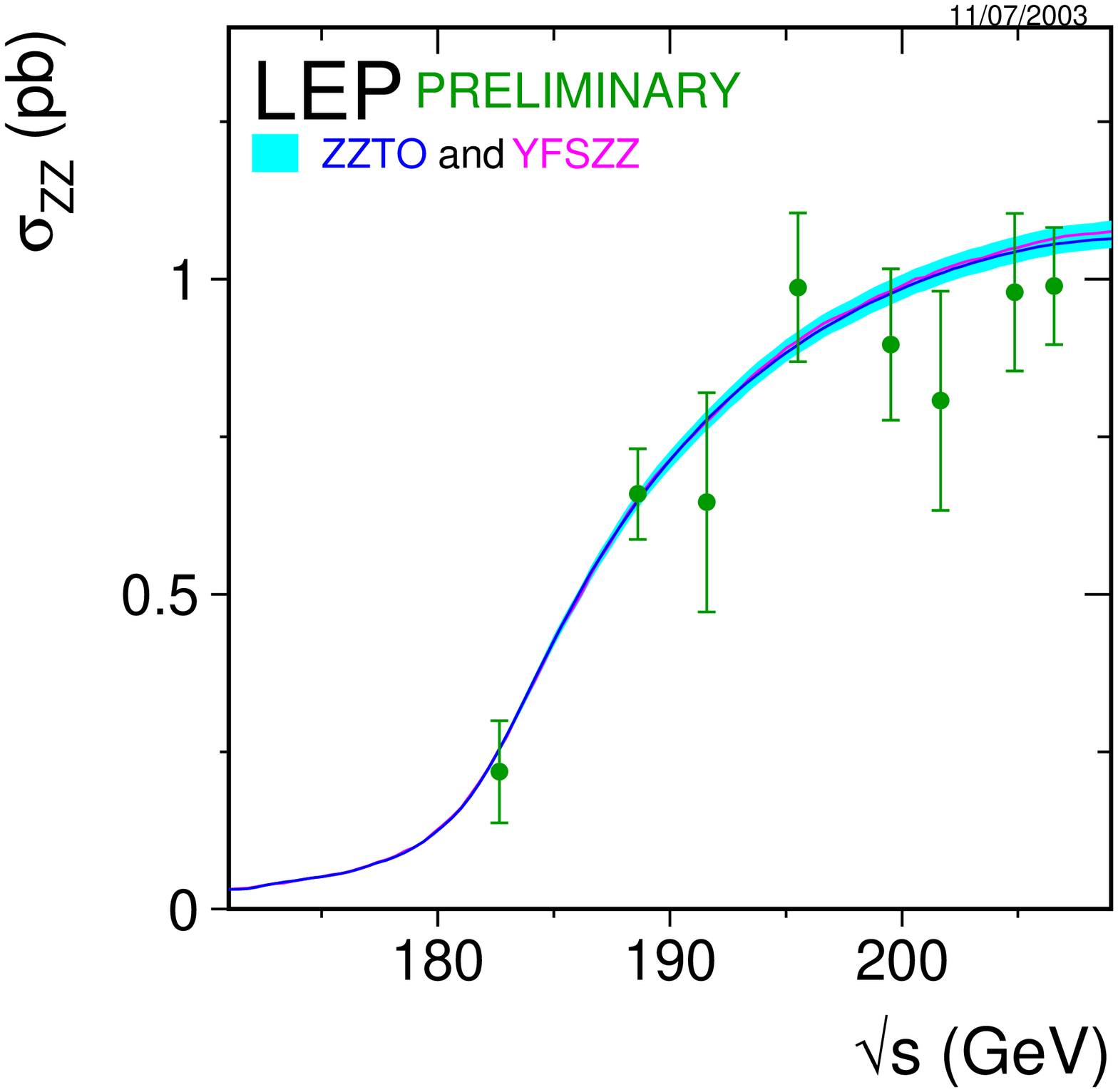}
\end{minipage}
\caption{Measured energy dependence of \ $\sigma(e^+e^-\to W^+W^-)$ \
(left) and \ $\sigma(e^+e^-\to ZZ)$ \ (right). The three curves shown
for the $W$--pair production cross-section correspond to only
the $\nu_e$--exchange contribution (upper curve), $\nu_e$ exchange
plus photon exchange (middle curve) and all contributions
including also the $ZWW$ vertex (lower curve). Only the $e$--exchange
mechanism contributes to $Z$--pair production \cite{LEPEWWG:04}.}
\label{fig:sigma_eeWW_ZZ}
\end{figure}
%%%%%%%%%%%%%%%%%%%%%%%%%%%%%%%%%%%%%%%%%%%%%

Since the $Z$ is electrically neutral, it does not interact
with the photon. Moreover, the SM does not include any
local $ZZZ$ vertex. Therefore, the \ $e^+e^-\to ZZ$ \
cross-section only involves the contribution from $e$ exchange.
The agreement of the SM predictions with the experimental
measurements in both production channels, $W^+W^-$ and $ZZ$,
provides a test of the gauge self-interactions.
There is a clear signal of the presence of a $ZWW$ vertex, with the
predicted strength, and no evidence for any $\gamma ZZ$ or
$ZZZ$ interactions. The gauge structure of the
$SU(2)_L\otimes U(1)_Y$ theory is nicely confirmed
by the data.

\subsection{Higgs decays}

%%%%%%%%%%%%%%%  FIGURE %%%%%%%%%%%%%%%%%%%%%%%%%
\begin{figure}[tbh]\centering
\begin{minipage}[c]{.48\linewidth}\centering
\includegraphics[width=8cm]{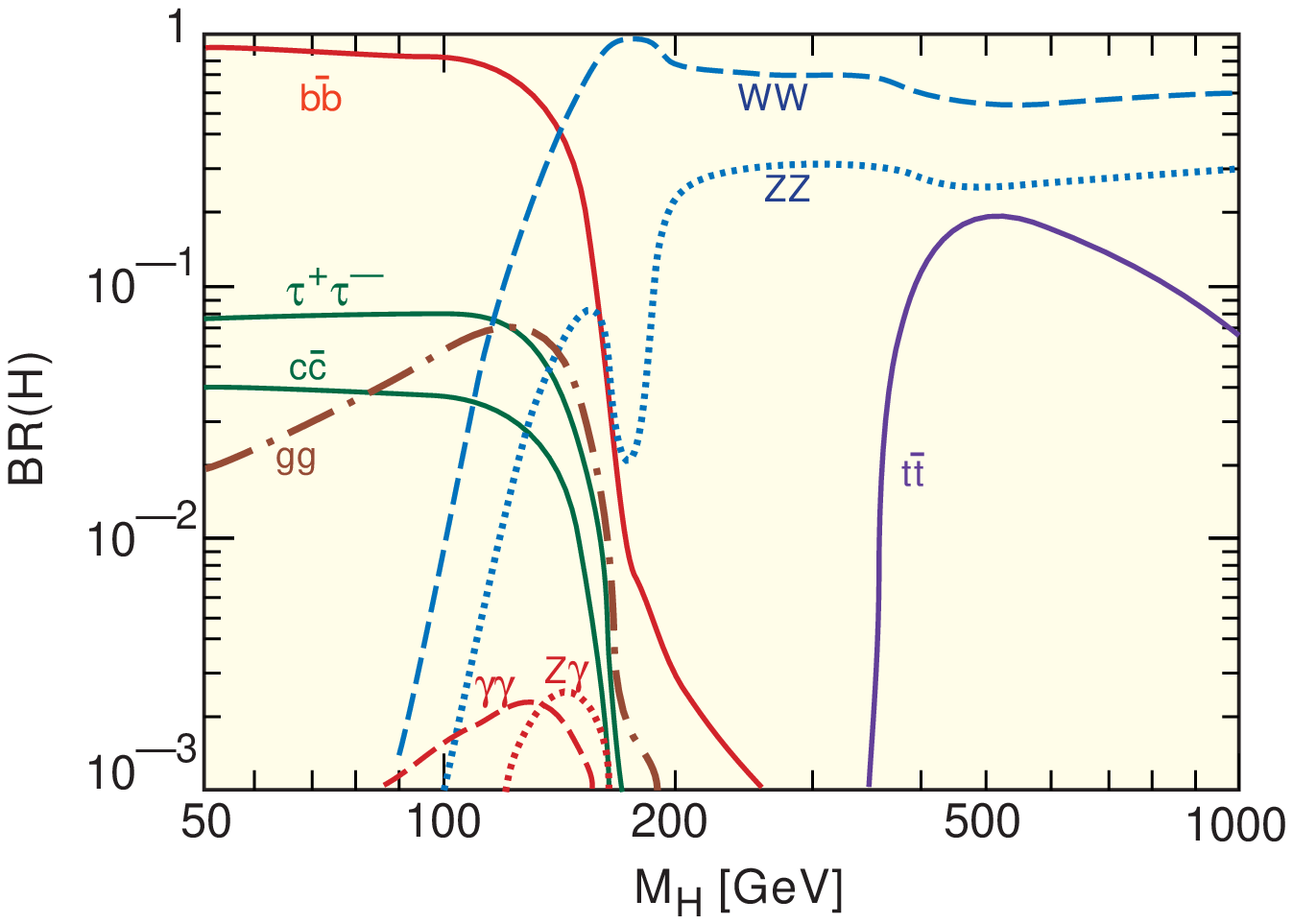}
\end{minipage}
\hfill
\begin{minipage}[c]{.48\linewidth}\centering
\includegraphics[width=8cm]{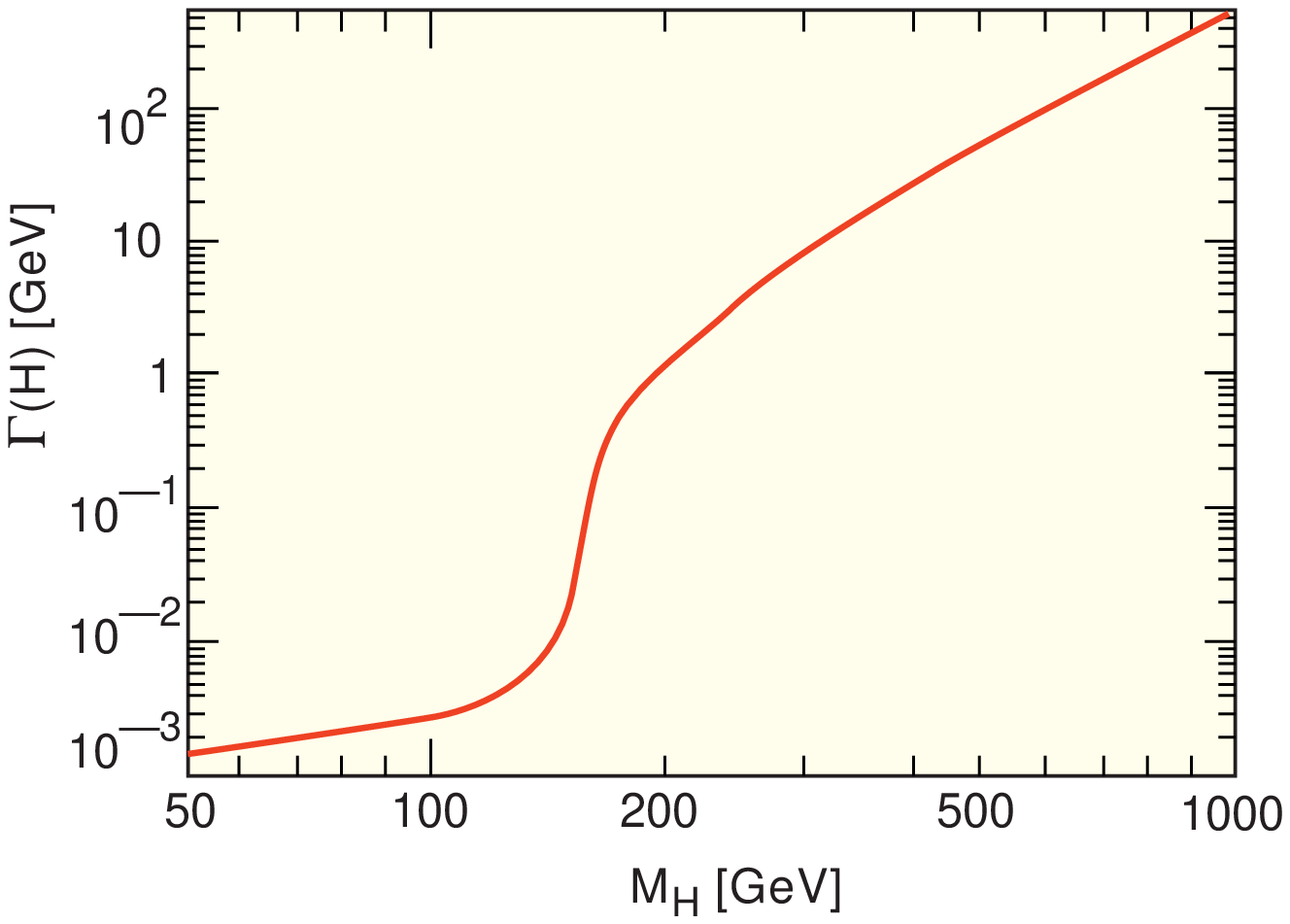}
\end{minipage}
\caption{Branching fractions of the different Higgs decay modes (left)
and total decay width of the Higgs boson (right) as function
of $M_H$ \cite{Denegri}.}
\label{fig:HiggsDecays}
\end{figure}
%%%%%%%%%%%%%%%%%%%%%%%%%%%%%%%%%%%%%%%%%%%%%

The couplings of the Higgs boson are always proportional to some mass
scale. The $Hf\bar f$ interaction grows linearly with the fermion mass,
while the $HWW$ and $HZZ$ vertices are proportional to $M_W^2$ and
$M_Z^2$, respectively. Therefore, the most probable decay mode of
the Higgs will be the one into the heaviest possible final state.
This is clearly illustrated in Fig.~\ref{fig:HiggsDecays}.
The $H\to b\bar b$ decay channel is by far the dominant one below
the $W^+W^-$ production threshold.
When $M_H$ is large enough to allow the production of a pair of
gauge bosons, $H\to W^+W^-$ and \ $H\to ZZ$ \ become dominant.
For $M_H>2m_t$, the $H\to t\bar t$ \ decay width is also sizeable,
although smaller than the $WW$ and $ZZ$ ones due to the different
dependence of the corresponding Higgs coupling with the mass scale
(linear instead of quadratic).

The total decay width of the Higgs grows with increasing values of $M_H$.
The effect is very strong above the $W^+W^-$ production threshold.
A heavy Higgs becomes then very broad. At $M_H\sim 600\;\mathrm{GeV}$,
the width is around $100\;\mathrm{GeV}$; while for
$M_H\sim 1\;\mathrm{TeV}$, $\Gamma_H$ is already of the same size
as the Higgs mass itself.

The design of the LHC detectors has taken into account all
these very characteristic properties in order to optimize the
future search for the Higgs boson.

%%%%%%%%%% FLAVOUR DYNAMICS %%%%%%%%%%

\setcounter{equation}{0}
\section{FLAVOUR \ DYNAMICS}
\label{sec:flavour}

We have learnt experimentally that there are 6 different quark flavours
\ $u\,$, $d\,$, $s\,$, $c\,$, $b\,$, $t\,$, 3 different charged leptons
\ $e\,$, $\mu\,$, $\tau$ \
and their corresponding neutrinos \ $\nu_e\,$, $\nu_\mu\,$, $\nu_\tau\,$.
We can nicely include all these particles into the SM framework,
by organizing them into 3 families of quarks and leptons, as
indicated in Eqs.~\eqn{eq:families} and \eqn{eq:structure}.
Thus, we have 3 nearly identical copies of the same
$SU(2)_L\otimes U(1)_Y$ structure, with masses as the only difference.

Let us consider the general case of $N_G$ generations of fermions,
and denote $\nup_j$, $\lp_j$, $\up_j$, $\dop_j$
the members of the weak family $j$ \ ($j=1,\ldots,N_G$),
with definite transformation properties under the gauge group.
Owing to the fermion replication, a large variety of
fermion--scalar couplings are allowed by the gauge symmetry.
The most general Yukawa Lagrangian has the form
\beqn\label{eq:N_Yukawa}
\cL_Y &=&\sum_{jk}\;\left\{
\left(\bar \up_j , \bar \dop_j\right)_L \left[\, c^{(d)}_{jk}\,
\left(\ba \phi^{(+)}\\ \phi^{(0)}\ea\right)\, \dop_{kR} \; +\;
c^{(u)}_{jk}\,
\left(\ba \phi^{(0)*}\\ -\phi^{(-)}\ea\right)\, \up_{kR}\,
\right]
\right.\no\\ && \qquad\!\left. +\;\;
\left(\bar \nup_j , \bar \lp_j\right)_L\, c^{(l)}_{jk}\,
\left(\ba \phi^{(+)}\\ \phi^{(0)}\ea\right)\, \lp_{kR}
\,\right\}
\; +\; \mathrm{h.c.},
\eeqn
where $c^{(d)}_{jk}$, $c^{(u)}_{jk}$ and $c^{(l)}_{jk}$
are arbitrary coupling constants.

After SSB, the Yukawa Lagrangian can be written as
\bel{eq:N_Yuka}
\cL_Y\, =\, - \left(1 + {H\over v}\right)\,\left\{\,
\overline{\bmd}\hskip .6pt'_L \,\bM_d'\,
\bmd\hskip .6pt'_R \; + \;
\overline{\bmu}\hskip .6pt'_L \,\bM_u'\,
\bmu\hskip .6pt'_R
\; + \;
\overline{\bml}\hskip .2pt'_L \,\bM'_l\,
\bml\hskip .2pt'_R \; +\;
\mathrm{h.c.}\right\} .
\ee
Here, $\bmd\hskip .6pt'$, $\bmu\hskip .6pt'$
and $\bml\hskip .2pt'$ denote vectors in
the $N_G$-dimensional flavour
space, and the corresponding mass matrices are given by
\bel{eq:M_c_relation}
(\bM'_d)^{}_{ij}\,\equiv\,
-\, c^{(d)}_{ij}\, {v\over\sqrt{2}}\, , \qquad
(\bM'_u)^{}_{ij}\,\equiv\,
-\, c^{(u)}_{ij}\, {v\over\sqrt{2}}\, , \qquad
(\bM'_l)^{}_{ij}\,\equiv\,
-\, c^{(l)}_{ij}\, {v\over\sqrt{2}}\, .
\ee
The diagonalization of these mass matrices determines the mass
eigenstates $d_j$, $u_j$ and $l_j$,
which are linear combinations of the corresponding weak eigenstates
$\dop_j$, $\up_j$ and $\lp_j$, respectively.

The matrix $\bM_d'$ can be decomposed as\footnote{
%%%%%%%%%%%%%%%%%
The condition $\det{\bM'_f}\not=0$ \
($f=d,u,l$)
guarantees that the decomposition
$\bM'_f=\bH_f\bU_f$ is unique:
$\bU_f\equiv\bH_f^{-1}\bM_f'$.
The matrices $\bS_f$
are completely determined (up to phases)
only if all diagonal elements of $\bM_f$
are different.
If there is some degeneracy, the arbitrariness of
$\bS_f$
reflects the freedom to define the physical fields.
If $\det{\bM'_f}=0$,
the matrices $\bU_f$ and
$\bS_f$ are not
uniquely determined, unless their unitarity is explicitly imposed.}
%%%%%%%%%%%%%%%%%
$\bM_d'=\bH^{}_d\,\bU^{}_d=\bS_d^\dagger\,
\mathbf{\cM}^{}_d \,\bS^{}_d\,\bU^{}_d$, where \
$\bH^{}_d\equiv
\sqrt{\bM_d'\bM_d'^{\dagger}}$
is an hermitian positive-definite matrix,
while $\bU^{}_d$ is unitary.
$\bH^{}_d$ can be diagonalized
by a unitary matrix $\bS^{}_d$; the resulting matrix
$\mathbf{\cM}^{}_d$ is diagonal, hermitian and positive definite.
Similarly, one has \
$\bM_u'= \bH^{}_u\,\bU^{}_u= \bS_u^\dagger\,
\mathbf{\cM}^{}_u\, \bS^{}_u\,\bU^{}_u$ \ and \
$\bM_l'= \bH^{}_l\,\bU^{}_l= \bS_l^\dagger\,
\mathbf{\cM}^{}_l \,\bS^{}_l\,\bU^{}_l$.
In terms of the diagonal mass matrices
\bel{eq:Mdiagonal}
\mathbf{\cM}^{}_d =\mathrm{diag}(m_d,m_s,m_b,\ldots)\, ,\quad
\mathbf{\cM}^{}_u =\mathrm{diag}(m_u,m_c,m_t,\ldots)\, ,\quad
\mathbf{\cM}^{}_l=\mathrm{diag}(m_e,m_\mu,m_\tau,\ldots)\, ,
\ee
the Yukawa Lagrangian takes the simpler form
\bel{eq:N_Yuk_diag}
\cL_Y\, =\, - \left(1 + {H\over v}\right)\,\left\{\,
\overline{\bmd}\,\mathbf{\cM}^{}_d\,\bmd \; + \;
\overline{\bmu}\, \mathbf{\cM}^{}_u\,\bmu \; + \;
\overline{\bml}\,\mathbf{\cM}^{}_l\,\bml \,\right\}\, ,
\ee
where the mass eigenstates are defined by
\beqn\label{eq:S_matrices}
\bmd^{}_L &\!\!\!\!\equiv &\!\!\!\!
\bS^{}_d\, \bmd\hskip .6pt'_L \, ,
\qquad\,\,\,\,\,\,\,\,\,
\bmu^{}_L \equiv \bS^{}_u \,\bmu\hskip .6pt'_L \, ,
\qquad\,\,\,\,\,\,\,\,\,
\bml^{}_L \equiv \bS^{}_l \,\bml\hskip .2pt'_L \, ,
\no\\
\bmd^{}_R &\!\!\!\!\equiv &\!\!\!\!
\bS^{}_d \bU^{}_d\,\bmd\hskip .6pt'_R \, , \qquad
\bmu^{}_R \equiv \bS^{}_u\bU^{}_u\, \bmu\hskip .6pt'_R \, , \qquad
\bml^{}_R \equiv \bS^{}_l\bU^{}_l \, \bml\hskip .2pt'_R \, .
\eeqn
Note, that the Higgs couplings are proportional to the
corresponding fermions masses.

%%%%%%%%%%%%%%%  FIGURE %%%%%%%%%%%%%%%%%%%%%%%%%
\begin{figure}[tb]\centering
\begin{minipage}[c]{.45\linewidth}\centering
\includegraphics[width=4cm]{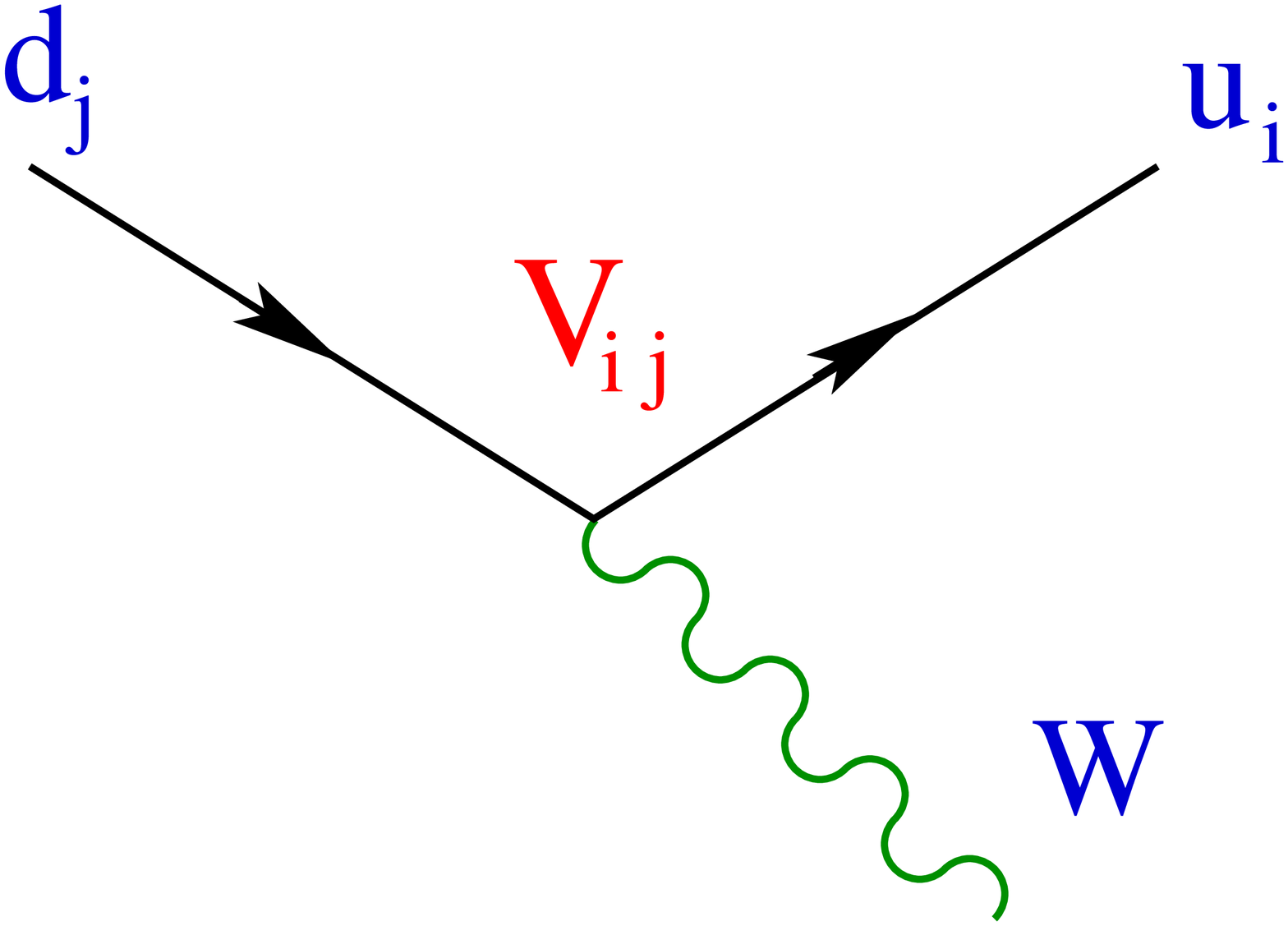}
\end{minipage}
\hskip 1cm
\begin{minipage}[c]{.45\linewidth}\centering
\includegraphics[width=4cm]{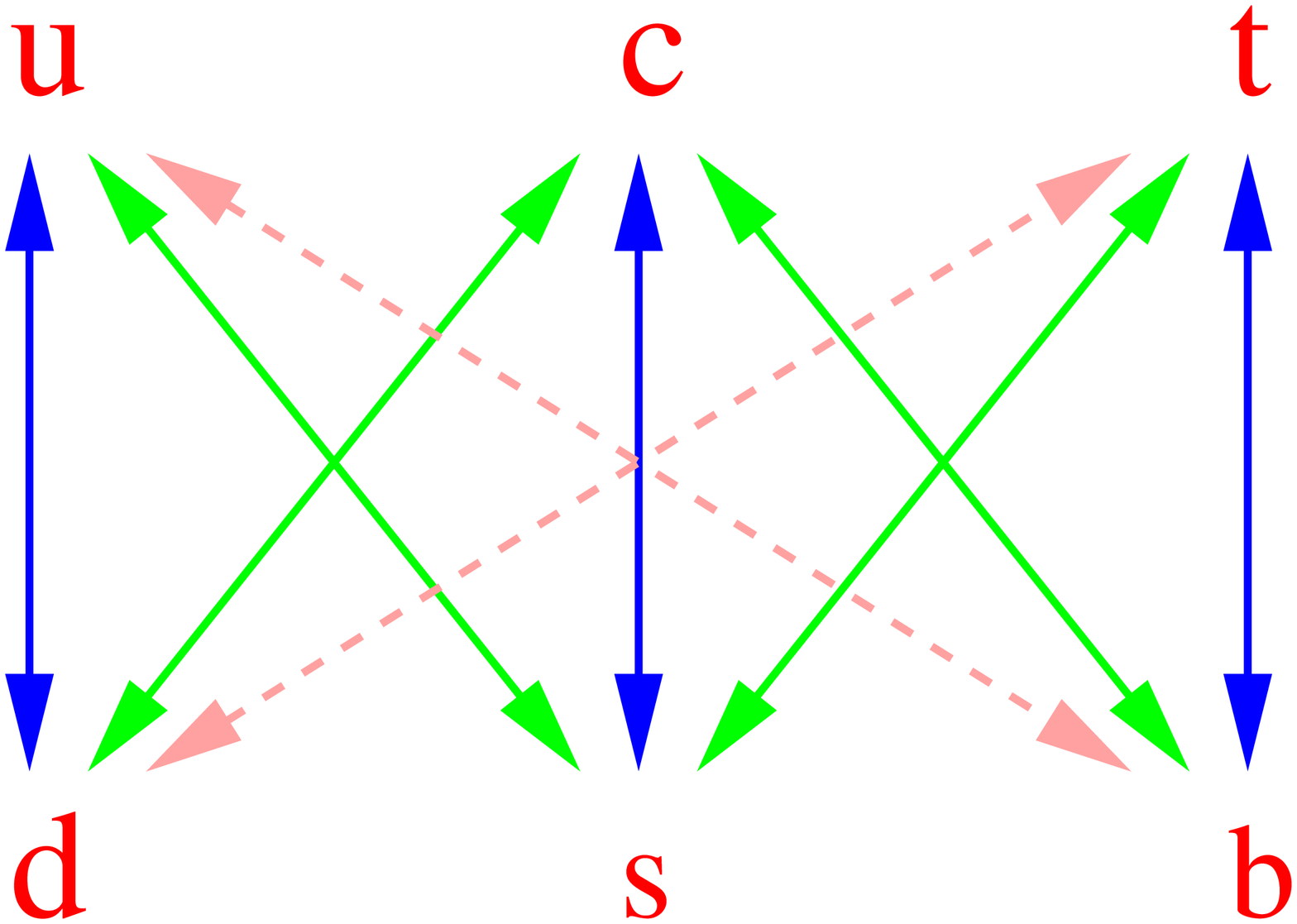}
\end{minipage}
\caption{Flavour-changing transitions through the charged-current
couplings of the $W^\pm$ bosons.}
\label{fig:CKM}
\end{figure}
%%%%%%%%%%%%%%%%%%%%%%%%%%%%%%%%%%%%%%%%%%%%%%%%%

Since, \ $\overline{\bmf}\hskip .7pt'_L\, \bmf\hskip .7pt'_L =
\overline{\bmf}^{}_L \,\bmf^{}_L$ \ and \
$\overline{\bmf}\hskip .7pt'_R \,\bmf\hskip .7pt'_R =
\overline{\bmf}^{}_R \,\bmf^{}_R$ \
($f=d,u,l$), the form of the neutral-current part of the 
$SU(2)_L\otimes U(1)_Y$ Lagrangian does not change when expressed
in terms of mass eigenstates. Therefore, there are no
flavour-changing neutral currents in the SM (GIM
mechanism \cite{GIM:70}). This
is a consequence of treating all equal-charge fermions
on the same footing.

However, $\overline{\bmu}\hskip .7pt'_L \,\bmd\hskip .7pt'_L =
\overline{\bmu}^{}_L \,\bS^{}_u\,\bS_d^\dagger\,\bmd^{}_L\equiv
\overline{\bmu}^{}_L \bV\,\bmd^{}_L$.
In general, $\bS^{}_u\not= \bS^{}_d\,$; thus
if one writes the weak eigenstates in terms of mass eigenstates,
a $N_G\times N_G$ unitary mixing matrix $\bV$,
called the Cabibbo-Kobayashi-Maskawa (CKM) matrix
\cite{cabibbo,KM:73}, appears in
the quark charged-current sector:
\bel{eq:cc_mixing}
\cL_{\rms CC}\, = \, {g\over 2\sqrt{2}}\,\left\{
W^\dagger_\mu\,\left[\,\sum_{ij}\;
\bar u_i\,\gamma^\mu(1-\gamma_5) \,\bV_{\! ij}\, d_j
\; +\;\sum_l\; \bar\nu_l\,\gamma^\mu(1-\gamma_5)\, l
\,\right]\, + \, \mathrm{h.c.}\right\}\, .
\ee
The matrix $\bV$ couples any ``up-type'' quark with all
``down-type'' quarks.

If neutrinos are assumed to be massless, we can always
redefine the neutrino flavours, in such a way as to eliminate
the analogous mixing in the lepton sector:
$\overline{\mbox{\boldmath $\nu$}}\hskip .7pt'_L\, \bml\hskip .7pt'_L =
\overline{\mbox{\boldmath $\nu$}}\hskip .7pt_L'\, \bS^\dagger_l\,\bml^{}_L\equiv
\overline{\mbox{\boldmath $\nu$}}^{}_L\, \bml^{}_L$.
Thus, we have lepton flavour conservation in the minimal SM
without right-handed neutrinos.
If sterile $\nu^{}_R$ fields are included in the model, one would have an additional
Yukawa term in Eq.~\eqn{eq:N_Yukawa}, giving rise to a neutrino mass matrix \
$(\bM'_\nu)_{ij}\equiv - c^{(\nu)}_{ij}\, {v/\sqrt{2}}\,$. Thus, the model could
accommodate non-zero neutrino masses and lepton flavour violation through
a lepton mixing matrix $\bV^{}_{\! L}$ analogous to the one present in the quark sector.
Note however that the total lepton number \ $L\equiv L_e + L_\mu + L_\tau$ \ would
still be conserved. We know experimentally that neutrino masses are tiny
and there are strong bounds on lepton flavour violating decays: \
$\mathrm{Br}(\mu^\pm\to e^\pm e^+ e^-) < 1.0\cdot 10^{-12}\,$, \
$\mathrm{Br}(\mu^\pm\to e^\pm\gamma) < 1.2\cdot 10^{-11}\,$, \
$\mathrm{Br}(\tau^\pm\to \mu^\pm\gamma) < 3.1\cdot 10^{-7}\,$, \ \ldots
\cite{PDG:04,TAU:04}. However, we do have a clear evidence of
neutrino oscillation phenomena \cite{GO:04}.

The fermion masses and the quark mixing matrix $\bV$
are all determined by the Yukawa couplings in Eq.~\eqn{eq:N_Yukawa}. 
However, the coefficients $c_{ij}^{(f)}$ are not known;
therefore we have a bunch of arbitrary parameters.
A general $N_G\times N_G$ unitary matrix is characterized by $N_G^2$ real
parameters: \ $N_G (N_G-1)/2$ \ moduli and \ $N_G (N_G+1)/2$ \ phases.
In the case of $\,\bV$, many of these parameters are
irrelevant, because we can always choose arbitrary quark phases.
Under the phase redefinitions \ $u_i\to \e^{i\phi_i}\, u_i$ \ and \
$d_j\to\e^{i\theta_j}\, d_j$, the mixing matrix changes as \
$\bV_{\! ij}\to \bV_{\! ij}\,\e^{i(\theta_j-\phi_i)}$;
thus, $2 N_G-1$ phases are unobservable.
The number of physical free parameters in the quark-mixing matrix
gets then reduced to $(N_G-1)^2$: \
$N_G(N_G-1)/2$ moduli and $(N_G-1)(N_G-2)/2$ phases.

In the simpler case of two generations, $\bV$
is determined by a single parameter. One recovers then the
Cabibbo rotation matrix \cite{cabibbo}
\bel{eq:cabibbo}
\bV\, = \,
\left(\bat \cos{\theta_C} &\sin{\theta_C} \\[2pt] -\sin{\theta_C}& \cos{\theta_C}\ea
\right)\, .
\ee
With $N_G=3$, the CKM matrix is described by 3 angles and 1 phase.
Different (but equivalent) representations can be found in the literature.
The Particle data Group \cite{PDG:04} advocates the use of the
following one as the ``standard'' CKM parametrization:
\bel{eq:CKM_pdg}
\bV\, = \, \left[
\begin{array}{ccc}
c_{12}\, c_{13}  & s_{12}\, c_{13} & s_{13}\, \e^{-i\delta_{13}} \\[2pt]
-s_{12}\, c_{23}-c_{12}\, s_{23}\, s_{13}\, \e^{i\delta_{13}} &
c_{12}\, c_{23}- s_{12}\, s_{23}\, s_{13}\, \e^{i\delta_{13}} &
s_{23}\, c_{13}  \\[2pt]
s_{12}\, s_{23}-c_{12}\, c_{23}\, s_{13}\, \e^{i\delta_{13}} &
-c_{12}\, s_{23}- s_{12}\, c_{23}\, s_{13}\, \e^{i\delta_{13}} &
c_{23}\, c_{13}
\ea
\right] .
\ee
Here \ $c_{ij} \equiv \cos{\theta_{ij}}$ \ and \
$s_{ij} \equiv \sin{\theta_{ij}}\,$, with $i$ and $j$ being
``generation'' labels ($i,j=1,2,3$).
The real angles $\theta_{12}$, $\theta_{23}$ and $\theta_{13}$
can all be made to lie in the first quadrant, by an appropriate
redefinition of quark field phases; then, \
$c_{ij}\geq 0\,$, $s_{ij}\geq 0$ \ and \ $0\leq \delta_{13}\leq 2\pi\,$.

Notice that $\delta_{13}$ is the only complex phase in the SM
Lagrangian. Therefore, it is the only possible source of $\cCP$-violation
phenomena. In fact, it was for this reason that the third generation
was assumed to exist \cite{KM:73},
before the discovery of the $b$ and the $\tau$.
With two generations, the SM could not explain the observed
$\cCP$ violation in the $K$ system.

\subsection{Quark mixing}

%%%%%%%%%%%%%%%  FIGURE %%%%%%%%%%%%%%%%%%%%%%%%%
\begin{figure}[tbh]\centering
\includegraphics[width=12cm]{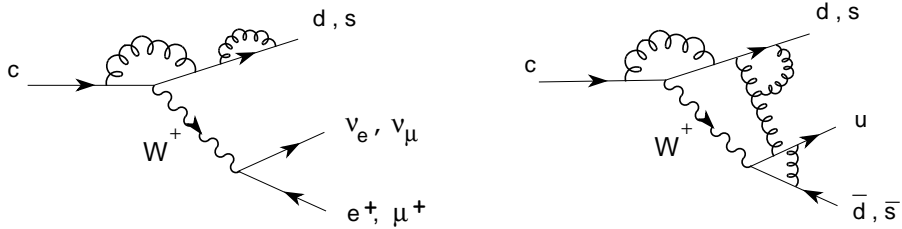}
\vskip -.3cm
\caption{Determinations of $\bV_{\! ij}$ are done
in semileptonic quark decays (left), where
a single quark current is present.
Hadronic decay modes (right) involve two
different quark currents and are more affected
by QCD effects
(gluons can couple everywhere).}
\label{fig:cDecay}
\end{figure}
%%%%%%%%%%%%%%%%%%%%%%%%%%%%%%%%%%%%%%%%%%%%%%%%%

Our knowledge of the charged-current parameters is unfortunately not so
good as in the neutral-current case. In order to measure the CKM matrix
elements, one needs to study hadronic weak decays of the type \
$H\to H'\, l^- \bar\nu_l$ \ or \ $H\to H'\, l^+ \nu_l$, which are
associated with the corresponding quark transitions
$d_j\to u_i\, l^-\bar\nu_l$ \  and \ $u_i\to d_j\, l^+\nu_l$.
Since quarks are confined within hadrons, the decay amplitude
\bel{eq:T_decay}
T[H\to H'\, l^- \bar\nu_l]\;\approx\; {G_F\over\sqrt{2}} \;\bV_{\! ij}\;\,
\langle H'|\, \bar u_i\, \gamma^\mu (1-\gamma_5)\, d_j\, | H\rangle \;\,
\left[\,\bar l \,\gamma_\mu (1-\gamma_5) \,\nu_l\,\right]
\ee
always involves an hadronic matrix element of the weak left current.
The evaluation of this matrix element is a non-perturbative QCD problem
and, therefore, introduces unavoidable theoretical uncertainties.

Usually, one looks for a semileptonic transition
where the matrix element can be fixed at some kinematical point, by
a symmetry principle. This has the virtue of reducing the theoretical
uncertainties to the level of symmetry-breaking corrections and
kinematical extrapolations.
The standard example is a \ $0^-\to 0^-$ \ decay such as \ $K\to\pi l\nu\,$,
$D\to K l\nu$ \ or \ $B\to D l\nu\,$.
Only the vector current can contribute in this case:
\bel{eq:vector_me}
\langle P'(k')| \,\bar u_i \,\gamma^\mu\, d_j\, | P(k)\rangle \; = \; C_{PP'}\,
\left\{\, (k+k')^\mu\, f_+(t)\, +\, (k-k')^\mu\, f_-(t)\,\right\}\, .
\ee
Here, $C_{PP'}$ is a Clebsh-Gordan factor and $t=(k-k')^2\equiv q^2$.
The unknown strong dynamics is fully contained in the form factors
$f_\pm(t)$.
In the massless quark limit, the divergence of
the vector current is zero; thus \
$q_\mu\left(\bar u_i \gamma^\mu d_j\right) = 0$,
which implies \ $f_-(t)=0$ \ and, moreover, $f_+(0)=1$ \ to all orders
in the strong coupling
because the associated flavour charge is a conserved quantity.\footnote{
%%%%%%%% FOOTNOTE %%%%%%%%%%%%%%%%%%%%%%%
This is completely analogous to the
electromagnetic charge conservation in QED.
The conservation of the
electromagnetic current implies that the proton electromagnetic
form factor does not get any QED or QCD correction at $q^2=0$ and,
therefore,
$Q(p)=2\, Q(u)+Q(d)=|Q(e)|$. A detailed proof can be found in 
Ref.~\cite{PI:96}.}
%%%%%%%% END OF FOOTNOTE %%%%%%%%%%%%%%%%%
Therefore, one only needs to estimate the corrections induced by the
finite values of the quark masses.

Since
$q^\mu\,\left[\bar l\gamma_\mu (1-\gamma_5)\nu_l\right]\sim m_l$, the contribution
of $f_-(t)$ is kinematically suppressed in the electron and muon modes.
The decay width can then be written as
\bel{eq:decay_width}
\Gamma(P\to P' l \nu)
\; =\; {G_F^2 M_P^5\over 192\pi^3}\; |\bV_{\! ij}|^2\; C_{PP'}^2\;
|f_+(0)|^2\; \cI\; \left(1+\delta_{\rms RC}\right)\, ,
\ee
where $\delta_{\rms RC}$ is an electroweak radiative
correction factor and $\cI$ denotes a phase-space integral,
which in the $m_l=0$ limit takes the form
\bel{eq:ps_integral}
\cI\;\approx\;\int_0^{(M_P-M_{P'})^2} {dt\over M_P^8}\;
\lambda^{3/2}(t,M_P^2,M_{P'}^2)\;
\left| {f_+(t)\over f_+(0)}\right|^2\, .
\ee
The usual procedure to determine $|\bV_{\! ij}|$ involves three steps:
\begin{enumerate}
\item Measure the shape of the $t$ distribution. This fixes %the ratio
$|f_+(t)/f_+(0)|$ and therefore determines $\cI$.
\item Measure the total decay width $\Gamma$. Since $G_F$ is already known
from $\mu$ decay, one gets then an
experimental value for the product $|f_+(0)|\, |\bV_{\! ij}|$.
\item Get a theoretical prediction for $f_+(0)$.
\end{enumerate}
The important point to realize is that theoretical input is
always needed. Thus, the accuracy of the $|\bV_{\! ij}|$
determination is limited by our ability to  calculate the relevant
hadronic input.

%%%%%%%%%%%%%%%%%%%%  TABLE  %%%%%%%%%%%%%%%%%%
\begin{table}[bth]
\caption{Direct determinations of the CKM matrix elements $\bV_{\! ij}$.}
\begin{center}
\renewcommand{\arraystretch}{1.2}
\begin{tabular}{|c|c|c|}
\hline
CKM entry & Value & Source
\\ \hline
$|\mathbf{V}_{\! ud}|$ & $0.9740\pm 0.0005$ & Nuclear $\beta$ decay
\cite{PDG:04}
\\
& $0.9729\pm 0.0012$ & $n\to p\, e^-\bar\nu_e$ \cite{CMS:04}
\\
& $0.9749\pm 0.0026$ & $\pi^+\to \pi^0\, e^+\nu_e$ \cite{CMS:04}
\\
& $0.9739\pm 0.0005$ & average
\\ \hline
$|\mathbf{V}_{\! us}|$ & $0.2220\pm 0.0025$ & $K\to\pi e^+\nu_e$ \cite{Kl3,Kl3_TH}
\\
& $0.2199\pm 0.0026$ & Hyperon decays \cite{FL:04}
\\
& $0.2208\pm 0.0034$ & $\tau$ decays \cite{GJPPS:04}
\\
& $0.2219\pm 0.0025$ & $K^+/\pi^+\to\mu^+\nu_\mu$, Lattice \cite{Kl2}
\\
& $0.2212\pm 0.0025$ & average
\\ \hline
$|\mathbf{V}_{\! cd}|$ & $0.224\pm 0.012$ & $\nu\, d \to c\, X$ \cite{PDG:04}
\\ \hline
$|\mathbf{V}_{\! cs}|$ & $0.97\pm 0.11$ & $W^+\to c\bar s$ \cite{PDG:04}
%%% $1.04\pm 0.16$ & $D \to \bar K\, e^+\nu_e$ \cite{PDG}
\\
& $0.975\pm 0.013$ & $W^+\to \mathrm{had.}\,$, $\bV_{\! uj}\,$, $\bV_{\! cd}\,$,
$\bV_{\! cb}\,$ \cite{LEPEWWG:04}
\\ \hline
%$|\mathbf{V}_{\! cb}|$ & $0.0420\pm 0.0022$ & $B\to D^* l\bar\nu_l$ \cite{PDG:04}
$|\mathbf{V}_{\! cb}|$ & $0.0414\pm 0.0021$ & $B\to D^* l\bar\nu_l$ \cite{HFAG:04}
\\
& $0.0410\pm 0.0015$ & $b\to c\, l\,\bar\nu_l$ \cite{PDG:04}
\\
%& $0.0413\pm 0.0015$ & average
& $0.0411\pm 0.0015$ & average
\\ \hline
$|\mathbf{V}_{\! ub}|$ & $0.0033\pm 0.0006$ &
 $B\to\rho\, l\,\bar\nu_l , \pi\, l\,\bar\nu_l$ \cite{PDG:04}
\\
& $0.0047\pm 0.0009$ & $b\to u\, l\,\bar\nu_l$ \cite{PDG:04}
\\
& $0.0037\pm 0.0005$ & average
\\ \hline
\raisebox{-.75ex}{
$|\mathbf{V}_{\! tb}|\, / \sqrt{\sum_q |\mathbf{V}_{\! tq}|^2}$}
& \raisebox{-.75ex}{$0.97\, {}^{+0.16}_{-0.12}$} & 
\raisebox{-.75ex}{$t\to b\, W / q\, W$ \cite{CDF:01}}
\\[7pt]  \hline
\end{tabular}
\end{center}
\label{tab:V_CKM}
\end{table}
%%%%%%%%%%%%% End Table %%%%%%%%%%%%%%%%%%%%%

The conservation of the vector and axial-vector QCD currents in the
massless quark limit allows for accurate determinations of the
light-quark mixings $|\bV_{\! ud}|$ and $|\bV_{\! us}|$.
The present values are shown in Table~\ref{tab:V_CKM},
which takes into account the recent changes in the
$K\to\pi e^+\nu_e$ data \cite{Kl3,Kl3_TH} 
and the new $|\bV_{\! us}|$ determinations
from semileptonic hyperon decays \cite{FL:04}, Cabibbo suppressed tau decays
\cite{GJPPS:04} and from the lattice evaluation of the ratio of decay constants
$f_K/f_\pi$ \cite{Kl2}.
Since $|\bV_{\! ub}|^2$ is tiny, these two light quark entries
provide a sensible test of the unitarity of the CKM matrix:
\bel{eq:unitarity_test}
|\bV_{\! ud}|^2 + |\bV_{\! us}|^2 + |\bV_{\! ub}|^2 \, = \,
0.9974\pm 0.0021 \, .
\ee
It is important to notice that at the quoted level of uncertainty
radiative corrections play a crucial role.

In the limit of very heavy quark masses, QCD has additional symmetries \cite{HQET}
which can be used to make rather precise determinations of $|\bV_{\! cb}|$,
either from exclusive decays such as $B\to D^* l\bar\nu_l$ \cite{NE:91,LU:90}
or from the inclusive analysis of $b\to c\, l\,\bar\nu_l$ transitions.
The control of theoretical uncertainties is much more
difficult for $|\bV_{\! ub}|$, $|\bV_{\! cd}|$ and $|\bV_{\! cs}|$,
because the symmetry arguments associated with the light and heavy
quark limits are not very helpful.

The best direct determination of $|\bV_{\! cs}|$ comes from charm-tagged
$W$ decays at LEP2. Moreover, the ratio of the total hadronic decay width
of the $W$ to the leptonic one provides the sum \cite{LEPEWWG:04}
\bel{eq:unitarity_test2}
\sum_{\ba\scriptstyle i\, =\, u,c \\[-6pt]
\scriptstyle j\,  =\,  d, s, b\ea}\;
|\bV_{\! ij}|^2\; = \; 1.999\pm 0.025\, .
\ee
Although much less precise than \eqn{eq:unitarity_test}, this result
test unitarity at the 2.5\% level.
From Eq.~\eqn{eq:unitarity_test2} one can also
obtain a tighter determination of
$|\bV_{\! cs}|$, using the experimental knowledge on the other CKM
matrix elements, i.e \
$|\bV_{\! ud}|^2 + |\bV_{\! us}|^2 + |\bV_{\! ub}|^2 + |\bV_{\! cd}|^2 +
|\bV_{\! cb}|^2 = 1.0493\pm 0.0076\,$. This gives the second
value of $|\bV_{\! cs}|$ quoted in Table~\ref{tab:V_CKM}.

The measured entries of the CKM matrix show a hierarchical pattern, with the
diagonal elements being very close to one, the ones connecting the
two first generations having a size
\bel{eq:lambda}
\lambda\approx |\bV_{\!\! us}| = 0.2212\pm 0.0025 \, ,
\ee
the mixing between the second and third families being of order
$\lambda^2$, and the mixing between the first and third quark generations
having a much smaller size of about $\lambda^3$.
It is then quite practical to use the
approximate parametrization \cite{WO:83}:

\bel{eq:wolfenstein}
\bV\; =\; \left[ \bath\displaystyle
1- {\lambda^2 \over 2} & \lambda & A\lambda^3 (\rho - i\eta)
\\[8pt]
-\lambda &\displaystyle 1 -{\lambda^2 \over 2} & A\lambda^2
\\[8pt]
A\lambda^3 (1-\rho -i\eta) & -A\lambda^ 2 &  1
\ea\right]\; +\; O\left(\lambda^4 \right) \, ,
\ee
where 
\bel{eq:circle}
A\approx {|\bV_{\! cb}|\over\lambda^2} = 0.84\pm 0.03 \, , \qquad\qquad
\sqrt{\rho^2+\eta^2} \,\approx\,
\left|{\bV_{\! ub}\over \lambda \bV_{\! cb}}\right|
\, =\, 0.41\pm 0.06 \, .
\ee
Defining to all orders in $\lambda$ \cite{BLO:94}
$s_{12}\equiv\lambda$, $s_{23}\equiv A\lambda^2$ and
$s_{13}\, \e^{-i\delta_{13}}\equiv A\lambda^3 (\rho-i\eta)$,
Eq.~\eqn{eq:wolfenstein} just corresponds to a Taylor expansion of
\eqn{eq:CKM_pdg} in powers of $\lambda$.

\subsection{CP Violation}
\label{subsec:CP-Violation}

Parity and charge conjugation are violated by the weak interactions
in a maximal way, however the product of the two discrete transformations
is still a good symmetry (left-handed fermions $\leftrightarrow$
right-handed antifermions). In fact, $\cCP$ appears to be a
symmetry of nearly all observed phenomena.
However, a slight violation
of the $\cCP$ symmetry at the level of $0.2\%$ is observed in the
neutral kaon system and more sizeable signals of $\cCP$ violation
have been recently established at the B factories.
Moreover, the huge matter-antimatter asymmetry present in our
Universe is a clear manifestation of $\cCP$ violation and its
important role in the primordial baryogenesis.

The $\cCPT$ theorem guarantees that the product of the three
discrete transformations is an exact symmetry of any local and
Lorentz-invariant quantum field theory preserving micro-causality.
Therefore, a violation of $\cCP$ requires a corresponding
violation of time reversal. Since $\cT$ is an antiunitary
transformation, this requires the presence of relative complex
phases between different interfering amplitudes.

The electroweak SM Lagrangian only contains a single
complex phase $\delta_{13}$ ($\eta$). This is
the only possible source of $\cCP$ violation and,
therefore,
the SM predictions for $\cCP$--violating phenomena are quite constrained.
The CKM mechanism requires several necessary conditions
in order to generate an observable $\cCP$--violation effect.
With only two fermion generations, the quark mixing mechanism cannot
give rise to $\cCP$ violation; therefore,
for $\cCP$ violation to occur in a particular process,
all 3 generations are required to play an active role.
In the kaon system, for instance, $\cCP$--violation effects can only
appear at the one-loop level, where the top quark is present.
In addition, all CKM matrix elements must be non-zero and the quarks
of a given charge must be non-degenerate in mass. If any of these
conditions were not satisfied, the CKM phase could be rotated away
by a redefinition of the quark fields. $\cCP$--violation effects
are then necessarily proportional to the product of all CKM angles, and
should vanish in the limit where any two (equal-charge) quark masses
are taken to be equal.
All these necessary conditions can be summarized in a very elegant
way as a single requirement
on the original quark mass matrices
$\bM'_u$ and $\bM'_d$ \cite{JA:85}:
\be
\cCP \:\mbox{\rm violation} \qquad \Longleftrightarrow \qquad
\mbox{\rm Im}\left\{\det\left[\bM_u^\prime \bM^{\prime\dagger}_u\, ,
  \,\bM^{\prime\phantom{\dagger}}_d \bM^{\prime\dagger}_d\right]\right\} \not=0 \, .
\ee

Without performing any detailed calculation, one can make the
following general statements on the implications of the CKM mechanism
of $\cCP$ violation:

\bi
\item
Owing to unitarity, for any choice of $i,j,k,l$ (between 1 and 3),
\beqn\label{eq:J_relation}
\mbox{\rm Im}\left[
\bV^{\phantom{*}}_{ij}\bV^*_{ik}\bV^{\phantom{*}}_{lk}\bV^*_{lj}\right]
\, =\, \cJ \sum_{m,n=1}^3 \epsilon_{ilm}\epsilon_{jkn}\, ,
\qquad\quad\\
\cJ \, =\, c_{12}\, c_{23}\, c_{13}^2\, s_{12}\, s_{23}\, s_{13}\, \sin{\delta_{13}}
\,\approx\, A^2\lambda^6\eta \, < \, 10^{-4}\, .
\eeqn
Any $\cCP$--violation observable involves the product
$\cJ$ \cite{JA:85}.
Thus, violations of the $\cCP$ symmetry are necessarily small.
\item In order to have sizeable $\cCP$--violating asymmetries
$\cA\equiv (\Gamma - \overline{\Gamma})/(\Gamma + \overline{\Gamma})$,
one should look
for very suppressed decays, where the decay widths already involve
small CKM matrix elements.
\item In the SM, $\cCP$ violation is a low-energy phenomena
in the sense that any
effect should disappear when the quark mass difference $m_c-m_u$ becomes
negligible.
\item $B$ decays are the optimal place for $\cCP$--violation signals to show up.
They involve small CKM matrix elements and are the lowest-mass processes
where the three quark generations play a direct (tree-level) role.
\ei

%%%%%%%%%%%%%%%  FIGURE %%%%%%%%%%%%%%%%%%%%%%%%%
\begin{figure}[tbh]\centering
\includegraphics[width=11cm]{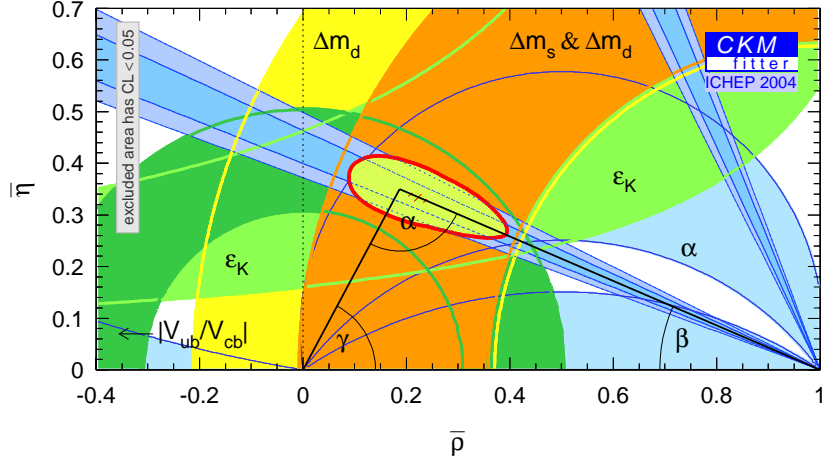}
\caption{Experimental constraints on the SM unitarity triangle
\cite{CKMfitter}.}
\label{fig:UTfit}
\end{figure}
%%%%%%%%%%%%%%%%%%%%%%%%%%%%%%%%%%%%%%%%%%%%%%%%%

The SM mechanism of $\cCP$ violation is based on the unitarity of the
CKM matrix. Testing the constraints implied by unitarity
is then a way to test the source of $\cCP$ violation.
The unitarity tests in Eqs.~\eqn{eq:unitarity_test} and
\eqn{eq:unitarity_test2} involve only the moduli of the CKM parameters,
while $\cCP$ violation has to do with their phases.
More interesting are the off-diagonal unitarity conditions:
\beqn\label{eq:triangles}
\bV^\ast_{\! ud}\bV^{\phantom{*}}_{\! us} \, +\,
\bV^\ast_{\! cd}\bV^{\phantom{*}}_{\! cs} \, +\,
\bV^\ast_{\! td}\bV^{\phantom{*}}_{\! ts} & = & 0 \, ,
\\[5pt]
\bV^\ast_{\! us}\bV^{\phantom{*}}_{\! ub} \, +\,
\bV^\ast_{\! cs}\bV^{\phantom{*}}_{\! cb} \, +\,
\bV^\ast_{\! ts}\bV^{\phantom{*}}_{\! tb} & = & 0 \, ,
\\[5pt]
\bV^\ast_{\! ub}\bV^{\phantom{*}}_{\! ud} \, +\,
\bV^\ast_{\! cb}\bV^{\phantom{*}}_{\! cd} \, +\,
\bV^\ast_{\! tb}\bV^{\phantom{*}}_{\! td} & = & 0 \, .
\eeqn
These relations can be visualized by triangles in a complex
plane which, owing to Eq.~\eqn{eq:J_relation}, have the
same area $|\cJ|/2$.
In the absence of $\cCP$ violation, these triangles would degenerate
into segments along the real axis.

In the first two triangles, one side is much shorter than the other
two (the Cabibbo suppression factors of the three sides are
$\lambda$, $\lambda$ and $\lambda^5$ in the first triangle,
and $\lambda^4$, $\lambda^2$ and $\lambda^2$ in the second one).
This is the reason why $\cCP$ effects are so small for $K$ mesons
(first triangle), and why certain  asymmetries in $B_s$ decays are
predicted to be tiny (second triangle).
The third triangle looks more interesting, since the
three sides have a similar size of about $\lambda^3$.
They are small, which means that the relevant $b$--decay branching ratios
are small, but once enough $B$ mesons have been produced, the $\cCP$--violation
asymmetries are sizeable.
The present experimental constraints on this triangle are shown in
Fig.~\ref{fig:UTfit}, where it has
been scaled dividing its sides by
$\bV^\ast_{\! cb}\bV^{\phantom{*}}_{\! cd}$.
This aligns one
side of the triangle along the real axis and makes its length equal to
1; the coordinates of the 3 vertices are then
$(0,0)$, $(1,0)$ and $(\bar\rho,\bar\eta)\equiv (1-\lambda^2/2) (\rho,\eta)$.

One side of the unitarity triangle has been already determined in
Eq.~\eqn{eq:circle} from the ratio $|\bV_{\! ub}/\bV_{\! cb}|$.
The other side can be obtained from the measured mixing between
the $B^0_d$ and $\bar B^0_d$ mesons, $\Delta M_d$,
which provides information on $|\bV_{\! tb}|$.
A more direct constraint on the parameter $\eta$ is given by
the observed $\cCP$ violation in $K^0\to 2\pi$ decays. The measured
value of $\varepsilon_K$ determines the parabolic region shown
in Fig.~\ref{fig:UTfit}.

$B^0$ decays into $\cCP$ self-conjugate final states
provide independent ways to determine the angles of the unitarity
triangle \cite{CPself}.
The $B^0$ (or $\bar B^0$) can decay directly to the given final state $f$,
or do it after the meson has been changed to its antiparticle via the
mixing process. $\cCP$--violating effects can then result from the interference
of these two contributions.
The time-dependent $\cCP$--violating rate asymmetries
contain direct information on the CKM parameters.
The gold-plated decay mode is $B^0_d\to J/\psi K_S$, which
gives a clean measurement of
$\beta\equiv -\arg(\mathbf{V}^{\phantom{*}}_{\! cd}\mathbf{V}^*_{\! cb}/
\mathbf{V}^{\phantom{*}}_{\! td}\mathbf{V}^*_{\! tb})$,
without strong-interaction uncertainties. Including the information
obtained from other final states, one gets \cite{HFAG:04}:
\begin{equation}\label{eq:beta}
\sin{2\beta} = 0.726\pm 0.037\, .
\end{equation}
%

%%%%%%%%%%%%%%%  FIGURE %%%%%%%%%%%%%%%%%%%%%%%%%
\begin{figure}[tbh]\centering
\includegraphics[width=9cm]{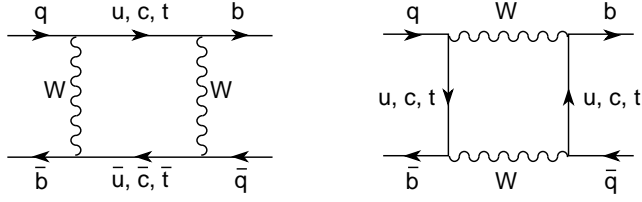}
\caption{$B^0$--$\bar B^0$ mixing diagrams.
Owing to the unitarity of the CKM matrix, the mixing vanishes for equal
up-type quark masses (GIM mechanism). The mixing amplitude is then proportional
to the  mass (squared) splittings between the $u$, $c$ and $t$ quarks,
and is completely dominated by the top contribution.}
\label{fig:Bmixing}
\end{figure}
%%%%%%%%%%%%%%%%%%%%%%%%%%%%%%%%%%%%%%%%%%%%%%%%%

Additional tests of the CKM matrix are underway.
Approximate determinations of the other two angles,
$\alpha\equiv -\arg(
   \mathbf{V}^{\phantom{*}}_{\! td}\mathbf{V}^*_{\! tb}/
   \mathbf{V}^{\phantom{*}}_{\! ud}\mathbf{V}^*_{\! ub})$
and
$\gamma\equiv -\arg(
   \mathbf{V}^{\phantom{*}}_{\! ud}\mathbf{V}^*_{\! ub}/
   \mathbf{V}^{\phantom{*}}_{\! cd}\mathbf{V}^*_{\! cb})$,
from different $B$ decay modes are being pursueded at the
$B$ factories.
Complementary and very valuable information could be also
obtained from the kaon decay modes
$K^\pm\to\pi^\pm\nu\bar\nu$, $K_L\to\pi^0\nu\bar\nu$ and
$K_L\to\pi^0e^+e^-$ \cite{Buras}.

\subsection{Lepton mixing}

The so-called ``solar neutrino problem'' has been a long-standing
question, since the very first chlorine experiment at the Homestake mine
\cite{DA:68}. The flux of solar $\nu_e$ neutrinos reaching the earth
has been measured by several experiments
to be significantly below the standard solar model prediction \cite{BP:04}.
More recently, the Sudbury Neutrino Observatory has provided
strong evidence that neutrinos do change flavour as they propagate
from the core of the Sun \cite{SNO}, independently of solar model
flux predictions. These results have been further reinforced with
the KamLAND data, showing that $\bar\nu_e$ from nuclear reactors
disappear over distances of about 180 Km \cite{KamLAND}.

Another important evidence of neutrino oscillations has been
provided by the Super-Kamiokande measurements of atmospheric neutrinos \cite{SKatm}.
The known discrepancy between the experimental observations and the
predicted ratio of muon to electron neutrinos has become much stronger
with the high precision and large statistics of Super-Kamiokande.
The atmospheric anomaly has been identified to originate in a reduction
of the $\nu_\mu$ flux, and the data strongly favours the
$\nu_\mu\to\nu_\tau$ hypothesis.
This result has been confirmed by K2K \cite{K2K}, observing the
disappearance of accelerator $\nu_\mu$'s at distances of 250 Km,
and will be further studied with the MINOS experiment at Fermilab.
The direct detection of the produced $\nu_\tau$ is the main goal
of the future CERN to Gran Sasso neutrino program.

Thus, we have now clear experimental evidence that neutrinos
are massive particles and there is mixing in the lepton sector.
Figures \ref{fig:SolarNu} and \ref{fig:AtmosNu} show the present
information on neutrino oscillations, from solar and atmospheric
experiments.
A global analysis, combining the full set of solar, atmospheric and reactor
neutrino data, leads to the following preferred ranges for the
oscillation parameters \cite{GO:04}:
\bel{nu_mix1}
7.3 \cdot 10^{-5}\, <\,\Delta m^2_{21}\, /\, \mathrm{eV}^2
\, <\, 9.3 \cdot 10^{-5}\; ,
\qquad
1.6 \cdot 10^{-3}\, <\,\Delta m^2_{32}\, /\, \mathrm{eV}^2
\, <\, 3.6 \cdot 10^{-3} \; ,
\ee
\bel{nu_mix2}
0.28 \, <\,\tan^2{\theta_{12}}\, <\, 0.60 \; ,\qquad
0.5 \, <\,\tan^2{\theta_{23}}\, <\, 2.1 \; ,\qquad
\sin^2{\theta_{13}}\, <\, 0.041 \; ,
\ee
where $\Delta m^2_{ij}\equiv m^2_i - m^2_j$ are the mass squared
differences between the neutrino mass eigenstates $\nu_{i,j}$
and $\theta_{ij}$ the corresponding mixing angles (in the standard
three-flavour parametrization \cite{PDG:04}).
% The angle $\theta_{13}$ is strongly constrained by the CHOOZ reactor
% experiment \cite{CHOOZ}.
In the limit $\theta_{13}=0$, solar and
atmospheric neutrino oscillations decouple because
$\Delta m^2_\odot \ll\Delta m^2_\mathrm{atm}$. Thus,
$\Delta m^2_{21}$, $\theta_{12}$ and $\theta_{13}$ are constrained by
solar data, while atmospheric experiments constrain
$\Delta m^2_{32}$, $\theta_{23}$ and $\theta_{13}$.

%%%%%%%%%%%%%%%  FIGURE %%%%%%%%%%%%%%%%%%%%%%%%%
\begin{figure}[tbh]
\begin{minipage}[c]{.55\linewidth}\centering
\includegraphics[height=5cm]{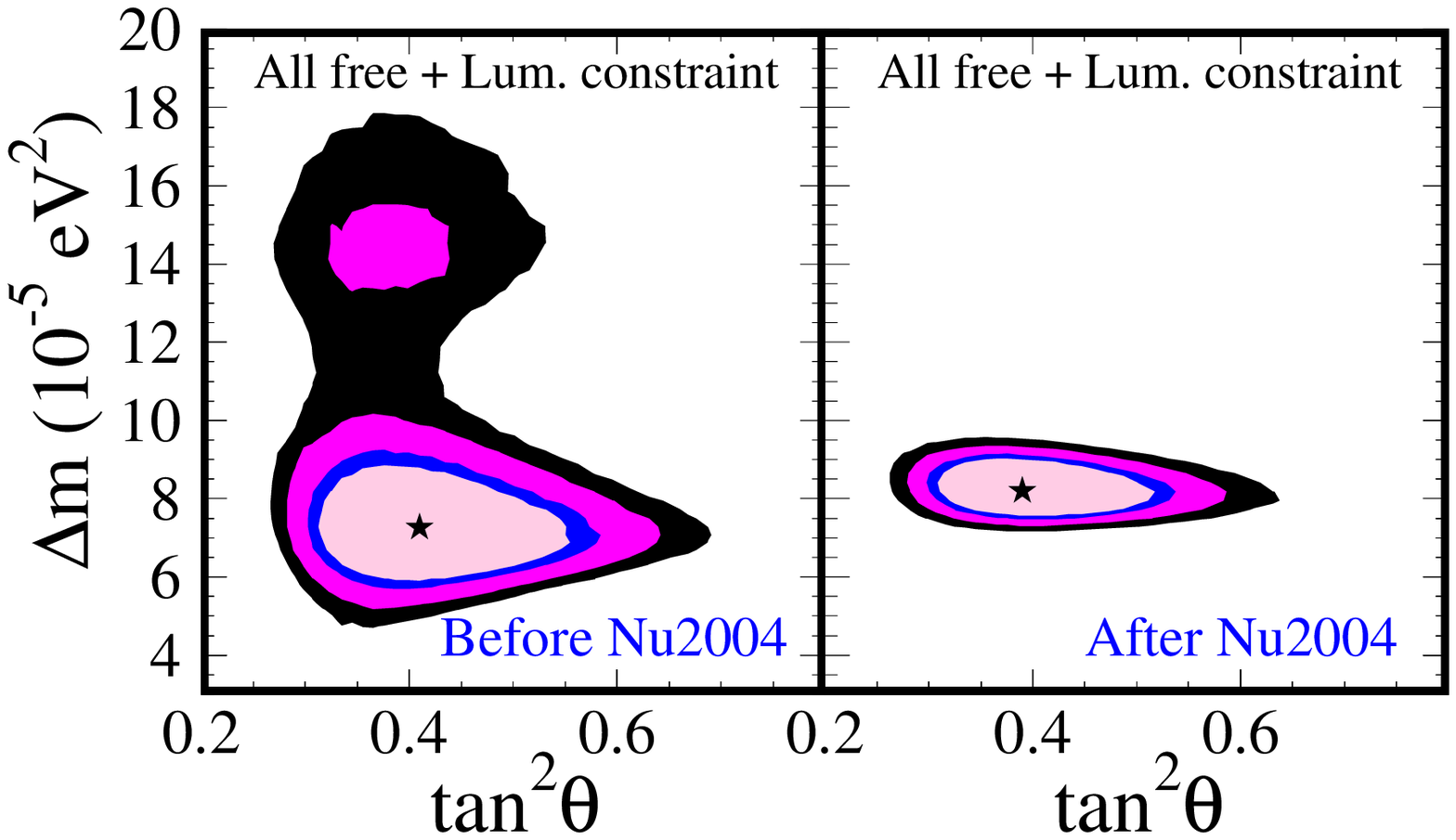}
\caption{Allowed regions for $2\nu$ oscillations for the combination of
solar ($\nu_e$) and KamLAND  ($\bar\nu_e$) data, assuming $\cCPT$ symmetry
\cite{BGP:04}.}
\label{fig:SolarNu}
\end{minipage}
\hfill
\begin{minipage}[c]{.37\linewidth}\centering
\includegraphics[height=5cm]{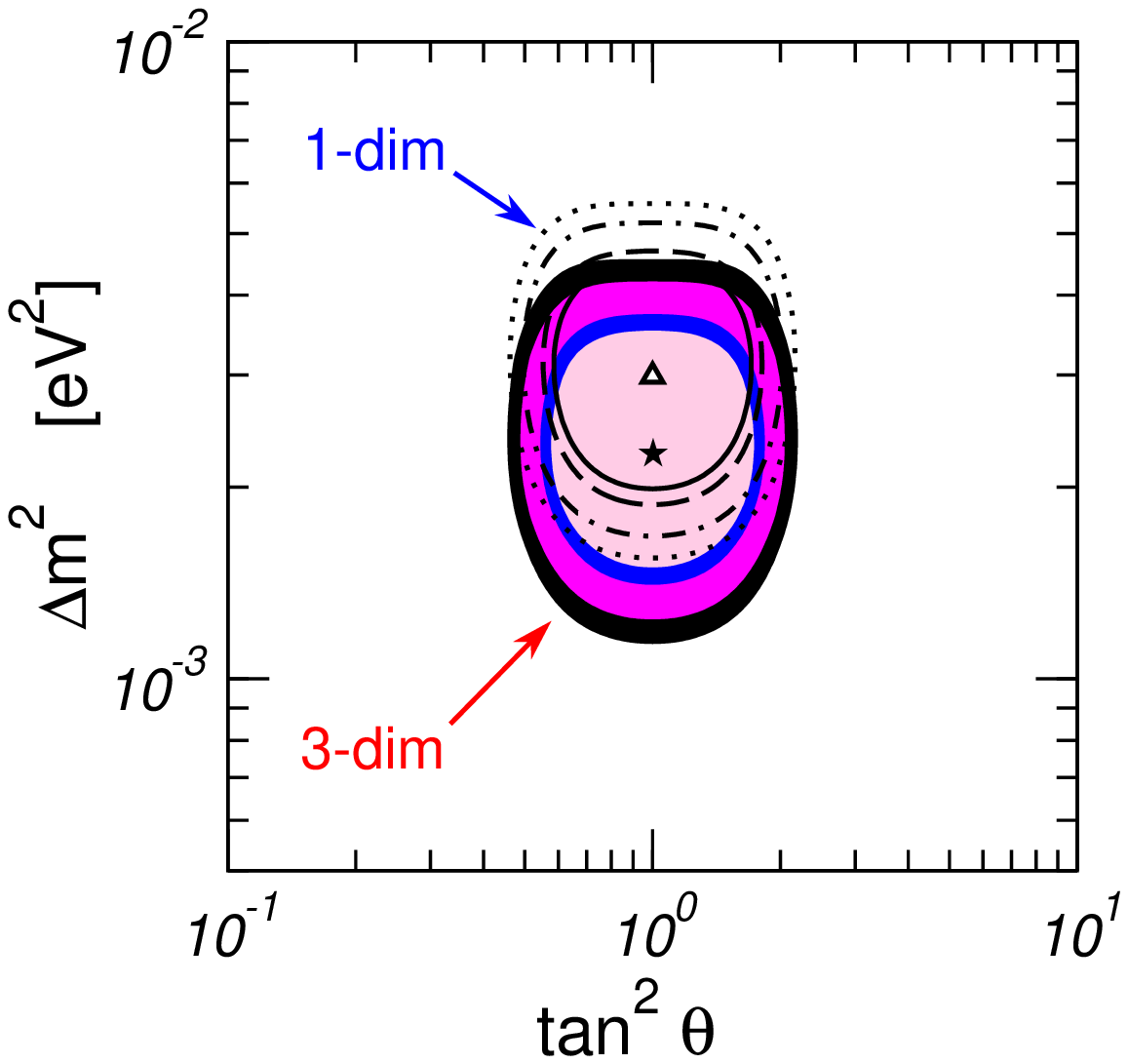}
\caption{Allowed regions for $2\nu$ oscillations
from the analysis of the atmospheric data \cite{GM:04}.}
\label{fig:AtmosNu}
\end{minipage}
\end{figure}
%%%%%%%%%%%%%%%%%%%%%%%%%%%%%%%%%%%%%%%%%%%%%%%%%

Neglecting possible $\cCP$-violating phases, the present data
on neutrino oscillations implies the mixing structure:
\be
\bV^{}_{\! L} \; \approx \; \left[  \begin{array}{ccc}
{1\over\sqrt{2}}\, (1+\lambda) & {1\over\sqrt{2}}\, (1-\lambda) & \epsilon \\[8pt]
-{1\over 2}\, (1-\lambda+\epsilon) & {1\over 2}\, (1+\lambda-\epsilon) &
{1\over\sqrt{2}} \\[8pt]
{1\over 2}\, (1-\lambda-\epsilon) & -{1\over 2}\, (1+\lambda+\epsilon) &
{1\over\sqrt{2}} \end{array}\right] ,
\ee
with $\lambda\sim 0.2$ and $\epsilon < 0.2$ \cite{GO:04}.
Therefore, the mixing among leptons appears to be very different
from the one in the quark sector.

The non-zero value of neutrino masses constitutes a clear indication
of new physics beyond the SM framework.
The simplest modification would be to add the needed right-handed
neutrino components to allow for Dirac neutrino mass terms,
generated through the electroweak SSB
mechanism. However, those $\nu^{}_{iR}$ fields would be
$SU(3)_C\otimes SU(2)_L\otimes U(1)_Y$ singlets and, therefore, would
not have any SM interaction.
If such objects do exist, it would seem natural to expect that they
are able to communicate with the rest of the world through some still
unknown dynamics. Moreover, the SM gauge symmetry
would allow for a right-handed Majorana neutrino mass term,
\bel{eq:Majorana}
\cL_M = -{1\over 2}\, \overline{\nu_{iR}^c}\, M_{ij}\, \nu^{}_{jR}
\, +\,\mathrm{h.c.} \, ,
\ee
where $\nu_{iR}^c\equiv\cC\,\bar\nu_{iR}^T$ denotes the charge-conjugated
field. The Majorana mass matrix $M_{ij}$ could have an arbitrary size,
because it is not related to the ordinary Higgs mechanism.
Moreover, since both fields $\nu^{}_{iR}$ and $\overline{\nu_{iR}^c}$
absorb $\nu$ and create $\bar\nu$,
the Majorana mass term mixes neutrinos and
anti-neutrinos, violating lepton number by two units.
Clearly, new physics is called for.
If the Majorana masses are well above the electroweak
symmetry breaking scale, the see-saw mechanism \cite{RGMY:79}
leads to three light neutrinos at low energies.

In the absence of right-handed neutrino fields, it is still possible
to have non-zero Majorana neutrino masses if the left-handed neutrinos are
Majorana fermions, i.e. with neutrinos equal to antineutrinos.
The number of relevant phases characterizing the lepton mixing
matrix $\bV^{}_{\! L}$ depends on the Dirac or Majorana nature of neutrinos,
because if one rotates a Majorana neutrino by a phase, this phase
will appear in its mass term which will no longer be real.
With only three Majorana (Dirac) neutrinos, the $3\times 3$ matrix $\bV^{}_{\! L}$
involves six (four) independent parameters: three mixing angles
and three (one) phases.

At present, we still ignore whether neutrinos are Dirac
or Majorana fermions.
Another important question to be addressed in the future concerns the
possibility of leptonic $\cCP$ violation and
its relevance for explaining the baryon asymmetry of our universe
through a leptogenesis mechanism.

%%%%%%%%%% SUMMARY %%%%%%%%%%

\setcounter{equation}{0}
\section{SUMMARY}

The SM provides a beautiful theoretical framework which is able to
accommodate all our present knowledge on electroweak and strong
interactions. It is able to explain any single experimental fact
and, in some cases, it has successfully passed very precise
tests at the 0.1\% to 1\% level.
In spite of this impressive phenomenological success, the SM leaves
too many unanswered questions to be considered as a complete
description of the fundamental forces. We do not understand yet
why fermions are replicated in three (and only three)
nearly identical copies. Why the pattern of masses and mixings
is what it is?  Are the masses the only difference among the three
families? What is the origin of the SM flavour structure?
Which dynamics is responsible for the observed $\cCP$ violation?

In the gauge and scalar sectors, the SM Lagrangian contains only
4 parameters: $g$, $\gp$, $\mu^2$ and $h$. We can trade them by
$\alpha$, $M_Z$, $G_F$ and $M_H$; this has the advantage of using the 3
most precise experimental determinations to fix the interaction.
In any case, one describes a lot of physics with only 4 inputs.
In the fermionic flavour sector, however, the situation is very different.
With $N_G=3$, we have 13 additional free parameters in the minimal SM:
9 fermion masses, 3 quark mixing angles and 1 phase. Taking into account
non-zero neutrino masses, we have 3 more mass parameters plus the leptonic
mixings: 3 angles and 1 phase (3 phases) for Dirac (or Majorana) neutrinos.

Clearly, this is not very satisfactory. The source of this proliferation
of parameters is the set of unknown Yukawa couplings in Eq.~\eqn{eq:N_Yukawa}.
The origin of masses and mixings, together with the reason
for the existing family replication, constitute at present 
the main open problem in electroweak physics. The problem of fermion mass
generation is deeply related with the mechanism responsible for the
electroweak SSB.
Thus, the origin of these parameters lies in the most obscure part of
the SM Lagrangian: the scalar sector. 
The dynamics of flavour appears to be ``terra incognita''
which deserves a careful investigation.

The SM incorporates a mechanism to generate $\cCP$ violation, through the
single phase naturally occurring in the CKM matrix.
Although the present laboratory experiments are well described,
this mechanism is unable to explain the
matter-antimatter asymmetry of our universe.
A fundamental explanation of the origin of $\cCP$--violating phenomena is
still lacking.

The first hints of new physics beyond the SM
have emerged recently, with convincing evidence of
neutrino oscillations from both solar and atmospheric experiments,
showing that
$\nu_e\to\nu_{\mu,\tau}$ and $\nu_\mu\to\nu_\tau$
transitions do occur.
The existence of lepton flavour violation opens a very interesting
window to unknown phenomena.

The Higgs particle is the main missing block of the SM framework.
The successful tests of the SM quantum corrections
with precision electroweak data provide a confirmation of the
assumed pattern of SSB, but do not prove
the minimal Higgs mechanism embedded in the SM.
The present experimental bounds \eqn{eq:MH_limits} put the Higgs
hunting within the reach of the new generation of detectors.
LHC should find out whether such scalar field indeed exists,
either confirming the SM Higgs mechanism or
discovering completely new phenomena.

Many interesting experimental signals are expected to be seen
in the near future. New experiments will probe
the SM to a much deeper level of sensitivity
and will explore the frontier of its possible extensions.
Large surprises may well be expected, probably
establishing the existence of new physics beyond the SM and
offering clues to the problems of mass generation, fermion
mixing and family replication.

\vskip1cm
\noindent

\section*{ACKNOWLEDGEMENTS}

I would like to thank the organizers for the charming atmosphere
of this school and all the students for their many interesting
questions and comments.
I am also grateful to Vicent Mateu and Ignasi Rosell for their
useful comments on the manuscript.
This work has been supported
by EU HPRN-CT2002-00311 (EURIDICE),  MEC
(FPA2004-00996) and Generalitat Valenciana
(GRUPOS03/013, GV04B-594).

%%%%%%%%%% APPENDIX %%%%%%%%%%
\newpage

\appendix
\renewcommand{\theequation}{A.\arabic{equation}}
\setcounter{equation}{0}
\section{BASIC INPUTS FROM QUANTUM FIELD THEORY}

\subsection{Wave equations}

The classical Hamiltonian of a non-relativistic free particle is given
by $H = \vec{p}^{\: 2}/(2m)$. In quantum mechanics energy and momentum
correspond to operators acting on the particle wave function. The
substitutions \ $H = i \hbar\, {\partial\over\partial\, t}$ \ and \
$\vec{p} = -i \hbar\,\vec{\nabla}$ \ lead then to the Schr\"odinger
equation:
\bel{eq:Schrodinger}
i \hbar\, {\partial\over\partial t}\, \psi\left(\vec{x},t\right)
\, =\, -{\hbar^2\over 2 m}\, \vec{\nabla}^2\psi\left(\vec{x},t\right)\, .
\ee
We can write the energy and momentum operators in a relativistic covariant
way as \
$p^\mu = i \,\partial^\mu \equiv i \,
{\partial\over\partial x_\mu}$, where we have adopted the usual
natural units convention $\hbar = c = 1$. The relation
$E^{\, 2} =\vec{p}^{\: 2} + m^2$ determines the Klein-Gordon
equation for a relativistic free particle:
\bel{eq:KG}
\left( \Box + m^2 \right) \phi(x) = 0 \qquad\qquad , \qquad\qquad
\Box \equiv \partial^\mu \partial_\mu
= {\partial^2\over\partial t^2} - \vec{\nabla}^2\, .
\ee

The Klein-Gordon equation is quadratic on the time derivative because
relativity puts the space and time coordinates on an equal footing.
Let us investigate whether an equation linear in derivatives could
exist. Relativistic covariance and dimensional analysis restrict
its possible form to
\bel{eq:Dirac}
\left( i\,\gamma^\mu\partial_\mu - m\right) \psi(x) = 0\, .
\ee
Since the r.h.s. is identically zero, we can fix the coefficient of the
mass term to be $-1$; this just determines the normalization of the four
coefficients $\gamma^\mu$. Notice that $\gamma^\mu$ should transform
as a Lorentz four-vector. The solutions of Eq.~\eqn{eq:Dirac} should
also satisfy the Klein-Gordon relation~\eqn{eq:KG}.
Applying an appropriate differential operator to Eq.~\eqn{eq:Dirac},
one can easily obtain the wanted quadratic equation:
\bel{trick}
- \left( i\,\gamma^\nu\partial_\nu + m\right)
\left( i\,\gamma^\mu\partial_\mu - m\right) \psi(x) = 0
\;\equiv \;\left( \Box + m^2 \right) \psi(x)\, .
\ee
Terms linear in derivatives cancel identically, while the term with two
derivatives reproduces the operator $\Box \equiv\partial^\mu \partial_\mu$
\ provided the coefficients $\gamma^\mu$ satisfy the
algebraic relation
\bel{eq:GammaAlgebra}
\left\{\gamma^\mu , \gamma^\nu\right\}\equiv
\gamma^\mu\gamma^\nu + \gamma^\nu\gamma^\mu = 2\, g^{\mu\nu}\, ,
\ee
which defines the so-called Dirac algebra. Eq.~\eqn{eq:Dirac} is known
as the Dirac equation.

Obviously the components of the four-vector $\gamma^\mu$ cannot
simply be numbers.
The three $2\times 2$ Pauli matrices satisfy \
$\left\{\sigma^i ,\sigma^j\right\} = 2 \,\delta^{ij}$, which is very
close to the relation \eqn{eq:GammaAlgebra}. The lowest-dimensional
solution to the Dirac algebra is obtained with \ $D=4$ \ matrices.
An explicit representation is given by:
\bel{eq:DiracMatrices}
\gamma^0 = \left(\bat I_2 & 0\cr 0 & -I_2\ea\right)
\qquad , \qquad
\gamma^i = \left(\bat 0 & \sigma^i\cr -\sigma^i & 0\ea\right)\, .
\ee
Thus, the wave function $\psi(x)$ is a column vector with
4 components in the Dirac space. The presence of the Pauli matrices
strongly suggests that it contains two spin--$\frac{1}{2}$ components.
A proper physical analysis of its solutions shows that the Dirac
equation describes simultaneously a fermion of spin $\frac{1}{2}$ and its
own antiparticle \cite{Bjorken}.

It turns useful to define the following combinations of gamma matrices:
\bel{eq:gamma5}
\sigma^{\mu\nu} \equiv {i\over 2}\,\left[\gamma^\mu , \gamma^\nu\right]
\qquad ,\qquad
\gamma_5 \equiv \gamma^5 \equiv i\,\gamma^0\gamma^1\gamma^2\gamma^3
= -{i\over 4!}\,\epsilon_{\mu\nu\rho\sigma}
\gamma^\mu\gamma^\nu\gamma^\rho\gamma^\sigma\, .
\ee
In the explicit representation \eqn{eq:DiracMatrices},
\bel{eq:sigmaForm}
\sigma^{ij} = \epsilon^{ijk}\,\left(\bat \sigma^k & 0\cr 0 & \sigma^k\ea\right)
\quad , \quad
\sigma^{0i} = i\,\left(\bat 0 & \sigma^i \cr \sigma^i & 0\ea\right)
\quad , \quad
\gamma_5 =\left(\bat 0 & I_2 \cr I_2 & 0\ea\right)
\, .
\ee
The matrix $\sigma^{ij}$ is then related to the spin operator.
Some important properties are:
\bel{eq:Gproperties}
\gamma^0\gamma^\mu\gamma^0 = {\gamma^\mu}^\dagger
\quad ,\quad
\gamma^0\gamma_5\gamma^0 = -{\gamma_5}^\dagger = -\gamma_5
\quad ,\quad
\left\{\gamma_5 , \gamma^\mu\right\} = 0
\quad ,\quad
(\gamma_5)^2 = I_4
\, .
\ee
Specially relevant for weak interactions are the chirality projectors \
($P_L+P_R=1$)
\bel{eq:PLPR}
P_L\equiv {1-\gamma_5\over 2}
\quad ,\quad
P_R\equiv {1+\gamma_5\over 2}
\quad ,\quad
P_R^2 = P_R
\quad ,\quad
P_L^2 = P_L
\quad ,\quad
P_L P_R = P_R P_L = 0
\, ,
\ee
which allow to decompose the Dirac spinor in its left-handed and right-handed
chirality parts:
\bel{eq:chirality}
\psi(x) = \left[ P_L + P_R \right]\, \psi(x)
\equiv \psi_L(x) + \psi_R(x)\, .
\ee
In the massless limit, the chiralities correspond to the fermion
helicities.

\subsection{Lagrangian formalism}

The Lagrangian formulation of a physical system
provides a compact dynamical description
and makes easier to discuss the underlying symmetries. Similarly
to classical mechanics, the dynamics is encoded in the action
\bel{eq:action}
S\, = \int d^{\, 4} x\quad \cL\left[\phi_i(x),\partial_\mu\phi_i(x)\right]\, .
\ee
The integration is over the four space-time coordinates to preserve
relativistic invariance. The Lagrangian density $\cL$ is a
Lorentz-invariant functional of the fields $\phi_i(x)$ and their
derivatives. The space integral \
$L = \int d^{\, 3} x\;\cL$ \ would correspond to the usual non-relativistic
Lagrangian.

The principle of stationary action requires the variation $\delta S$
of the action to be zero under small fluctuations\ $\delta\phi_i$\
of the fields. Assuming that the variations\ $\delta\phi_i$\ are
differentiable and vanish outside some bounded region of space-time
(which allows an integration by parts), the condition $\delta S = 0$
determines the Euler-Lagrange equations of motion for the fields:
\bel{eq:EulerLagrange}
{\partial\cL\over\partial\phi_i}\, - \,\partial^\mu\!\left(
{\partial\cL\over\partial\left(\partial^\mu\phi_i\right)}\right)
\, =\, 0\, .
\ee

One can easily find appropriate Lagrangians to generate the
Klein-Gordon and Dirac equations. They should be quadratic
on the fields and Lorentz invariant, which determines their
possible form up to irrelevant total derivatives.
The Lagrangian
\bel{eq:KGcomplex}
\cL\, =\,\partial^\mu\phi^*\partial_\mu\phi - m^2\, \phi^*\phi
\ee
describes a complex scalar field without interactions.
Both the field $\phi(x)$ and its complex conjugate $\phi^*(x)$
satisfy the Klein-Gordon equation. Thus, $\phi(x)$ describes
a particle of mass $m$ without spin and its antiparticle.
Particles which are their own antiparticles (i.e. with no
internal charges) have only one degree of freedom and are
described through a real scalar field. The appropriate
Klein-Gordon Lagrangian is then
\bel{eq:KGreal}
\cL\, =\,\frac{1}{2}\,\partial^\mu\phi\,\partial_\mu\phi -
\frac{1}{2}\, m^2\, \phi^2\, .
\ee

The Dirac equation can be derived from the Lagrangian density
\bel{eq:DiracL}
\cL\, =\, \overline\psi\,\left(
i\,\gamma^\mu\partial_\mu - m\right) \psi\, .
\ee
The adjoint spinor \ $\overline\psi(x) = \psi^\dagger(x)\,\gamma^0$
closes the Dirac indices. The matrix $\gamma^0$ is included
to guarantee the proper behaviour under Lorentz transformations:
$\overline\psi\psi$ is a Lorentz scalar, while
$\overline\psi\gamma^\mu\psi$ transforms as a four-vector
\cite{Bjorken}. Therefore, $\cL$ is Lorentz invariant as it
should.

Using the decomposition \eqn{eq:chirality} of the Dirac field in its
two chiral components, the fermionic Lagrangian adopts the form:
\bel{eq:DiracL_LR}
\cL\, =\, \overline\psi_L\, i\,\gamma^\mu\partial_\mu\psi_L
\, +\,\overline\psi_R\, i\,\gamma^\mu\partial_\mu\psi_R
\, -\, m\,\left(\overline\psi_L\psi_R +\overline\psi_R\psi_L
\right)\, .
\ee
Thus, the two chiralities decouple if the fermion is massless.

\subsection{Symmetries and conservation laws}

Let us assume that the Lagrangian of a physical system is invariant
under some set of continuous transformations
\bel{eq:PhiTransf}
\phi_i(x)\;\to\; \phi'_i(x) = \phi_i(x) + \epsilon\:\delta_\epsilon\phi_i(x)
+ O(\epsilon^2)\, ,
\ee
i.e. \ $\cL\left[\phi_i(x),\partial_\mu\phi_i(x)\right]
=\cL\left[\phi'_i(x),\partial_\mu\phi'_i(x)\right]$.
One finds then that
\bel{eq:VarS}
\delta_\epsilon\cL\; =\; 0\; =\;
\sum_i\left\{\left[
{\partial\cL\over\partial\phi_i}\, - \,\partial^\mu\!\left(
{\partial\cL\over\partial\left(\partial^\mu\phi_i\right)}\right)
\right]\delta_\epsilon\phi_i
\, +\, \partial^\mu\left[
{\partial\cL\over\partial\left(\partial^\mu\phi_i\right)}\,
\delta_\epsilon\phi_i\right]\right\}\, .
\ee
If the fields satisfy the Euler-Lagrange equations of motion
\eqn{eq:EulerLagrange}, the first term is identically zero.
Therefore, the system has a conserved current:
\bel{eq:Noether}
J_\mu \equiv\sum_i
{\partial\cL\over\partial\left(\partial^\mu\phi_i\right)}
\,\delta_\epsilon\phi_i
\qquad\qquad , \qquad\qquad
\partial^\mu J_\mu = 0\, .
\ee
This allows us to define a conserved charge
\bel{eq:NoetherCharge}
\cQ \equiv \int d^{\, 3} x\; J^0\, .
\ee
The condition \ $\partial^\mu J_\mu = 0$ \ guarantees that \
${d\cQ\over dt} = 0\,$, i.e. that $\cQ$ is a constant of motion.

This result, known as Noether's theorem, can be easily extended to
general transformations involving also the space-time
coordinates. For every continuous symmetry transformation which
leaves the Lagrangian invariant, there is a corresponding divergenceless
Noether's current and, therefore, a conserved charge.
The selection rules observed in nature, where there exist several
conserved quantities (energy, momentum, angular momentum,
electric charge, \ldots), correspond to dynamical symmetries
of the Lagrangian.

\subsection{Classical electrodynamics}

The well-known Maxwell equations,
\beqn\label{eq:Maxwell_1}
\vec{\nabla}\cdot\vec{B} = 0
\qquad\qquad &,& \qquad\qquad
\vec{\nabla}\times\vec{E} + {\partial\vec{B}\over\partial\, t} = 0\, ,
\\[3pt] \label{eq:Maxwell_2}
\vec{\nabla}\cdot\vec{E} = \rho
\qquad\qquad &,& \qquad\qquad
\vec{\nabla}\times\vec{B} - {\partial\vec{E}\over\partial\, t} = \vec{J}\, ,
\eeqn
summarize a big amount of experimental and theoretical work and provide
a unified description of the electric and magnetic forces.
The first two equations in \eqn{eq:Maxwell_1} are easily solved, writing
the electromagnetic fields in terms of potentials:
\bel{eq:potentials}
\vec{E} = -\vec{\nabla} V - {\partial\vec{A}\over\partial\, t}
\qquad\qquad , \qquad\qquad
\vec{B} = \vec{\nabla}\times\vec{A}\, .
\ee

It is very useful to rewrite these equations in a Lorentz covariant
notation. The charge density $\rho$ and the electromagnetic current $\vec{J}$
transform as a four-vector \ $J^\mu \equiv\left(\rho,\vec{J}\right)$.
The same is true for the potentials which combine into \
$A^\mu \equiv\left(V,\vec{A}\right)$.
% $V$ and $\vec{A}$:
% %
% \bel{eq:4-potentials}
% J^\mu \equiv\left(\rho,\vec{J}\right)
% \qquad\qquad , \qquad\qquad
% A^\mu \equiv\left(V,\vec{A}\right)\, .
% \ee
% %
The relations \eqn{eq:potentials} between the potentials and the fields
take then a very simple form, which defines the field strength tensor
\bel{eq:Fmunu}
F^{\mu\nu}\equiv\partial^\mu A^\nu-\partial^\nu A^\mu =
\left(\begin{array}{cccc}
0 & -E_1 & -E_2 & -E_3 \\
E_1 & 0 & -B_3 & B_2 \\
E_2 & B_3 & 0 & -B_1 \\
E_3 & -B_2 & B_1 & 0
\ea\right)
\qquad , \qquad
\tilde{F}^{\mu\nu}\equiv\frac{1}{2}\, \epsilon^{\mu\nu\rho\sigma}\,
F_{\rho\sigma}\, .
\ee
In terms of the tensor $F^{\mu\nu}$,
the covariant form of Maxwell equations turns out to be very transparent:
\bel{eq:Maxwell_Cov}
\partial_\mu \tilde{F}^{\mu\nu} = 0
\qquad\qquad , \qquad\qquad
\partial_\mu F^{\mu\nu} = J^\nu\, .
\ee
The electromagnetic dynamics is clearly a relativistic phenomena, but Lorentz
invariance was not very explicit in the original formulation of
Eqs.~\eqn{eq:Maxwell_1} and \eqn{eq:Maxwell_2}. Once a covariant formulation
is adopted, the equations become much simpler.
The conservation of the electromagnetic current appears now as a natural
compatibility condition:
\bel{eq:CurrentCons}
\partial_\nu J^\nu = \partial_\nu\partial_\mu F^{\mu\nu} = 0\, .
\ee
In terms of potentials, $\partial_\mu \tilde{F}^{\mu\nu}$ is identically zero
while $\partial_\mu F^{\mu\nu} = J^\nu$ adopts the form:
\bel{eq:Maxwell_Cov2}
\Box\, A^\nu -\partial^\nu\left(\partial_\mu A^\mu\right) = J^\nu\, .
\ee

The same dynamics can be described by many different electromagnetic
four-potentials, which give the same field strength tensor $F^{\mu\nu}$.
Thus, Maxwell equations are invariant under gauge transformations:
\bel{eq:GaugeTransf}
A^\mu\;\longrightarrow\; A'^\mu = A^\mu + \partial^\mu\Lambda\, .
\ee
Taking the {\it Lorentz gauge} \ $\partial_\mu A^\mu = 0$,
Eq.~\eqn{eq:Maxwell_Cov2} simplifies to
\bel{eq:Maxwell_Cov3}
\Box\, A^\nu = J^\nu\, .
\ee
In the absence of an external current, i.e. with $J^\mu = 0$, the
four components of $A^\mu$ satisfy then a Klein-Gordon
equation with $m= 0$.
The photon is therefore a massless particle.

The Lorentz condition \ $\partial_\mu A^\mu = 0$ \ still allows for
a residual gauge invariance under transformations of the type
\eqn{eq:GaugeTransf}, with the restriction \ $\Box\,\Lambda = 0$.
Thus, we can impose a second constraint on the electromagnetic field
$A^\mu$, without changing $F^{\mu\nu}$. Since $A^\mu$ contains
four fields ($\mu = 0, 1, 2, 3$) and there are two arbitrary
constraints, the number of physical degrees of freedom is just two.
Therefore, the photon has two different physical polarizations

\renewcommand{\theequation}{B.\arabic{equation}}
\setcounter{equation}{0}
\section{SU(N) \ Algebra}

$SU(N)$ is the group of\ $N\times N$ unitary matrices,
$U U^\dagger = U^\dagger U =1$, with \ $\det U=1$.
Any $SU(N)$ matrix can be written in the form
\bel{eq:Uexp}
U = \exp{\left\{i\, T^a \theta_a\right\}}
\qquad\qquad , \qquad\qquad a=1,2,\ldots,N^2-1\, ,
\ee
with \ $T^a = \lambda^a/2$ \ hermitian,
traceless matrices. Their commutation relations
\bel{eq:T_com}
[T^a, T^b] \, = \, i\, f^{abc}\, T^c
\ee
define the $SU(N)$ algebra.
The $N\times N$ matrices $\lambda^a/2$ generate the fundamental
representation of the $SU(N)$ algebra.
The basis of generators $\lambda^a/2$ can be chosen so that
the structure constants $f^{abc}$ are real and totally antisymmetric.

For $N=2$,\ $\lambda^a$ are the usual Pauli matrices,
\bel{eq:pauli}
\sigma_1=\left(\bat  0 & 1 \\ 1 & 0 \ea\right) , \qquad
  \sigma_2=\left(\bat 0 & -i \\ i & 0 \ea\right) , \qquad
  \sigma_3=\left(\bat 1 & 0 \\ 0 & -1 \ea\right)\, ,
\ee
which satisfy the commutation relation
\bel{eq:SU(2)}
\left[ \sigma_i,\sigma_j\right]  =  2\, i \,\epsilon_{ijk}\,\sigma_k \, .
\ee
Other useful properties are: \
$\left\{ \sigma_i,\sigma_j\right\}  =  2\,\delta_{ij}$ \
and \ $\mbox{\rm Tr}\left(\sigma_i\sigma_j\right)  =  2\,\delta_{ij}$.

For $N=3$, the fundamental representation
corresponds to the eight Gell-Mann matrices:
\beqn\label{eq:GM_matrices}
\lambda^1 =\left( \bath 0 & 1 & 0 \\ 1 & 0 & 0 \\ 0 & 0 & 0 \ea \right) ,
\quad\,
\lambda^2 & = &
\left( \bath 0 & -i & 0 \\ i & 0 & 0 \\ 0 & 0 & 0 \ea \right) ,
\quad
\lambda^3 = \left( \bath 1 & 0 & 0 \\ 0 & -1 & 0 \\ 0 & 0 & 0 \ea \right) ,
\quad
\lambda^4 = \left( \bath 0 & 0 & 1 \\ 0 & 0 & 0 \\ 1 & 0 & 0 \ea \right) ,
 \no\\   && \\
\lambda^5 =
\left( \bath 0 & 0 & -i \\ 0 & 0 & 0 \\ i & 0 & 0 \ea \right) \! ,
\;\;
\lambda^6 & = &
\left( \bath 0 & 0 & 0 \\ 0 & 0 & 1 \\ 0 & 1 & 0 \ea \right)\! , \;\;
\lambda^7 = \left( \bath 0 & 0 & 0 \\ 0 & 0 & -i \\ 0 & i & 0 \ea \right)
\! , \;\;
\lambda^8 = 
{1\over\sqrt{3}}
\left( \bath 1 & 0 & 0 \\ 0 & 1 & 0 \\ 0 & 0 & -2 \ea \right) \! . \no
\eeqn
They satisfy the anticommutation relation
\bel{eq:anticom}
\left\{\lambda^a,\lambda^b\right\} \, = \,
{4\over N} \,\delta^{ab} \, I_N \,
+ \, 2\, d^{abc} \, \lambda^c \, ,
\ee
where $I_N$ denotes the $N$--dimensional unit matrix and the constants
$d^{abc}$ are totally symmetric in the three indices.

For $SU(3)$, the only non-zero (up to permutations)
$f^{abc}$ and $d^{abc}$ constants are
\beqn\label{eq:constants}
&&{1\over 2}\, f^{123} = f^{147} = - f^{156} = f^{246} = f^{257} = f^{345}
= - f^{367} 
= {1\over\sqrt{3}}\, f^{458} = {1\over\sqrt{3}}\, f^{678} =
{1\over 2}\, ,
\CR
&&d^{146} = d^{157} = -d^{247} = d^{256} = d^{344} =
d^{355} = -d^{366} = - d^{377} = {1\over 2}\, ,  
\\
&&d^{118} = d^{228} = d^{338} = -2\, d^{448} = -2\, d^{558} = -2\, d^{668} =
-2\, d^{778} = -d^{888} = {1\over \sqrt{3}}\, .
\no
\eeqn

The adjoint representation of the $SU(N)$ group is given by
the $(N^2-1)\!\times\! (N^2-1)$ matrices
$(T^a_A)_{bc} \equiv - i f^{abc}$.
The relations
\beqn\label{eq:invariants}
{\rm Tr}\left(\lambda^a\lambda^b\right) =  4 \, T_F \, \delta_{ab}
\qquad\qquad\quad ,\quad\quad\quad && T_F = {1\over 2} \, ,
\no\\
\left(\lambda^a\lambda^a\right)_{\alpha\beta} = 4\,
C_F \, \delta_{\alpha\beta} \qquad\qquad\quad ,\quad\quad\quad
&& C_F = {N^2-1\over 2N} \, ,
\\[5pt] \;
{\rm Tr}(T^a_A T^b_A) = f^{acd} f^{bcd} = C_A \,\delta_{ab}
\qquad\quad ,\quad\qquad
&& C_A = N \, , \qquad\no
\eeqn
define the $SU(N)$ invariants $T_F$, $C_F$ and $C_A$.
Other useful properties are:
$$
\left(\lambda^a\right)_{\alpha\beta}
\left(\lambda^a\right)_{\gamma\delta} 
= 2 \,\delta_{\alpha\delta}\delta_{\beta\gamma}
 -{2\over N} \,\delta_{\alpha\beta}\delta_{\gamma\delta}
\qquad ,\qquad
{\rm Tr}\left(\lambda^a\lambda^b\lambda^c\right)
 =  2\, (d^{abc} + i f^{abc})\, ,
$$
\be
{\rm Tr}(T^a_A T^b_A T^c_A)  =  i \, {N\over 2}\, f^{abc}
\qquad ,\qquad \sum_b d^{abb} = 0\qquad , \qquad
d^{abc} d^{ebc}  =  \left( N - {4\over N}\right) \delta_{ae} \, ,
\ee
$$
f^{abe} f^{cde} + f^{ace} f^{dbe} + f^{ade} f^{bce} = 0
\qquad , \qquad
f^{abe} d^{cde} + f^{ace} d^{dbe} + f^{ade} d^{bce} = 0 \, .
$$

\renewcommand{\theequation}{C.\arabic{equation}}
\setcounter{equation}{0}
\section{ANOMALIES}
\label{sec:anomalies}

%%%%%%%%%%%%%%%
\begin{figure}[htb]\centering
\includegraphics[width=9cm]{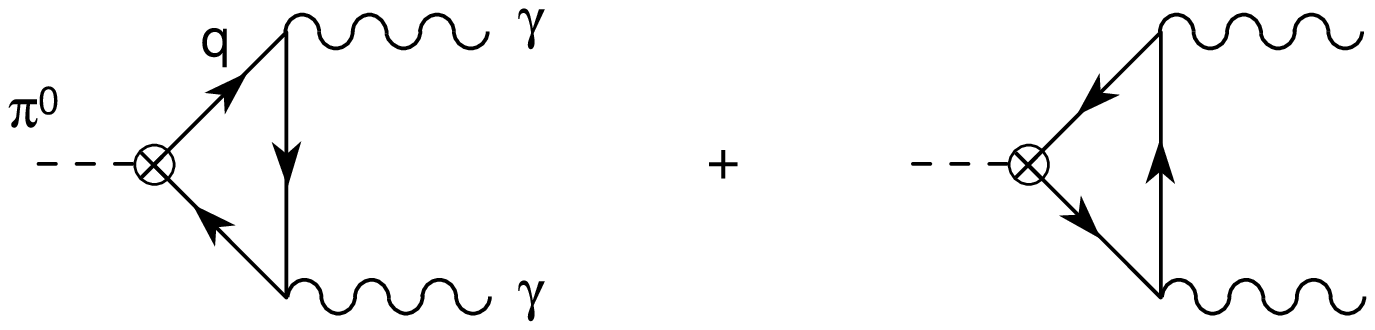}
\caption{Triangular quark loops generating the decay $\pi^0\to\gamma\gamma$.}
\label{fig:triangle}
\end{figure}
%%%%%%%%%%%%%%%

Our theoretical framework is based on the local gauge symmetry.
However, we have only
discussed so far the symmetries of the classical Lagrangian.
It happens sometimes that a symmetry of $\cL$ gets broken by quantum effects,
i.e. it is not a symmetry of the quantized theory;
one says then that there is an ``anomaly''.
Anomalies appear in those symmetries involving both axial
($\overline{\psi}\gamma^\mu\gamma_5\psi$) and vector
($\overline{\psi}\gamma^\mu\psi$)
currents, and reflect the impossibility of regularizing the quantum
theory (the divergent loops) in a way which preserves the chiral
(left / right) symmetries.

A priori there is nothing wrong with having an anomaly. In fact, sometimes
they are even welcome. A good example is provided by the decay
$\pi^0\to\gamma\gamma$. There is a chiral symmetry of the QCD Lagrangian
which forbids this transition; the $\pi^0$ should then be a stable particle,
in contradiction with the experimental evidence.
Fortunately, there is an anomaly generated by a triangular quark loop
which couples the axial current
$A_\mu^3\equiv (\bar u\gamma_\mu\gamma_5 u - \bar d\gamma_\mu\gamma_5 d)$
to two electromagnetic currents
and breaks the conservation of the axial current at the quantum level:
\bel{eq:div_A}
\partial^\mu A_\mu^3 \,=\,  {\alpha\over 4\pi}\,
\epsilon^{\alpha\beta\sigma\rho}\,
F_{\alpha\beta}
\, F_{\sigma\rho} \, + \, \cO\left(m_u + m_d\right) .
\ee
Since the $\pi^0$ couples to $A_\mu^3\,$, \
$\langle 0|A_\mu^3|\pi^0\rangle = 2\, i\, f_\pi\, p_\mu\,$,
the $\pi^0\to\gamma\gamma$
decay does finally occur, with a predicted rate
\bel{eq:pi_decay}
\Gamma(\pi^0\to\gamma\gamma)\, = \, \left({N_C\over 3}\right)^2
{\alpha^2 m_\pi^3\over 64\pi^3 f_\pi^2}
\, =\, 7.73\, \mbox{\rm eV} ,
\ee
where $N_C=3$ denotes the number of quark colours and $f_\pi=92.4$~MeV
is known from the $\pi^-\to\mu^-\bar\nu_\mu$ decay rate
(assuming isospin symmetry).
The agreement with the measured value, $\Gamma = 7.7\pm 0.6$ eV
\cite{PDG:04}, is excellent.

Anomalies are, however, very dangerous in the case of local gauge symmetries,
because they destroy the renormalizability of the Quantum Field Theory.
Since the $SU(2)_L\otimes U(1)_Y$ model is chiral (i.e. it distinguishes
left from right), anomalies are clearly present.
The gauge bosons couple to vector and axial-vector currents; we can then
draw triangular diagrams
%like the one in Fig.~\ref{fig:anomaly}, but
with
three arbitrary gauge bosons ($W^\pm$, $Z$, $\gamma$) in the external legs.
Any such diagram involving one axial and two vector currents generates
a breaking of the gauge symmetry.
Thus, our nice model looks meaningless at the quantum level.

We have still one way out. What matters is not the value of a single
Feynman diagram, but the sum of all possible contributions.
The anomaly generated by the sum of all triangular diagrams
connecting the three gauge bosons $G_a$, $G_b$ and $G_c$ is proportional
to
\bel{eq:an_condition}
\cA\, = \, \mbox{\rm Tr}\left( \{ T^a , T^b \}\, T^c \right)_L -
 \mbox{\rm Tr}\left( \{ T^a , T^b \}\, T^c \right)_R,
\ee
where the traces sum over all possible left- and right-handed
fermions, respectively, running along the internal lines
of the triangle.
The matrices $T^a$ are the generators associated with the corresponding
gauge bosons; in our case, $T^a = \sigma_a/2\, ,\, Y$.

In order to preserve the gauge symmetry, one needs a cancellation of all
anomalous contributions, i.e. $\cA=0$.
Since $\mbox{\rm Tr}(\sigma_k)=0$, we have an automatic cancellation in two
combinations of generators: \
$\mbox{\rm Tr}\left(\{ \sigma_i , \sigma_j \}\, \sigma_k \right)=2\,\delta^{ij}\,
\mbox{\rm Tr}(\sigma_k)=0\, $ \ and \
$\mbox{\rm Tr}\left(\{ Y , Y\}\, \sigma_k \right)\propto\mbox{\rm Tr}(\sigma_k)=
0\,$.
However, the other two combinations, \
$\mbox{\rm Tr}\left(\{ \sigma_i , \sigma_j \}\, Y \right)$ \ and \
$\mbox{\rm Tr}(Y^3)$ \ turn out to be proportional to \
$\mbox{\rm Tr}(Q)\,$, i.e. to the sum of fermion electric charges:
\bel{eq:an_cancellation}
\sum_i Q_i \, = \, Q_e + Q_\nu + N_C \left( Q_u + Q_d \right) \, = \,
-1 + {1\over 3} N_C \, =\, 0\, .
\ee

Eq.~\eqn{eq:an_cancellation} is telling us a very important message: the gauge
symmetry of the $SU(2)_L\otimes U(1)_Y$ model does not have any quantum
anomaly, provided that $N_C=3$.
Fortunately, this is precisely the right number of colours to understand
strong interactions.
Thus, at the quantum level, the electroweak model seems to know something
about QCD. The complete SM gauge theory based on the group
$SU(3)_C\otimes SU(2)_L\otimes U(1)_Y$ is free of anomalies and, therefore,
renormalizable.
The anomaly cancellation involves one complete generation of leptons
and quarks: $\nu\, ,\, e\, ,\, u\, ,\, d$. The SM wouldn't make any sense
with only leptons or quarks.

%%%%%%%%%% REFERENCES %%%%%%%%%%

\end{document}